# TAXONOMY AND LIGHT-CURVE DATA OF 1000 SERENDIPITOUSLY OBSERVED MAIN-BELT ASTEROIDS

N. Erasmus,[1] A. McNeill,[2] M. Mommert,[3,2] D. E. Trilling,[2,1] A. A. Sickafoose,[1,4] and C. van Gend[1]

[1] South African Astronomical Observatory, Cape Town, 7925, South Africa.
[2] Department of Physics and Astronomy, Northern Arizona University, Flagstaff, AZ 86001, USA.
[3] Lowell Observatory, Flagstaff, AZ 86001, USA.
[4] Department of Earth, Atmospheric, and Planetary Sciences, Massachusetts Institute of Technology, Cambridge, MA 02139-4307, USA.

## ABSTRACT

We present *VRI* spectrophotometry of 1003 Main-Belt Asteroids (MBAs) observed with the Sutherland, South Africa, node of the Korea Microlensing Telescope Network (KMTNet). All of the observed MBAs were serendipitously captured in KMTNet's large $2° \times 2°$ field of view during a separate targeted near-Earth Asteroid study (Erasmus et al. 2017). Our broadband spectrophotometry is reliable enough to distinguish among four asteroid taxonomies and we confidently categorize 836 of the 1003 observed targets as either a S-, C-, X-, or D-type asteroid by means of a Machine Learning (ML) algorithm approach. Our data show that the ratio between S-type MBAs and (C+X+D)-type MBAs, with $H$ magnitudes between 12 and 18 (12 km $\gtrsim$ diameter $\gtrsim$ 0.75 km), is almost exactly 1:1. Additionally, we report 0.5- to 3-hour (median: 1.3-hour) light-curve data for each MBA and we resolve the complete rotation periods and amplitudes for 59 targets. Two out of the 59 targets have rotation periods potentially below the theoretical zero cohesion boundary limit of 2.2 hours. We report lower limits for the rotation periods and amplitudes for the remaining targets. Using the resolved and unresolved light curves we determine the shape distribution for this population using a Monte Carlo simulation. Our model suggests a population with an average elongation $b/a = 0.74 \pm 0.07$ and also shows that this is independent of asteroid size and taxonomy.

*Keywords:* minor planets, asteroids: individual (Main-Belt Asteroids) — techniques: photometric — surveys

Corresponding author: Nicolas Erasmus
nerasmus@saao.ac.za

# 1. INTRODUCTION

Main-Belt Asteroids (MBAs) are the most abundant designated objects in the Solar System, representing approximately 95% of all bodies listed in the Minor Planet Center (MPC)[1] catalog. There are estimated to be $1 \times 10^6$ asteroids in the main belt with a diameter larger than 1 km (Bottke et al. 2005). In this work we present a series of spectrophotometric light curves, both partial and full, for around 1000 MBAs of comparable diameter. In principle, this work is a successor to the Thousand Asteroid Light Curve Survey (Masiero et al. 2009) but with simultaneous taxonomy determination included. Similar to Masiero et al. (2009), by conducting an untargeted survey we have obtained photometry for a sample of MBAs that will be unbiased against objects with low-amplitude light curves or long rotation periods.

Photometric data is the most abundant data available for minor planets and through analysis of this it is possible to gain insight into the rotational behavior and shape properties of asteroids. The Asteroid Light Curve Database (LCDB; Warner et al. 2009, Updated 2017 September 11)[2] contains 16,837 MBA light curves with 3201 having accurately determined and unambiguous rotation periods. With sufficient observations at multiple observing geometries it is possible to estimate the shape and spin pole of the objects by light curve inversion (Kaasalainen & Torppa 2001). With sparse photometry or partial light curves this is much more difficult. Instead, statistical models are often used to determine the shape distribution of a population as a whole where sparse photometry is available, rather than attempting to determine the shape properties of an individual object. Previous work on this topic was carried out in McNeill et al. (2016) and Cibulková et al. (2017).

Asteroid composition can be remotely probed with spectroscopy. The reflected light spectrum of an asteroid can be compared to the reflectance spectra of known collected meteorite types and in this way the composition of asteroids can be inferred. The most widely used spectral system is the Bus-DeMeo taxonomic scheme (DeMeo et al. 2009) consisting of 24 spectral types and sub-types that are defined in the wavelength range of 0.45–2.45 $\mu$m. Although there are many Bus-DeMeo taxonomic classes, MBAs fall mainly in two groups: silicon-rich asteroids with a distinct spectral feature, which are classified as S-type asteroids, and those that have featureless flat or reddish spectral slopes in the visible region, which could be one of several classes. Asteroids with featureless spectra are most commonly C-type (carbonaceous) asteroids, but also include X-type (several different possible compositions) asteroids, and lastly D-type (even more primitive material) asteroids.

In some cases spectrophotometry where the reflectance spectra of an asteroid is sampled at strategic wavelengths using broadband filters is adequate to determine the spectral. Therefore spectrophotometry can not ascertain asteroid composition but still has the ability to distinguish between some of the taxonomic classes. For instance, *VRI* spectrophotometry in the visible region is sufficient to distinguish between spectral slopes of S-, X- and C- and D-type Bus-DeMeo taxonomic classes (Erasmus et al. 2017). This is the method used in this study to determine each MBA's likely taxonomy. Spectrophotometry has the benefit of being employable to investigate fainter asteroids. This significantly increases the observable population numbers since the size distribution is heavily skewed towards smaller asteroids.

Here we present *VRI* spectrophotometry observations and 0.5- to 3-hour light-curve data for 1003 MBAs. In Section 2 the observational details, the method for extracting serendipitously observed MBAs, and the photometry pipeline are explained. In Section 3 the photometric results of the 1003 observed MBAs are presented. Section 4 describes how the color, classification, and light curve of each MBA was determined from the photometric data. In Section 5 we give detail on the shape-modeling procedure. Sections 6 and 7 conclude this study by discussing the analyzed data and the derived shape distribution, comparing to other published work, and presenting our conclusions.

# 2. OBSERVATIONS AND DATA REDUCTION

Observations were made with the Sutherland, South Africa, node of the Korea Microlensing Telescope Network (KMTNet) (Kim et al. 2016). The telescope has a primary mirror 1.6 m in diameter and is fitted with four 9k $\times$ 9k CCDs, mosaicking the $2^\circ \times 2^\circ$ field of view. Each CCD covers $1^\circ \times 1^\circ$ of sky with a plate-scale of 0.40 arcsec/pixel. Observations were performed alternating among $V$ ($\lambda = 550$ nm, $\Delta\lambda = 90$ nm), $R$ ($\lambda = 660$ nm, $\Delta\lambda = 140$ nm), and $I$ ($\lambda = 805$ nm, $\Delta\lambda = 150$ nm) filters in the sequence *VRVI*, repeating the sequence continuously for the entire observing duration. The exposure time for each filter was 60 seconds, which limited this study to MBAs with magnitudes $V \lesssim 21$ to achieve the desired signal-to-noise ratio (SNR) of at least 10 in a single exposure.

[1] http://minorplanetcenter.net/

[2] http://www.MinorPlanet.info/lightcurvedatabase.html

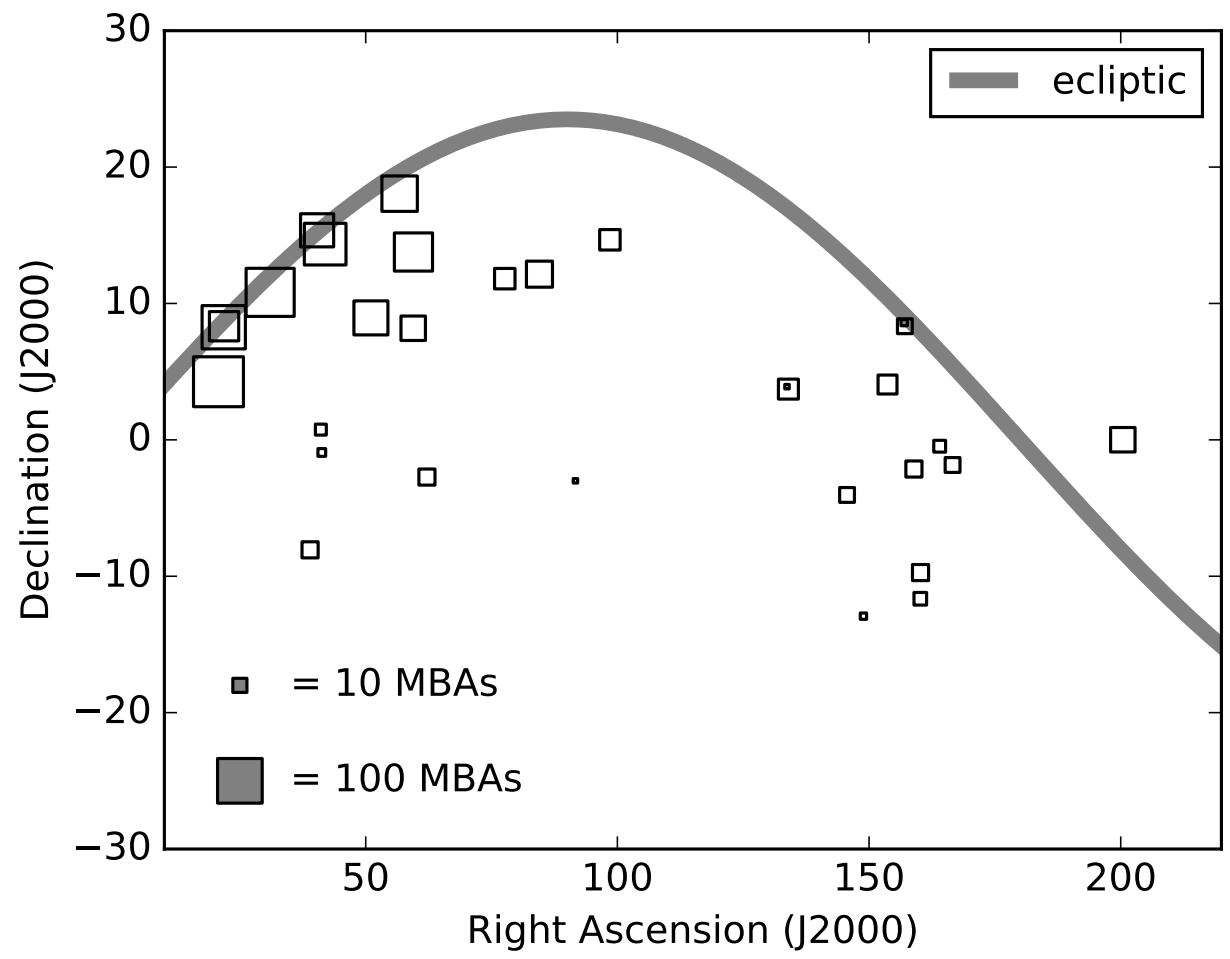

**Figure 1.** Sky-coordinates of all observed KMTNet fields used in this study (hollow squares). The size of the square is proportional to the number of serendipitous MBAs with acceptable SNR extracted in each field. As expected, fields observed on or close to the ecliptic yielded the highest number of MBAs, though some fields on the ecliptic only yielded a few MBAs because unfavorable observing conditions for that specific field meant a large number of MBAs had to be discarded.

Observations were originally performed for a near-Earth asteroid study that took place between 25 October 2016 – 20 February 2017 over four observing weeks (Erasmus et al. 2017). However, the large field of view of KMTNet meant that the 31 fields observed in that study also included many serendipitously observed MBAs. The squares in Figure 1 show the location of the 31 observed fields with the size of the square proportional to the number of MBAs extracted from each field.

To extract all serendipitously observed MBAs in the observed fields, the ephemerides for all ∼700,000 known asteroids in the MPC's database were calculated at the respective mid-time of each observation. Ephemerides were calculated using PyEphem[3] and any MBA falling within the observed field's coordinate perimeter was flagged. Flagged asteroids were queried with JPL's Horizons service[4] using the `Python` module CALLHORIZONS[5] and discarded if $V > 21$. In total ∼2000 serendipitous MBAs were captured and 1003 of them produced spectrophotometry data with acceptable data quality (SNR $\geq$ 10 in all bands).

Photometry and light-curve extraction was achieved using PHOTOMETRYPIPELINE (PP) developed by Mommert (2017). The pipeline utilizes the widely used Source Extractor software (Bertin & Arnouts 1996) for source identification and aperture photometry. SCAMP (Bertin 2006) is used for image registration. Both image registration and photometric calibration are based on matching field stars with star catalogs from the Sloan Digital Sky Survey, the AAVSO Photometric All-Sky Survey, Pan-STARRS, and GAIA.

## 3. RESULTS

Figure 2 shows the photometric data for nine of the 1003 observed MBAs (see the electronic edition of Figure 2 for all observed MBAs. For reviewer: see attached zipped file). We show the calibrated photometric results generated by PP from the $V$-, $R$-, and $I$-filter images. Adjusted $R$ and $I$ data are also shown (inter-spaced between $V$ data), which results in a final light curve (bottom set of data in each plot). Sections 4.1 & 4.3 describe details of the process for adjusting the $R$ and $I$ data points. Photometric uncertainties account for instrumental uncertainties as well as statistical uncertainties when compared to the photometric catalog.

[3] https://github.com/brandon-rhodes/pyephem
[4] https://ssd.jpl.nasa.gov/horizons.cgi
[5] https://github.com/mommermi/callhorizons

For a single exposure a SNR of 30–40 was achieved for targets with $V \approx 18$ and around 10 for $V \approx 21$. A SNR of 10 in a single exposure with a median observation duration of 1.3 hours yielded an error in calculated color of less than 0.1 magnitude.

For each image the PP also outputs cropped thumbnail images centered around the predicted position of the target. The aperture size and aperture position is also indicated. These thumbnail images were visually inspected for all outlying data points. Data points were manually deleted where in the corresponding image it was clear that the wrong source was selected, or that the target intersected a CCD defect or a star.

## 4. ANALYSIS

In this section we discuss how the colors, taxonomy, and rotational parameters for each observed MBA were determined. The results from this analysis are summarized in Table 1.

### 4.1. *Colors*

The $V-R$ and $V-I$ colors of each MBA are calculated by simply subtracting the calibrated $I$ and $R$ magnitudes from the $V$ magnitude. To account for variations in magnitude due to rotation, a linear interpolation was performed between adjacent $V$ data points. The interpolation was used to obtain a corrected $V$ magnitude at times of inter-spaced non-$V$ observations. Respective colors were derived by subtracting the non-$V$ magnitudes from these interpolated $V$ magnitudes. The final color is the weighted average (by error in respective data points) of all retrievable subtraction calculations in an observation and is shown in the top-right of each plot in Figure 2. The final colors were also corrected for solar colors by subtracting the respective $V-R = 0.41$ and $V-I = 0.75$ solar colors (Binney & Merrifield 1998). The solar corrected colors of all observed MBAs are shown in Table 1 and plotted in Figure 3 (b).

### 4.2. *Taxonomic Classification*

To classify each observed MBA as one of S-, X-, C- or D-type asteroid, a Machine Learning (ML) algorithm approach was employed as described by Mommert et al. (2016). The ML algorithm was trained using asteroid colors that were synthesized from measured asteroid spectra taken from the MIT-UH-IRTF Joint Campaign for NEO Spectral Reconnaissance[6]. The spectra were classified using the Bus-DeMeo taxonomy spectrum classification online routine[7] and in the end 74, 34, 28 and 6 reliable $V-R$ and $V-I$ colors of S-, X-, C- and D-type asteroids, respectively, were synthesized. Mommert et al. (2016) provides more detail on generating training data for the ML algorithm.

The synthesized colors (training data) are plotted and shown in Figure 3 (a) and the resultant classification boundaries generated by the ML algorithm are also indicated. The training of the ML algorithm was executed using the `scikit-learn` module (Pedregosa et al. 2011) for `Python`. The k-nearest neighbor (k=5) method was implemented, as in Mommert et al. (2016) and Erasmus et al. (2017).

In order to account for uncertainties in the measurement of each MBA's color, a Monte-Carlo approach was used where $10^4$ pseudo-measured colors were randomly generated with a Gaussian distribution within the corresponding $1\sigma$ uncertainties of each MBA's real measured color (see Erasmus et al. (2017) for example plots illustrating this process). Each of the $10^4$ instances of each MBA was classified using the trained ML algorithm (i.e., where they are located relative to the decision boundaries) and the overall frequency of each taxonomic type was tallied. The classification probability for each taxonomic type was then calculated. The most likely (i.e., highest probability) taxonomic type was assigned to the MBA only if this probability was larger than 50% (see Table 1 for probabilities and the symbol in adjacent column which indicates the most likely classification). The assigned classifications are indicated in Figure 3 (b) by means of the color of the data point (orange = S-type, blue = X-type, pink = C-type, yellow = D-type asteroid and black = undetermined).

### 4.3. *Light Curves*

Light curves were produced by combining the $V$ magnitude data points with adjusted $I$ and $R$ magnitude data points. Adjustments were implemented by normalizing the original $I$ and $R$ magnitude data points to $V$ magnitudes by adding the derived colors (see Section 4.1 for detail on how colors were derived). Examples of resultant light curves are shown as the bottom curves in Figure 2 (a)-(i) with the original $V$ data and adjusted $R$ and $I$ data inter-spaced.

[6] http://smass.mit.edu/minus.html

[7] http://smass.mit.edu/busdemeoclass.html

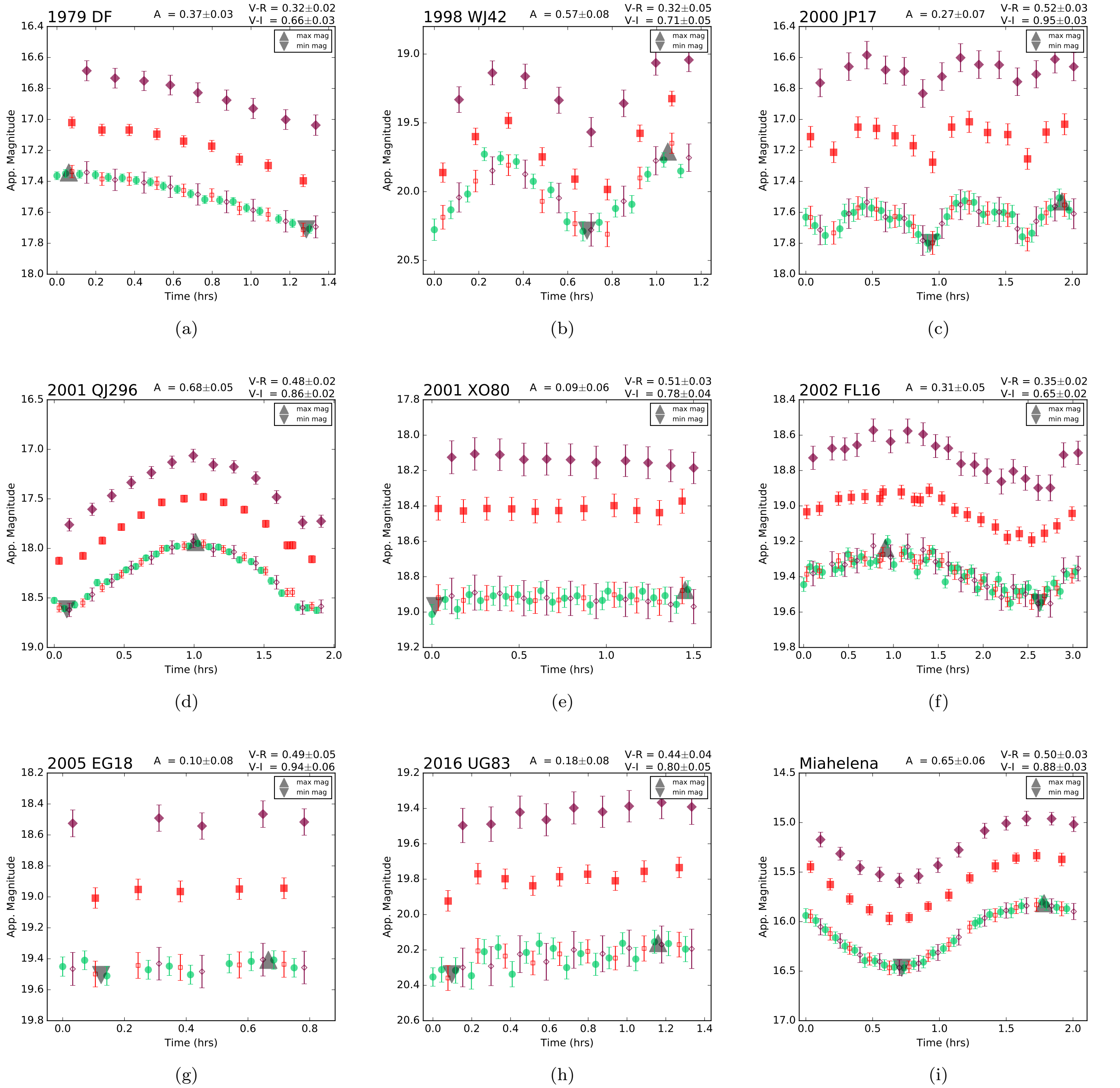

**Figure 2.** Example photometric data for nine of the observed MBAs. The data are the calibrated photometric results generated by PP developed by Mommert (2017). The $V$- (green circles), $R$- (red squares), and $I$- (burgundy diamonds) filter data are shown. A final light curve (bottom set of data in each plot) is constructed by offsetting $R$ and $I$ data with a constant value (see Section 4.1 and 4.3). The adjusted data points are displayed as smaller, hollow symbols interspersed with the $V$ data. The lower limit on the amplitude (A) is calculated using the difference between magnitudes highlighted with the gray triangle symbols. The triangles are situated where the mean of two adjacent data points have the minimum/maximum magnitude. See the electronic version for all 1003 plots. (For reviewer: see attached zipped file)

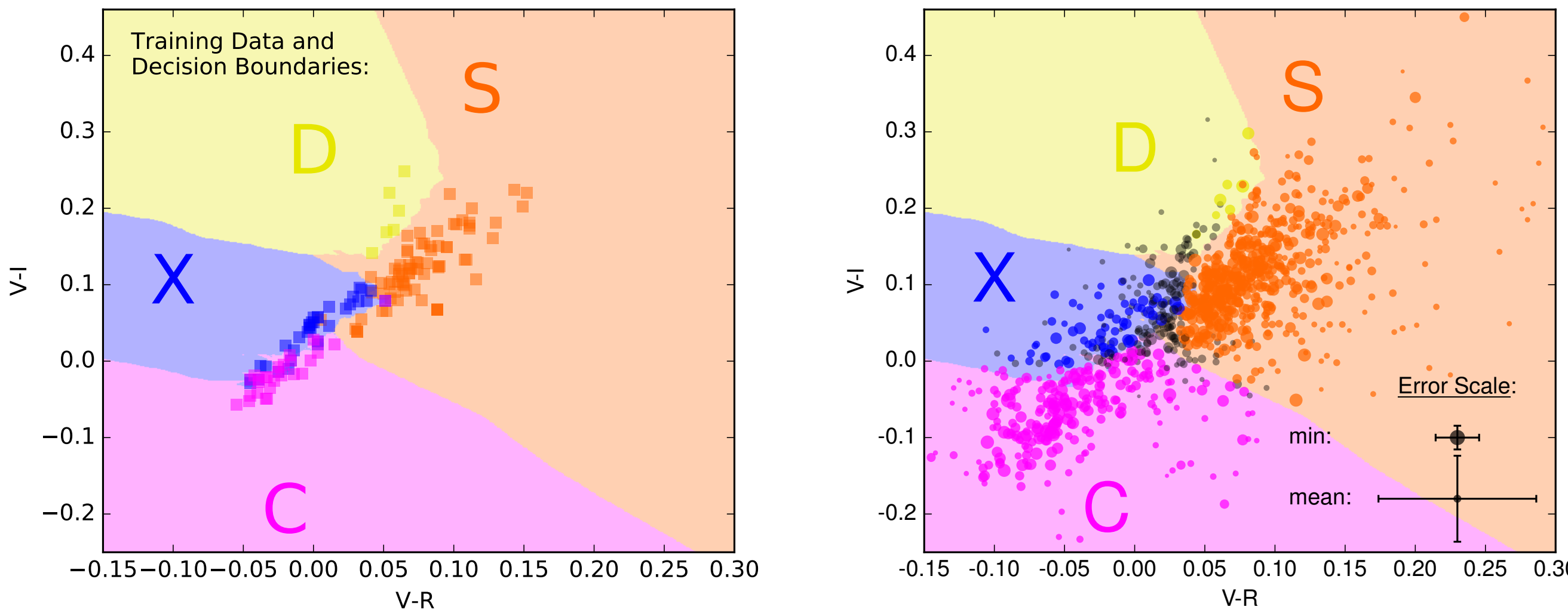

(a) (b)

**Figure 3.** (a) The training data synthesized from MIT-UH-IRTF spectra (see Section 4.2) and resultant decision boundaries generated by the ML algorithm. (b) The calculated colors of all observed MBAs (see Section 4.1 for detail) with the decision boundaries superimposed. MBAs in orange, blue, pink and yellow were classified by the ML algorithm (see Section 4.2) as S-, X- and C- and D-type asteroids, respectively. MBAs in black remain unclassified since the highest taxonomic probability was less than 50% . Sizes of the data points are inversely proportional to the uncertainty (scale is shown). The probabilities (see Section 4.2) for each type are given in Table 1.

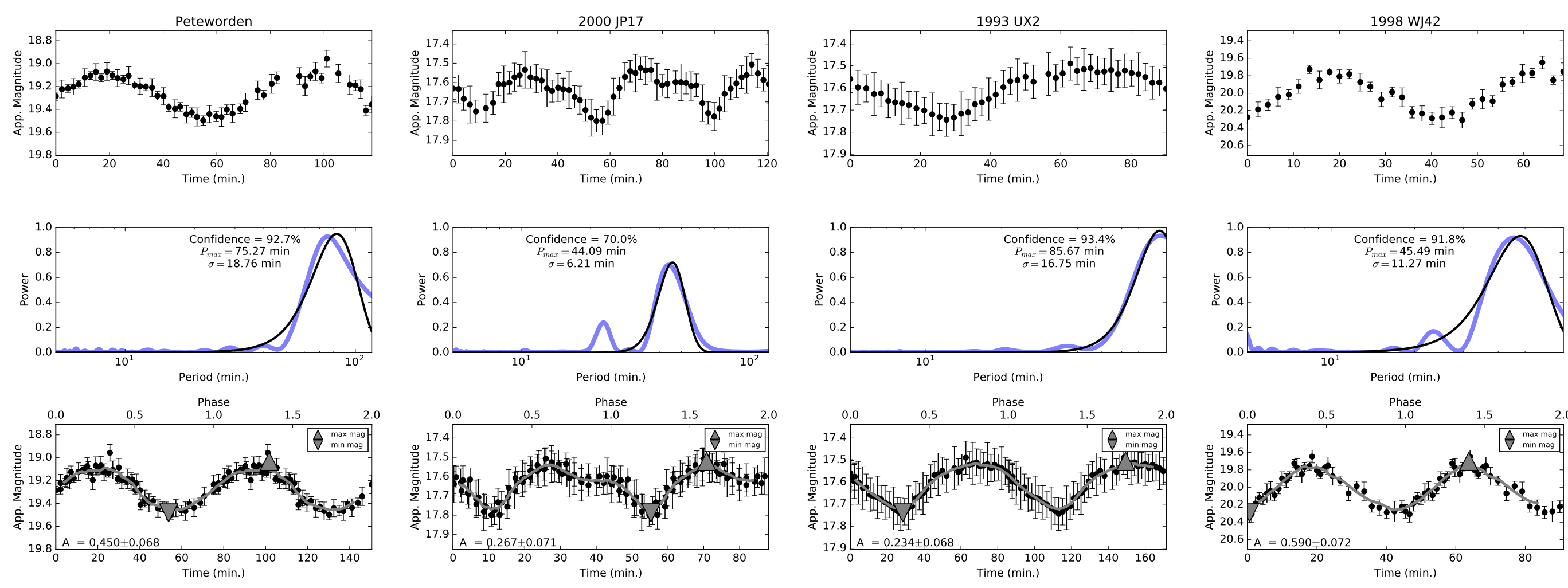

(a) (b) (c) (d)

**Figure 4.** Top plots: combined $V$, $R$ and $I$ calibrated photometric data of four observed MBAs producing a light curve. The periodogram of the light-curve data and folded representation of the original light-curve data are shown underneath each light-curve plot. $P_{max}$ is used as the folding interval. The bottom plots show folded data spanning two of the best-fit light-curve periods. The amplitude (A) is displayed with the gray triangles, situated where the mean of two adjacent data points have the minimum/maximum value, indicating the values used for determining A. See the electronic edition for data for all 59 MBAs that produced periodograms with a distinct peak. (For reviewer: see attached zipped file).

The majority of the light curves only cover a fraction of the rotational period (e.g., 1979 DF in Figure 2 (a)), and therefore in most cases only lower limits on both the light curve amplitude and asteroid rotation period could be reported in Table 1. However, we performed a least-squares spectral analysis (Lomb-Scargle, Lomb 1976; Scargle 1982) on all the light curves to estimate a frequency spectrum and thereby generate a periodogram of each observed MBA. Any observed MBA with a periodogram peak with a confidence larger than 60% was flagged as a potentially resolved light curve. After manual inspection, 59 observations were identified as potentially resolved light curves. Four examples are shown in Figure 4.

The uncertainty of this periodicity is determined by fitting a Gaussian function to the periodogram peak and using the RMS width ($\sigma$) as the uncertainty (see superimposed black curve and $\sigma$ of the fitted function). The light-curve period $P_{max}$ and the confidence in the periodogram peak is also indicated on the plot. The bottom set of plots in Figure 4 show the folded representation of the original light-curve data using $P_{max}$ as the folding interval. Using this folded data, the amplitude is calculated using the difference between magnitudes highlighted with the gray triangle symbols. The triangles are situated where the mean of two adjacent data points have the minimum/maximum magnitude. The asteroid rotation period (i.e., double the light-curve period obtained from the periodogram peak position) and the amplitudes determined from the folded data are listed in Table 1 for the 59 targets. For the remaining targets the lower limit of the rotation period (i.e., the observational duration) and the lower limit of the amplitude are used to complete Table 1. In the latter case, the amplitude lower limited is calculated in the same way as before but instead the original unfolded data are used (see gray triangles in Figure 2).

## 5. SHAPE-DISTRIBUTION MODELING

To obtain an estimate of the shape distribution of our MBA population we use an improved form of the model described in McNeill et al. (2016). This model generates a synthetic population of ellipsoidal asteroids with assumed shapes and spin-pole orientations. We assume a uniform spin frequency distribution, corresponding to rotational periods between 2.2 and 24 hours, across all applicable size ranges comparable to the uniform distribution of rotational frequencies at the size range present in our dataset (Pravec et al. 2002). The synthetic population is sampled using the same cadence as the KMTNet observations and for time-spans consistent with the lengths of the partial light curves measured. The model also assumes an uncertainty on each measurement based on the distribution of uncertainties in the measured magnitudes from the data. It is important to note that the magnitudes generated by the model are assumed purely to be due to geometric effects i.e., limb-scattering effects and albedo variance are not accounted for.

Using the model described above we determined the best-fit shape distribution to the observed data. To quantitatively determine the goodness of the fit for each model population we employ the two-sampled chi-squared test and seek the best fit by minimizing this value. Assuming that all asteroids are prolate spheroids ($a>b=c$) the value of $b/a$ was varied as a truncated Gaussian with a centre $\mu$ and standard deviation $\sigma$. The Gaussian distribution was truncated at the point where $b/a = 1$. The spin-pole distribution of the populations was also varied between a distribution in which all objects are aligned with the orbital plane and a distribution where the spin-pole axes are randomly oriented. For each shape and spin-pole configuration the population is sampled using the same cadence as our observations in order to obtain the best possible fit to the measured partial light curves. We obtained valid fits of different shape distributions across all possible spin distributions considered so we are unable to determine any information about the spin distribution of the objects in our sample.

## 6. DISCUSSION

### 6.1. *Taxonomic Distribution*

Figure 5 shows the taxonomic distribution of all our observed MBAs as a function of $H$ magnitude. Panel (a) gives a broad overview of the distribution and is based on the discrete classification, i.e., only the highest classification probability, as in Figure 3 (b), is used to assign a taxonomy to each MBA. Panel (b) shows the distribution in three $H$ magnitude bins and in this case the distribution is obtained by summation of the probability fractions in Table 1 for each respective class type. The plots suggest that the ratio between S-type MBAs and (C+X+D)-type MBAs with $H$ magnitudes between 12 and 18 (12 km $\gtrsim$ D $\gtrsim$ 0.75 km) is within uncertainties almost exactly 1:1 with perhaps a slight decrease in S-type asteroids for the larger MBAs in our size range.

To determine if any correlation exists between asteroid taxonomy and orbital parameters we performed Spearman correlation tests between our determined taxonomic class probability and three orbital parameters: semi-major axis, eccentricity and inclination. The only significant correlation the tests revealed was between semi-major axis and the probability of an MBA being a S- or C-type. This correlation can also be seen in Figure 6 where we show the taxonomic distribution of all our observed MBAs as a function of semi-major axis. Panels (a) and (b) show the same data but with inclination and $H$ magnitude the y-axis variable. As in panel (a) of Figure 5 only discrete classification is used here. The histograms show that C-type asteroids are more prevalent in the outer main-belt and S-type asteroids more prevalent in the inner main-belt, which agrees with previously reported findings (Bell et al. 1989; Carvano et al. 2003; DeMeo & Carry 2013). Panel (c) shows the distribution in six semi-major axis ranges (separated by the Kirkwood gaps) and here the trend is clear that the ratio between S-type MBAs and C-type MBAs decreases with increasing heliocentric distance.

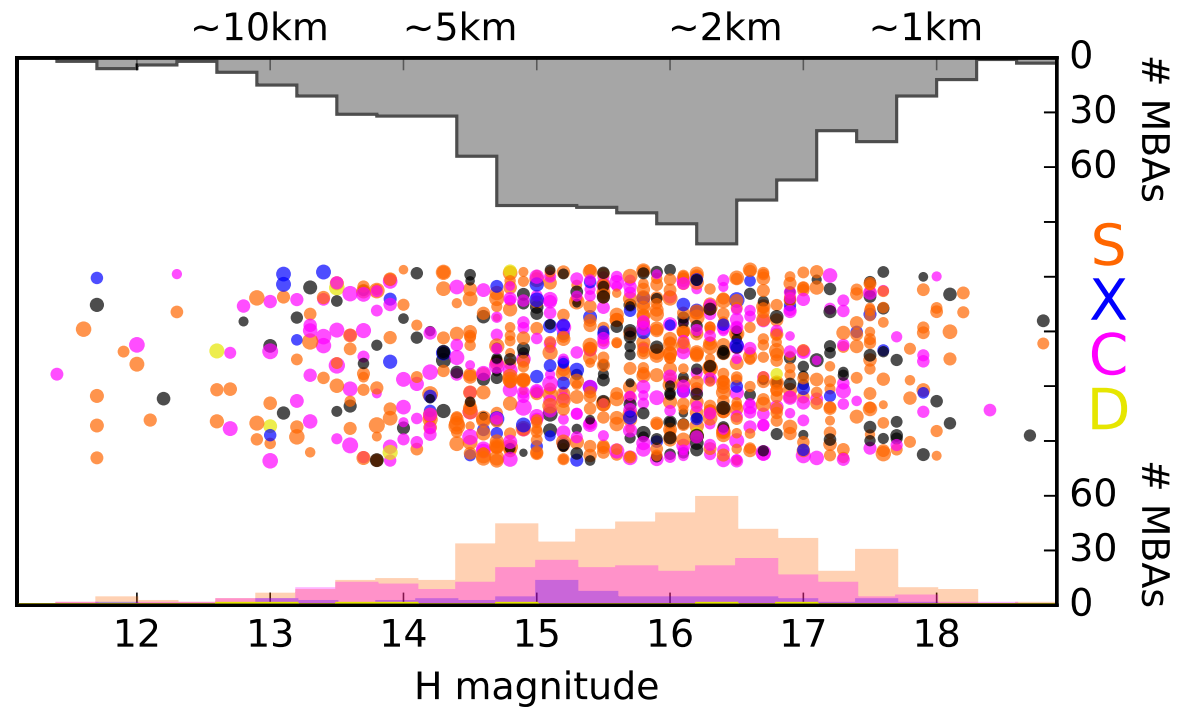

(a)

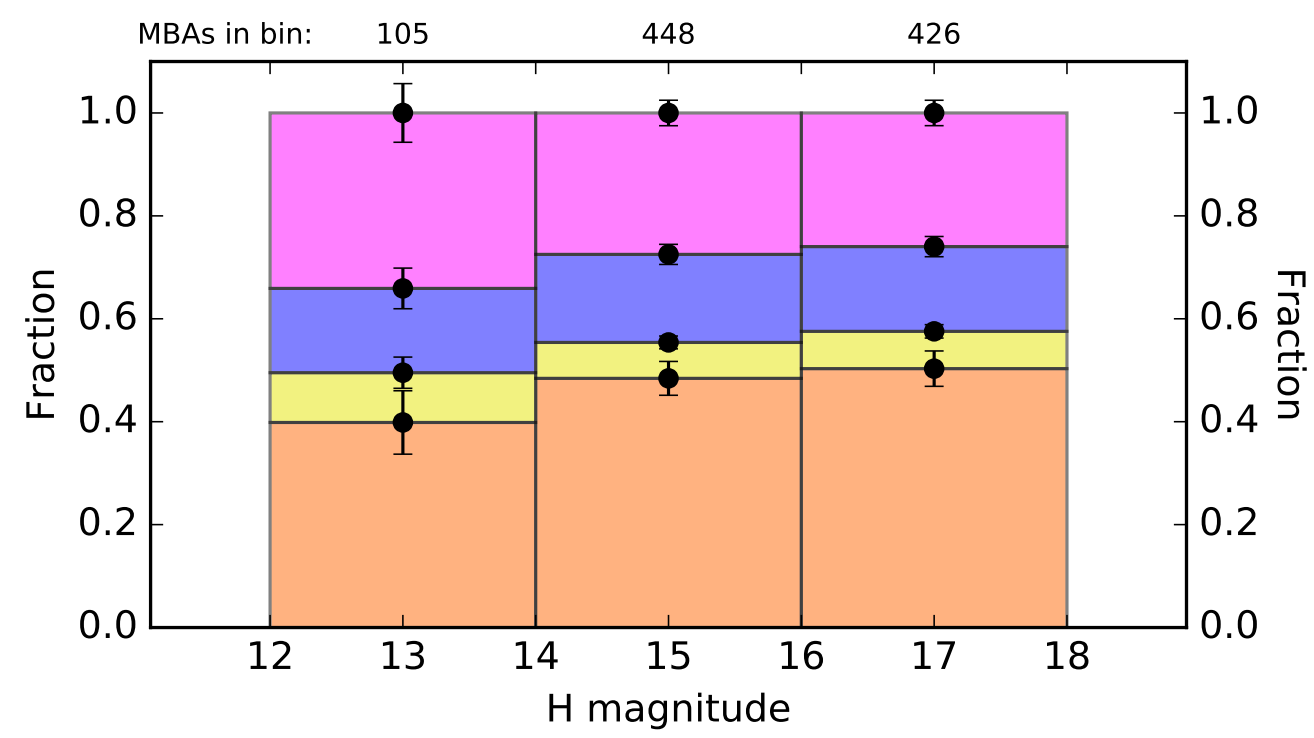

(b)

**Figure 5.** (a) Taxonomic distribution as a function of $H$ magnitude. Note that the data point colors and plotted histograms are based on discrete classification i.e., only the highest classification probability in Table 1 is considered. MBA size estimations (assuming an albedo of 0.12) are also indicated. (b) The distribution in three $H$-magnitude bins. Here the calculated fractions for each taxonomy are based on the summation of the probabilities in Table 1 for each respective class type.

We compare our taxonomic distribution result with DeMeo & Carry (2014) which is a summarized amalgamation of many published results on asteroid taxonomies. It should be noted that their distributions are based on mass rather than numbers. However, the densities (Carry 2012) of S-type asteroids (2 g/cm$^3 \lesssim \rho \lesssim$ 3 g/cm$^3$) and C-type asteroids (1.3 g/cm$^3 \lesssim \rho \lesssim$ 2.9 g/cm$^3$) are similar enough to make a direct comparison in rough trends permissible. Although their smallest size range (5-20 km) is slightly larger than our size range, their absolute taxonomic distribution ratios and the heliocentric-distance dependence on these ratios appears similar to our results.

We also cross-checked our MBAs with NASA's Small Bodies Data Ferret database (SBDF) (Nesvorny 2015) which allowed us to assign an asteroid family to roughly one third of our MBAs. In general we found a good agreement with our determined taxonomy and the taxonomy associated with the corresponding asteroid family (Nesvorny et al. 2015).

### 6.2. *Light-Curve Data*

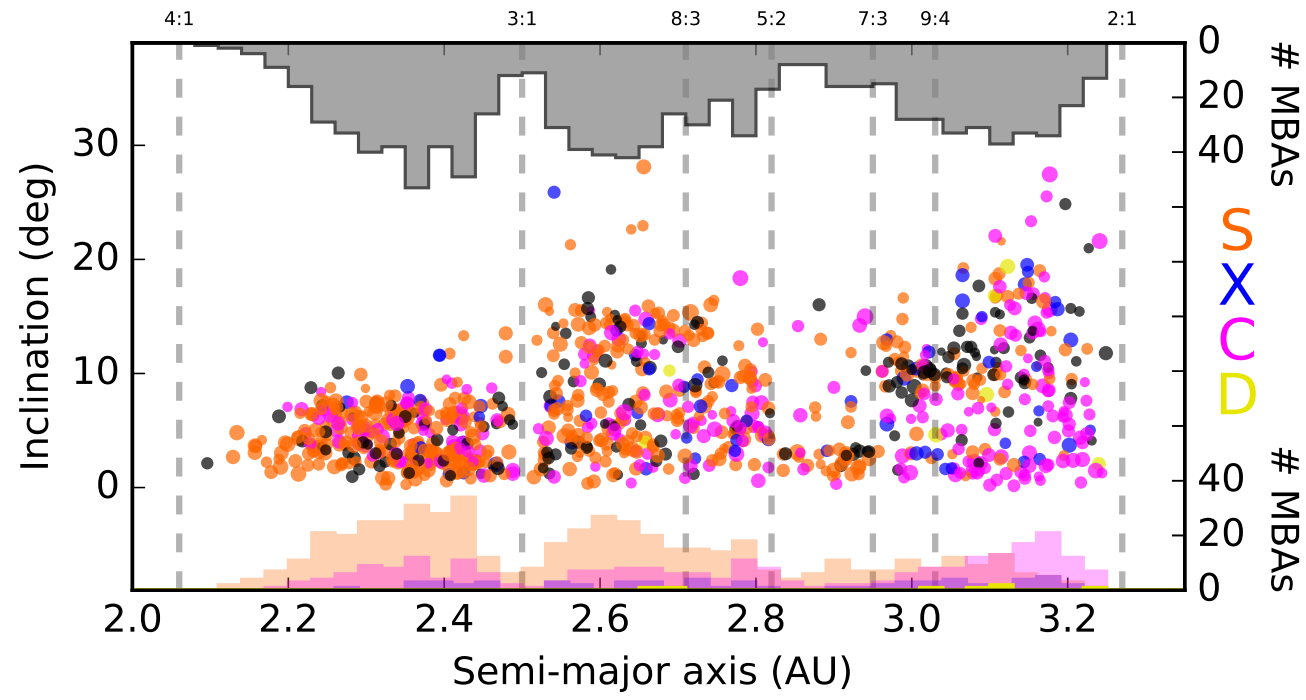

(a)

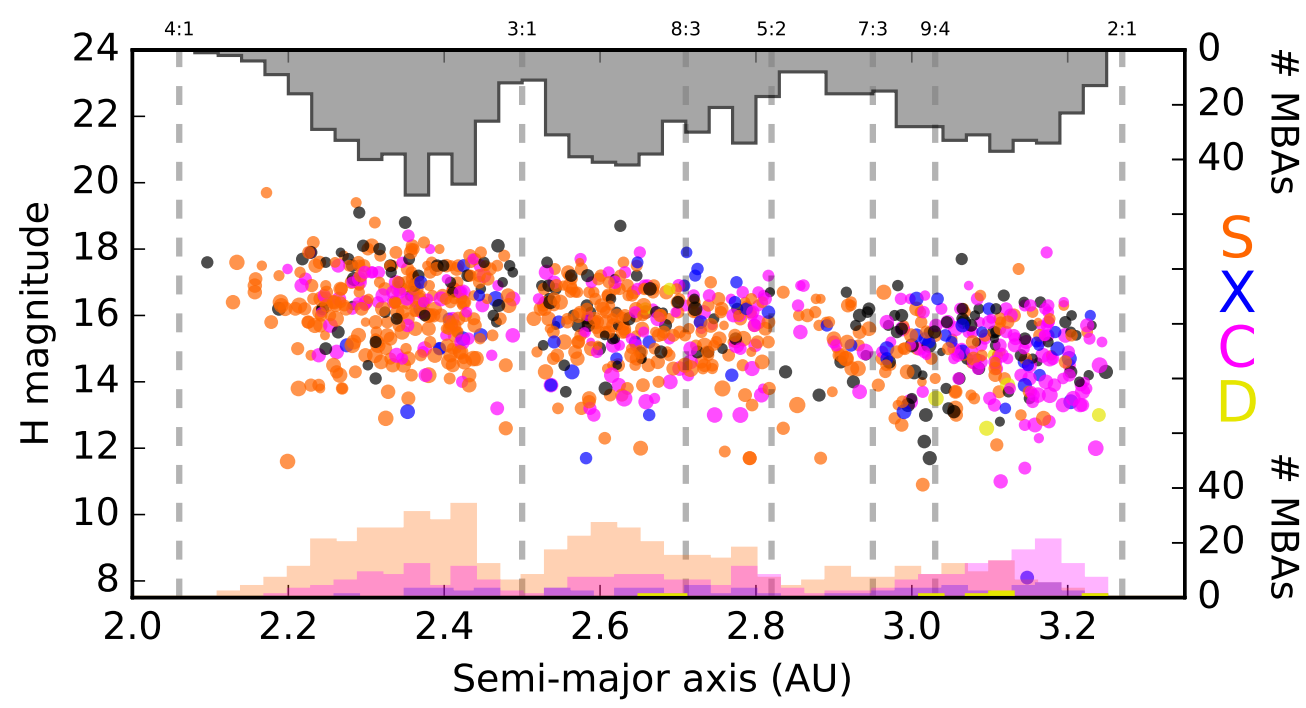

(b)

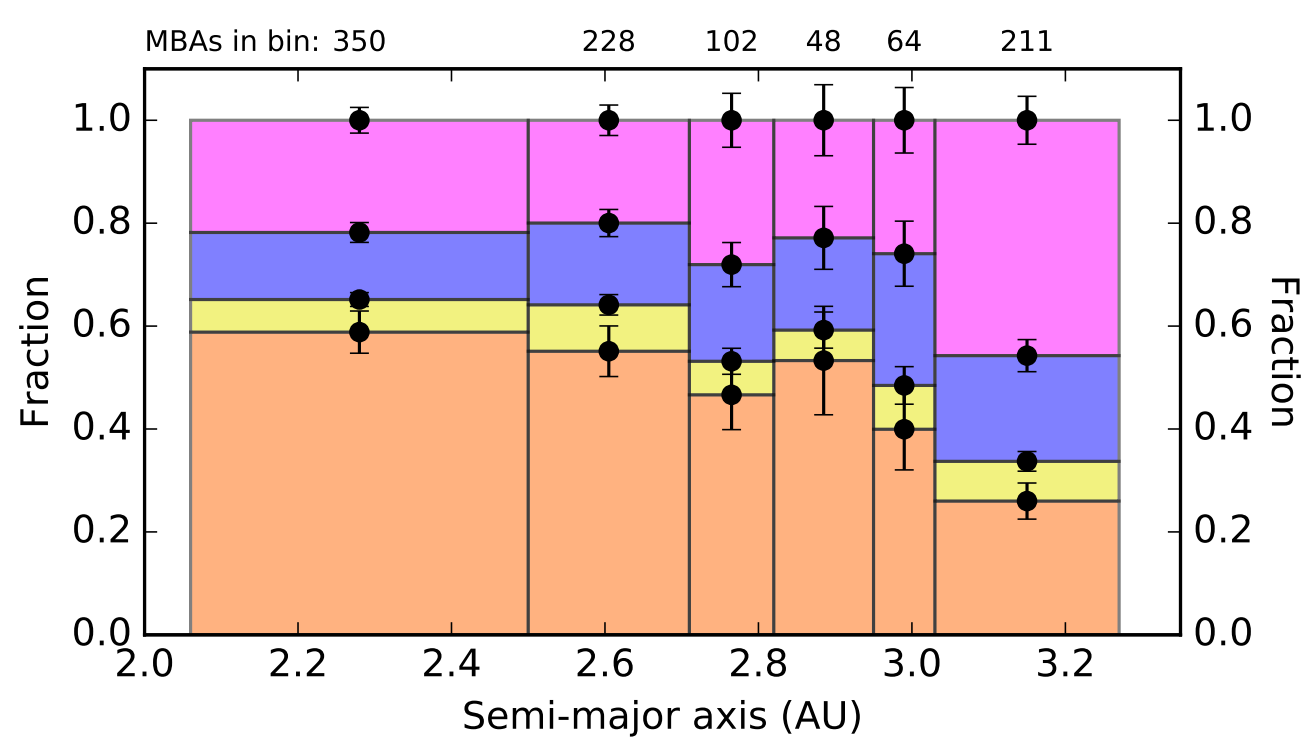

(c)

**Figure 6.** (a) and (b) Taxonomic distribution as a function of semi-major axis with inclination and $H$ magnitude on the y-axis variables respectively. Note that the data point colors and plotted histograms are based on discrete classification i.e., only the highest classification probability in Table 1 is considered. (c) The distribution in six semi-major axis ranges (separated by the Kirkwood gaps, see dotted lines in (a) and (b)). Here the calculated fractions for each taxonomy are based on the summation of the probabilities in Table 1 for each respective class type.

Figure 7 shows rotational periods plottted as a function of MBA size for the 59 targets observed that had potentially resolved light-curve periods (see Section 4.3). The known rotational periods for all asteroids (from the LCDB) are also shown and in most cases our data show similar trends to already reported rotational periods in our size range i.e., most of our MBAs have rotational periods below the theoretical zero cohesion boundary limit of 2.2 hours. However, there are two exceptions, 1998 WJ42 and 2000 JP17 (circled in 7 (b)) that both have light-curves (shown in Figure

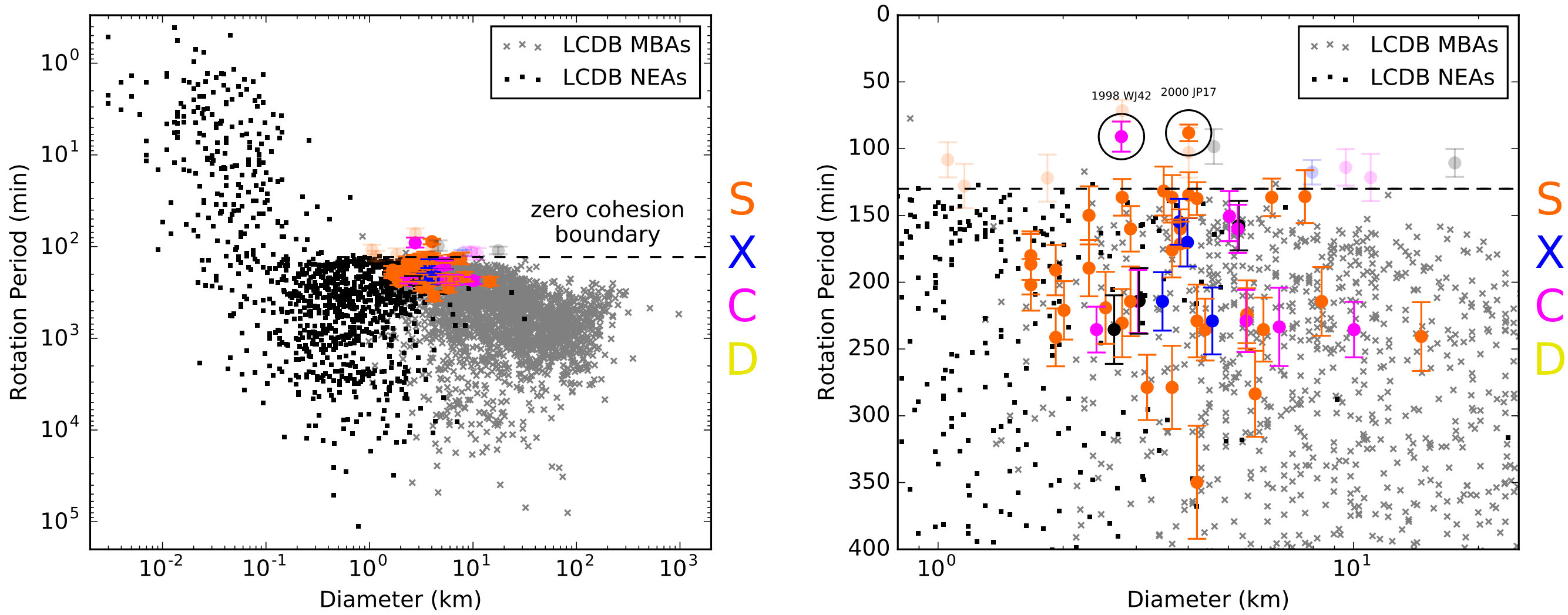

(a) (b)

**Figure 7.** (a) Rotation periods as a function of size of the 59 MBAs observed that had potentially resolved light-curve periods. For the size estimation we use albedos determined by Ryan & Woodward (2010) for the corresponding taxonomy. The LCDB (Warner et al. 2009, updated 2017 September 11) data are also plotted. (b) Zoom-in of panel (a) with the color of the data points showing the assigned taxonomy (see Section 4.2) with orange = S-type, blue = X-type, pink = C-type, yellow = D-type asteroid and black = undetermined. MBAs with a derived rotational period (see Section 4.3) below the theoretical zero cohesion boundary limit of 2.2 hours (dotted line) that had confidence in the periodicity determination below 70%, or where extracted light-curve period was close to the observation duration, are shown as lightly shaded data points because these conditions would mean there are justified doubts on the legitimacy of the calculated period. The light curves and periodograms of the two circled data points are shown in Figure 4 (b) and (d) with both showing fully-resolved rotation-period values suggesting that these two MBAs are potential Super-Fast Rotators (SFRs) for their size.

4 (b) and (d)) displaying rotation periods below the zero cohesion boundary. These two objects could be potential Super-Fast Rotators (SFRs) and further observations to confirm their SFR status are planned.

For the 59 targets we also tested for correlations between determined taxonomic class probability and the two calculated light-curve parameters: rotation period and light-curve amplitude. No statistical significant correlations were found.

### 6.3. *Shape-Distribution Modeling*

The result of the shape-distribution modeling and fitting explained in Section 5 is shown in Figure 8. The plot indicates that the best fit for the average measured elongation is $b/a = 0.74 \pm 0.07$. Using the taxonomic classifications determined in this work the shape distributions for both S- and C-type asteroids were calculated. There was no statistically significant difference between the shape distributions obtained for these two classifications, nor was there any difference between these two values and the value obtained for the data-set as a whole. The shape distribution at different size ranges was also generated and no statistically significant difference between the shape distributions was seen.

We compare our shape distribution result with Masiero et al. (2009) which used light-curve data obtained with the Canada-France-Hawaii Telescope (CFHT) and with results obtained by McNeill et al. (2016) and Cibulková et al. (2017), both using Pan-STARRS 1 (PS1) light-curve data with a similar asteroid size range as ours. Masiero et al. (2009) suggest a large contribution from asteroids with $b/a \approx 0.8$. McNeill et al. (2016) find an average axis ratio of $b/a = 0.83 \pm 0.12$ and Cibulková et al. (2017) find $b/a = 0.80$. Therefore, within uncertainties the average axis-ratio values derived in this work are in good agreement with previously published values.

### 6.4. *Duplicate Observations and Comparison to LCDB Values*

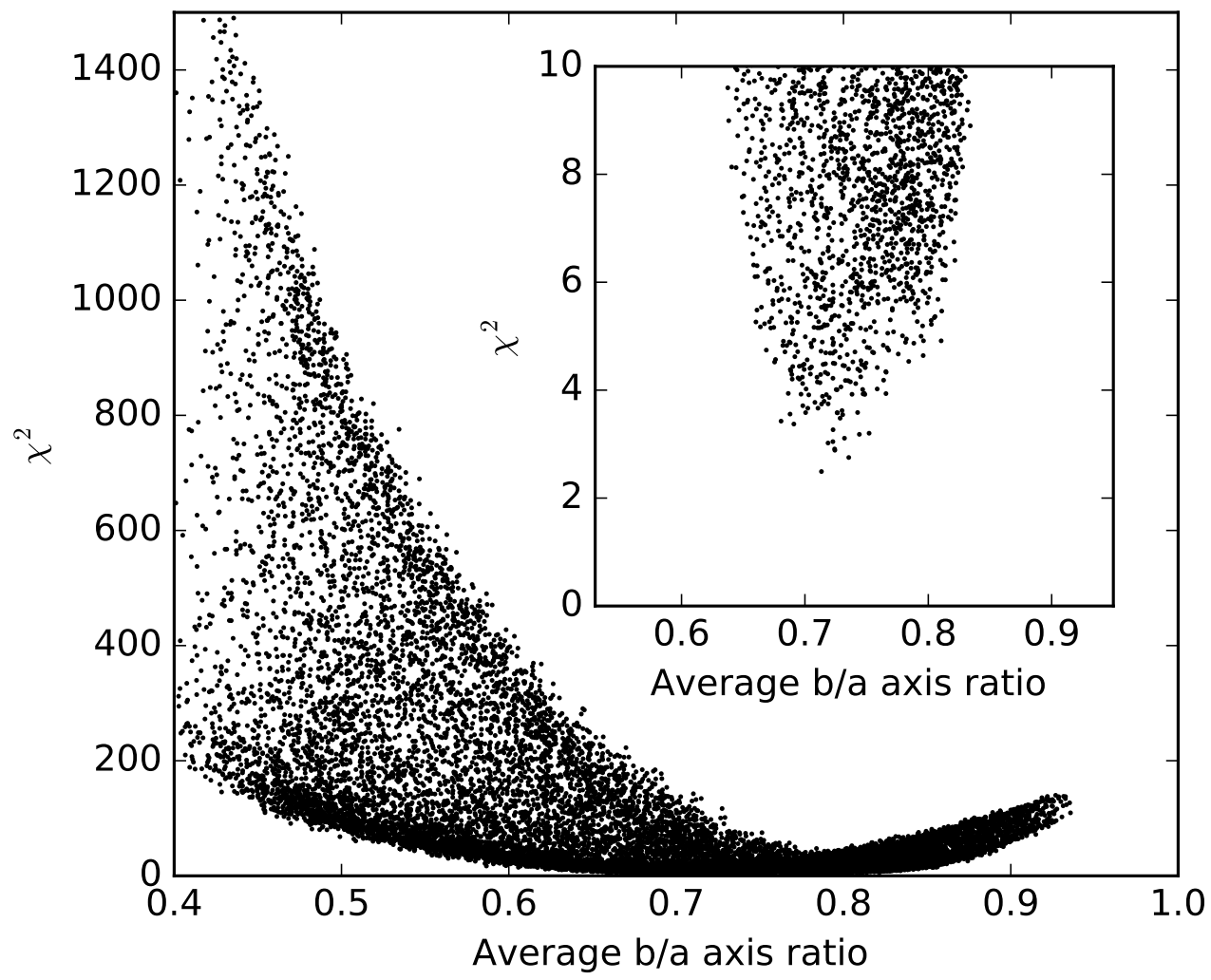

**Figure 8.** Chi-squared statistics from partial light curves generated from model asteroid populations compared to the observed data. The axis-ratio value corresponds to the mean shape for each model population with a best fit found for $b/a = 0.74 \pm 0.07$ assuming all distributions giving $\chi^2 < 5$ to be valid solutions.

**Table 2.** Comparison to LCDB values

| No. | Object[a] | Amplitude[b] (mag) | | Rotation Period[b] (minutes) | | Taxonomy[b] | |
|---|---|---|---|---|---|---|---|
| | | this study | LCDB | this study | LCDB | this study | LCDB |
| 01 | 2001 PV42 | ≥1.07 | - | ≥106 | 330.00 | S | S |
| 02 | Andronikov | ≥0.29 | 0.39 | ≥52 | 189.17 | X | S |
| 03 | Ara | ≥0.02 | 0.34 | ≥58 | 246.96 | X | M |
| 04 | Arthuradel | ≥0.17 | 0.28 | ≥58 | 388.56 | S | S |
| 05 | Billmclaughlin | ≥0.04 | 0.07 | ≥37 | 311.94 | X | X |
| 06 | Hirayama | ≥0.21 | 0.51 | ≥120 | 937.80 | C | C |
| 07 | Juliegrady | ≥0.42 | 0.90 | ≥114 | 1128.00 | S | S |
| 08 | Miahelena | ≥0.50 | 0.62 | ≥66 | 252.06 | S | S |
| 09 | Miahelena | 0.65±0.06 | 0.62 | 241±51 | 252.06 | S | S |
| 10 | Okabayashi | ≥0.10 | 0.10 | ≥53 | 153.00 | S | S |
| 11 | Pohjola | ≥0.12 | 0.11 | ≥54 | 175.20 | S | S |

[a] Miahelena appears twice since it was observed twice in this study

[b] First column are our derived values (see Table 1), second column are the values from the LCDB (Warner et al. 2009, updated 2017 September 11)

There are four MBAs that were by chance observed twice. They are 1981 SU2, 1999 RE190, 2000 UM3 and Miahelena. Within the error bars the derived rotational properties and colors for each instance of the four targets agree. In one case, there was discrepancy between assigned taxonomy. 1999 RE190 was classified as C and X for the two observing datasets but in both instances the calculated color lies in close proximity to a decision boundary generated by the ML algorithm and therefore the probability for the assigned classification is only in the 50% range.

For 1999 RE190 we were also able to resolve the rotation period in both observations. The determined period and amplitude agree within error bars but in both cases the observing duration was very close to the derived light-curve period so there is some doubt if this is the true rotation period. The rotation period and amplitude could only be determined in one of the observations for 1981 SU2 and Miahelena. In both cases the determined values are above the lower limit for the undetermined observation.

Table 2 shows the derived rotational properties and taxonomic classification of all our observed MBAs that were also present in the LCDB (Warner et al. 2009, updated 2017 September 11). In total there were ten MBAs present in both our dataset and in the LCDB. In general we see excellent agreement in both taxonomy and rotational properties.

## 7. CONCLUSIONS

The *VRI* spectrophotometric results of 1003 MBAs observed with the Sutherland, South Africa, node of KMTNet have been presented. Our broadband spectrophotometry in combination with a ML algorithm allowed us to classify 836 of the 1003 observed target as either a S-, C-, X-, or D-type asteroid. Our data suggest that the ratio between S-type MBAs and (C+X+D)-type MBAs with $H$ magnitudes between 12 and 18 is 1:1, with no discernible $H$ magnitude dependency within the range. This ratio is in agreement with previously published survey studies (DeMeo & Carry 2014). Our results show that the ratio is however dependent on the location within the main-belt with the ratio between S-type MBAs and C-type MBAs decreasing with increasing heliocentric distance. This is also in agreement with previously published survey studies (Bell et al. 1989; Carvano et al. 2003; DeMeo & Carry 2013).

Our light-curve data for each observed MBA allowed us to place lower-limit constraints on the rotational periods and light-curve amplitudes of $\sim$950 MBAs and resolve the rotational periods and light-curve amplitudes for 59 MBAs. The lower-limit amplitudes together with the resolved amplitudes allowed us to determine the best-fit shape distribution to the observed data and the average measured elongation is $b/a = 0.74 \pm 0.07$. For the 59 MBAs with resolvable periods our data are also consistent with previously published observations of MBAs with similar sizes with the exception of two targets. 1998 WJ42 and 2000 JP17 display rotation periods below the zero cohesion boundary, suggesting that these are potential Super-Fast Rotators (SFRs) for their size. These two targets require further investigation.

In total there were ten MBAs present in both our dataset and in the LCDB (Warner et al. 2009, updated 2017 September 11). In general we see excellent agreement in both taxonomy and rotational properties.

In the future we plan to continue with the work presented here but will increase our observational duration for each observing field in the hopes of resolving the rotation period for a higher fraction of our targets.

This research has made use of the KMTNet system operated by the Korea Astronomy and Space Science Institute (KASI) and the data were obtained by observations made at the South African Astronomical Observatory (SAAO). This work is partially supported by the South African National Research Foundation (NRF). This work is supported in part by the National Aeronautics and Space Administration (NASA) under grant No. NNX15AE90G issued through the SSO Near-Earth Object Observations Program and in part by a grant from NASA's Office of the Chief Technologist.

# APPENDIX

## A. ALL DATA

**Table 1**. Observations and Results

| No.[a] | Object | Obs. Start | Obs. Duration | H[b] | V | V-R[c] | V-I[c] | Amplitude[d] | Rot. Period[d] | S-type | X-type | C-type | D-type | Tax. | a[b] | i[b] |
|---|---|---|---|---|---|---|---|---|---|---|---|---|---|---|---|---|
| | | (JD) | (h:mm) | (mag) | (mag) | (mag) | (mag) | (mag) | (min) | | (probability) | | | | (au) | (deg) |
| 0001 | 1173 T-3 | 2457802.39311 | 0:45 | 15.0 | 19.28 | 0.02±0.03 | 0.06±0.04 | ≥0.11 | ≥46 | 0.417 | 0.435 | 0.131 | 0.017 | - | 2.6 | 9.1 |
| 0002 | 1356 T-2 | 2457722.38400 | 1:47 | 16.2 | 20.50 | -0.04±0.05 | -0.06±0.07 | ≥0.29 | ≥107 | 0.027 | 0.222 | 0.749 | 0.003 | C | 3.0 | 4.7 |
| 0003 | 1942 TB | 2457768.50255 | 2:00 | 13.2 | 17.89 | 0.05±0.06 | 0.05±0.07 | ≥0.18 | ≥120 | 0.560 | 0.188 | 0.202 | 0.051 | S | 2.5 | 9.4 |
| 0004 | 1975 SN1 | 2457724.34161 | 0:57 | 13.9 | 18.55 | 0.04±0.03 | 0.10±0.05 | ≥0.15 | ≥58 | 0.519 | 0.311 | 0.027 | 0.142 | S | 3.0 | 3.2 |
| 0005 | 1978 VE4 | 2457692.38892 | 1:20 | 17.0 | 18.48 | -0.08±0.02 | -0.10±0.03 | ≥0.10 | ≥80 | 0.000 | 0.004 | 0.996 | 0.000 | C | 2.3 | 3.9 |
| 0006 | 1979 DF | 2457692.38892 | 1:20 | 13.4 | 17.46 | -0.09±0.02 | -0.09±0.03 | ≥0.37 | ≥80 | 0.000 | 0.003 | 0.998 | 0.000 | C | 2.7 | 14.8 |
| 0007 | 1979 QT1 | 2457724.38316 | 1:08 | 14.7 | 18.68 | 0.05±0.02 | 0.10±0.03 | ≥0.14 | ≥69 | 0.760 | 0.159 | 0.002 | 0.080 | S | 2.4 | 1.3 |
| 0008 | 1981 EF21 | 2457690.42183 | 2:10 | 16.0 | 20.11 | 0.03±0.06 | 0.03±0.07 | ≥0.44 | ≥130 | 0.378 | 0.268 | 0.318 | 0.035 | - | 2.6 | 9.3 |
| 0009 | 1981 EX26 | 2457692.38892 | 1:20 | 14.6 | 17.76 | 0.05±0.02 | 0.07±0.03 | ≥0.04 | ≥80 | 0.763 | 0.224 | 0.010 | 0.002 | S | 2.4 | 3.0 |
| 0010 | 1981 SU2 | 2457766.50272 | 1:01 | 13.8 | 17.37 | 0.21±0.09 | 0.19±0.07 | ≥0.51 | ≥61 | 0.955 | 0.005 | 0.000 | 0.039 | S | 2.3 | 2.1 |
| 0011 | 1981 SU2 | 2457768.50255 | 2:00 | 13.8 | 17.50 | 0.07±0.05 | 0.14±0.05 | 0.53±0.10 | 225±41 | 0.666 | 0.094 | 0.004 | 0.235 | S | 2.3 | 2.1 |
| 0012 | 1986 QR2 | 2457766.28699 | 2:15 | 13.9 | 17.93 | 0.02±0.03 | -0.01±0.03 | ≥0.28 | ≥135 | 0.186 | 0.070 | 0.744 | 0.000 | C | 2.6 | 13.0 |
| 0013 | 1989 EE3 | 2457802.50174 | 0:55 | 14.5 | 19.35 | 0.05±0.07 | 0.12±0.08 | ≥0.22 | ≥55 | 0.527 | 0.194 | 0.057 | 0.222 | S | 3.1 | 16.6 |
| 0014 | 1990 EN1 | 2457724.38316 | 1:08 | 13.6 | 18.31 | -0.00±0.02 | 0.00±0.03 | ≥0.05 | ≥69 | 0.057 | 0.277 | 0.666 | 0.000 | C | 2.8 | 7.4 |
| 0015 | 1990 QF4 | 2457730.42038 | 1:30 | 13.5 | 16.67 | 0.07±0.03 | 0.04±0.04 | ≥0.09 | ≥90 | 0.876 | 0.070 | 0.054 | 0.001 | S | 2.4 | 7.3 |
| 0016 | 1990 QN7 | 2457691.37970 | 1:54 | 14.2 | 17.97 | 0.08±0.02 | 0.11±0.02 | ≥0.12 | ≥114 | 0.991 | 0.005 | 0.000 | 0.005 | S | 2.4 | 2.7 |
| 0017 | 1990 TD13 | 2457690.42183 | 2:21 | 16.0 | 18.75 | 0.11±0.03 | 0.12±0.03 | ≥0.17 | ≥142 | 0.978 | 0.005 | 0.000 | 0.017 | S | 2.6 | 7.5 |
| 0018 | 1990 UT10 | 2457692.38892 | 1:20 | 13.4 | 16.69 | -0.06±0.02 | -0.09±0.03 | ≥0.35 | ≥80 | 0.000 | 0.005 | 0.995 | 0.000 | C | 3.0 | 1.3 |
| 0019 | 1991 NE1 | 2457798.46669 | 0:53 | 12.8 | 17.48 | -0.04±0.03 | -0.01±0.04 | ≥0.04 | ≥53 | 0.007 | 0.450 | 0.543 | 0.000 | C | 3.2 | 8.5 |
| 0020 | 1991 PQ12 | 2457768.50255 | 2:00 | 12.7 | 17.96 | -0.05±0.05 | -0.10±0.06 | ≥0.20 | ≥120 | 0.009 | 0.073 | 0.918 | 0.000 | C | 3.1 | 1.1 |
| 0021 | 1991 RB12 | 2457768.50255 | 2:00 | 13.8 | 18.54 | -0.12±0.12 | -0.06±0.13 | ≥0.59 | ≥120 | 0.052 | 0.237 | 0.686 | 0.025 | C | 2.8 | 12.7 |
| 0022 | 1991 VG9 | 2457724.38316 | 1:08 | 14.8 | 19.43 | 0.06±0.03 | 0.04±0.04 | ≥0.15 | ≥69 | 0.778 | 0.123 | 0.096 | 0.003 | S | 2.9 | 3.4 |
| 0023 | 1991 VT3 | 2457800.27020 | 1:03 | 13.2 | 18.44 | 0.10±0.03 | 0.18±0.04 | ≥0.05 | ≥64 | 0.864 | 0.002 | 0.000 | 0.134 | S | 2.6 | 12.7 |
| 0024 | 1991 VY2 | 2457798.50513 | 1:09 | 14.3 | 18.16 | 0.14±0.04 | 0.15±0.05 | ≥0.24 | ≥69 | 0.964 | 0.003 | 0.000 | 0.033 | S | 2.3 | 6.6 |
| 0025 | 1992 DX9 | 2457802.42659 | 1:45 | 16.4 | 19.07 | -0.01±0.03 | 0.01±0.05 | ≥0.14 | ≥106 | 0.043 | 0.437 | 0.518 | 0.002 | C | 2.3 | 6.8 |
| 0026 | 1992 EO7 | 2457802.50174 | 0:55 | 15.8 | 19.19 | 0.08±0.04 | -0.02±0.06 | ≥0.23 | ≥55 | 0.587 | 0.019 | 0.394 | 0.001 | S | 2.3 | 5.9 |
| 0027 | 1992 UE3 | 2457690.33738 | 1:57 | 13.3 | 17.43 | 0.06±0.01 | 0.11±0.02 | ≥0.05 | ≥118 | 0.932 | 0.041 | 0.000 | 0.027 | S | 2.9 | 3.1 |
| 0028 | 1993 FO32 | 2457691.37970 | 1:54 | 16.4 | 20.25 | 0.04±0.06 | 0.09±0.07 | ≥0.22 | ≥114 | 0.515 | 0.274 | 0.081 | 0.131 | S | 2.4 | 1.1 |
| 0029 | 1993 FY46 | 2457691.37970 | 1:54 | 16.1 | 19.85 | 0.02±0.03 | 0.08±0.04 | ≥0.20 | ≥114 | 0.366 | 0.512 | 0.070 | 0.052 | X | 2.7 | 1.6 |
| 0030 | 1993 FZ43 | 2457796.48277 | 0:51 | 16.3 | 18.66 | 0.01±0.16 | 0.09±0.14 | ≥0.25 | ≥52 | 0.362 | 0.242 | 0.209 | 0.187 | - | 2.3 | 6.2 |
| 0031 | 1993 OC13 | 2457692.44645 | 1:00 | 15.9 | 18.86 | -0.00±0.04 | 0.06±0.06 | ≥0.09 | ≥60 | 0.200 | 0.566 | 0.155 | 0.079 | X | 2.4 | 1.2 |
| 0032 | 1993 OE3 | 2457798.46669 | 0:53 | 14.7 | 18.38 | 0.09±0.05 | 0.05±0.06 | ≥0.18 | ≥53 | 0.823 | 0.058 | 0.102 | 0.017 | S | 2.3 | 6.6 |
| 0033 | 1993 TC23 | 2457690.33738 | 1:05 | 14.5 | 19.76 | -0.04±0.04 | -0.10±0.04 | ≥0.11 | ≥66 | 0.001 | 0.024 | 0.975 | 0.000 | C | 3.2 | 5.6 |
| 0034 | 1993 TH23 | 2457690.33738 | 1:05 | 16.2 | 19.52 | 0.10±0.03 | 0.15±0.04 | ≥0.11 | ≥66 | 0.915 | 0.005 | 0.000 | 0.080 | S | 2.2 | 2.3 |
| 0035 | 1993 TN15 | 2457724.34161 | 0:57 | 15.9 | 18.80 | 0.09±0.04 | 0.10±0.05 | ≥0.11 | ≥58 | 0.908 | 0.046 | 0.007 | 0.040 | S | 2.4 | 2.0 |
| 0036 | 1993 UW6 | 2457802.50174 | 0:55 | 13.7 | 19.21 | -0.08±0.04 | -0.09±0.07 | ≥0.15 | ≥55 | 0.001 | 0.131 | 0.868 | 0.000 | C | 3.0 | 6.2 |
| 0037 | 1993 UX2 | 2457730.42038 | 1:30 | 14.6 | 17.59 | 0.09±0.04 | 0.12±0.04 | 0.23±0.07 | 171±34 | 0.893 | 0.045 | 0.001 | 0.061 | S | 2.7 | 14.4 |
| 0038 | 1994 AZ9 | 2457691.37970 | 1:50 | 16.0 | 20.50 | 0.11±0.07 | 0.14±0.07 | 0.92±0.12 | 221±44 | 0.808 | 0.055 | 0.008 | 0.129 | S | 2.9 | 2.7 |
| 0039 | 1994 EQ1 | 2457690.33738 | 1:05 | 14.2 | 17.19 | 0.06±0.02 | 0.05±0.03 | ≥0.07 | ≥66 | 0.937 | 0.053 | 0.009 | 0.001 | S | 2.3 | 3.8 |
| 0040 | 1994 GS6 | 2457691.37970 | 1:54 | 17.0 | 19.95 | 0.09±0.03 | 0.12±0.04 | ≥0.32 | ≥114 | 0.907 | 0.025 | 0.000 | 0.068 | S | 2.3 | 4.8 |
| 0041 | 1994 SP7 | 2457692.44645 | 1:00 | 15.5 | 19.95 | -0.04±0.05 | -0.09±0.07 | ≥0.13 | ≥60 | 0.014 | 0.097 | 0.889 | 0.000 | C | 3.1 | 1.2 |
| 0042 | 1994 TH14 | 2457724.34161 | 0:57 | 15.3 | 19.13 | -0.06±0.05 | -0.08±0.06 | ≥0.15 | ≥58 | 0.008 | 0.118 | 0.875 | 0.000 | C | 3.1 | 1.7 |
| 0043 | 1995 BO6 | 2457691.54720 | 0:52 | 14.8 | 18.23 | 0.09±0.04 | 0.15±0.09 | ≥0.08 | ≥52 | 0.804 | 0.019 | 0.022 | 0.155 | S | 2.7 | 13.6 |
| 0044 | 1995 BV | 2457724.28943 | 1:08 | 13.0 | 19.00 | 0.07±0.03 | 0.20±0.04 | ≥0.15 | ≥68 | 0.453 | 0.003 | 0.000 | 0.544 | D | 3.2 | 2.1 |

*Table 1 continued on next page*

**Table 1** *(continued)*

| No.[a] | Object | Obs. Start (JD) | Obs. Duration (h:mm) | H[b] (mag) | V (mag) | V-R[c] (mag) | V-I[c] (mag) | Amplitude[d] (mag) | Rot. Period[d] (min) | S-type (probability) | X-type | C-type | D-type | Tax. | a[b] (au) | i[b] (deg) |
|---|---|---|---|---|---|---|---|---|---|---|---|---|---|---|---|---|
| 0045 | 1995 SX37 | 2457691.37970 | 1:06 | 15.8 | 19.72 | 0.07±0.04 | 0.10±0.05 | ≥0.15 | ≥66 | 0.741 | 0.137 | 0.025 | 0.097 | S | 2.6 | 0.4 |
| 0046 | 1995 UR57 | 2457692.38892 | 1:20 | 16.2 | 19.42 | 0.05±0.02 | 0.12±0.03 | ≥0.05 | ≥80 | 0.706 | 0.145 | 0.001 | 0.149 | S | 3.0 | 4.7 |
| 0047 | 1995 WU23 | 2457691.37970 | 1:54 | 16.6 | 20.57 | 0.07±0.08 | 0.16±0.10 | ≥0.30 | ≥114 | 0.541 | 0.120 | 0.036 | 0.303 | S | 2.6 | 3.2 |
| 0048 | 1995 XO2 | 2457802.50174 | 0:55 | 15.4 | 20.25 | 0.12±0.10 | 0.01±0.16 | ≥0.34 | ≥55 | 0.603 | 0.053 | 0.281 | 0.063 | S | 2.2 | 5.5 |
| 0049 | 1995 YS25 | 2457692.44645 | 1:00 | 16.4 | 19.42 | 0.13±0.04 | 0.29±0.06 | ≥0.05 | ≥60 | 0.845 | 0.000 | 0.000 | 0.155 | S | 2.3 | 5.7 |
| 0050 | 1996 DT | 2457802.50174 | 0:55 | 13.3 | 18.26 | 0.03±0.03 | 0.09±0.04 | ≥0.32 | ≥55 | 0.430 | 0.457 | 0.034 | 0.080 | - | 3.0 | 10.6 |
| 0051 | 1996 HL20 | 2457692.38892 | 1:20 | 13.7 | 18.57 | -0.07±0.02 | -0.12±0.03 | ≥0.04 | ≥80 | 0.000 | 0.000 | 1.000 | 0.000 | C | 3.2 | 2.5 |
| 0052 | 1996 HO20 | 2457724.34310 | 0:55 | 15.0 | 19.46 | 0.10±0.05 | 0.07±0.07 | ≥0.16 | ≥55 | 0.851 | 0.055 | 0.058 | 0.036 | S | 2.4 | 3.3 |
| 0053 | 1996 JD6 | 2457690.33738 | 1:17 | 14.9 | 17.79 | 0.12±0.02 | 0.01±0.03 | ≥0.06 | ≥78 | 0.997 | 0.000 | 0.003 | 0.000 | S | 2.4 | 6.1 |
| 0054 | 1996 OK1 | 2457798.46669 | 0:49 | 14.2 | 19.13 | 0.03±0.11 | 0.11±0.11 | ≥0.51 | ≥49 | 0.415 | 0.253 | 0.118 | 0.214 | - | 3.2 | 21.0 |
| 0055 | 1997 AW2 | 2457691.37970 | 1:54 | 13.5 | 18.53 | 0.08±0.02 | 0.23±0.02 | ≥0.44 | ≥114 | 0.353 | 0.000 | 0.000 | 0.647 | D | 3.0 | 4.6 |
| 0056 | 1997 AX1 | 2457798.50513 | 1:09 | 13.6 | 17.91 | 0.03±0.04 | 0.07±0.04 | ≥0.09 | ≥69 | 0.471 | 0.443 | 0.063 | 0.023 | - | 2.9 | 16.0 |
| 0057 | 1997 CJ1 | 2457690.33738 | 1:57 | 15.4 | 19.50 | 0.08±0.03 | 0.15±0.03 | ≥0.32 | ≥118 | 0.790 | 0.008 | 0.000 | 0.202 | S | 2.3 | 3.2 |
| 0058 | 1997 EG46 | 2457724.28943 | 1:08 | 12.6 | 17.82 | 0.08±0.02 | 0.30±0.03 | ≥0.07 | ≥68 | 0.485 | 0.000 | 0.000 | 0.515 | D | 3.1 | 8.2 |
| 0059 | 1997 EH13 | 2457724.34608 | 0:51 | 15.8 | 20.02 | 0.08±0.07 | -0.10±0.14 | ≥0.48 | ≥51 | 0.302 | 0.038 | 0.645 | 0.015 | C | 2.6 | 1.8 |
| 0060 | 1997 FT1 | 2457730.42038 | 1:30 | 14.5 | 17.68 | 0.06±0.03 | 0.06±0.04 | ≥0.12 | ≥90 | 0.778 | 0.176 | 0.039 | 0.006 | S | 2.6 | 12.5 |
| 0061 | 1997 GD8 | 2457724.28943 | 1:08 | 15.0 | 19.46 | 0.08±0.03 | 0.16±0.06 | ≥0.52 | ≥68 | 0.733 | 0.020 | 0.001 | 0.246 | S | 2.3 | 0.3 |
| 0062 | 1997 GL23 | 2457692.44645 | 1:00 | 13.3 | 17.31 | -0.09±0.04 | -0.09±0.04 | ≥0.15 | ≥60 | 0.000 | 0.034 | 0.966 | 0.000 | C | 3.1 | 5.1 |
| 0063 | 1997 GQ22 | 2457724.28943 | 1:08 | 13.4 | 18.35 | -0.04±0.02 | 0.04±0.03 | ≥0.08 | ≥68 | 0.002 | 0.931 | 0.067 | 0.000 | X | 3.2 | 12.9 |
| 0064 | 1997 GV21 | 2457724.35824 | 0:33 | 14.1 | 17.07 | 0.02±0.04 | 0.04±0.07 | ≥0.05 | ≥34 | 0.334 | 0.301 | 0.299 | 0.067 | - | 2.3 | 7.4 |
| 0065 | 1997 GZ7 | 2457692.38892 | 1:20 | 13.6 | 17.89 | -0.09±0.02 | -0.14±0.03 | 0.07±0.04 | 122±35 | 0.000 | 0.000 | 1.000 | 0.000 | C | 3.2 | 13.9 |
| 0066 | 1997 JQ14 | 2457692.44645 | 1:00 | 16.6 | 19.98 | 0.04±0.05 | 0.16±0.07 | ≥0.53 | ≥60 | 0.380 | 0.169 | 0.015 | 0.436 | - | 2.4 | 2.8 |
| 0067 | 1997 JS12 | 2457692.44645 | 1:00 | 14.4 | 19.64 | -0.09±0.04 | -0.08±0.06 | ≥0.10 | ≥60 | 0.000 | 0.109 | 0.890 | 0.000 | C | 3.1 | 0.7 |
| 0068 | 1997 KX | 2457722.42167 | 0:52 | 14.2 | 19.35 | 0.10±0.03 | 0.17±0.04 | ≥0.12 | ≥53 | 0.857 | 0.004 | 0.000 | 0.139 | S | 3.1 | 18.3 |
| 0069 | 1997 NG1 | 2457691.37970 | 0:55 | 15.3 | 14.12 | 0.20±0.02 | 0.34±0.03 | ≥0.07 | ≥56 | 1.000 | 0.000 | 0.000 | 0.000 | S | 2.8 | 2.5 |
| 0070 | 1997 PD1 | 2457690.33738 | 1:05 | 15.4 | 19.00 | 0.09±0.02 | 0.12±0.03 | ≥0.10 | ≥66 | 0.978 | 0.006 | 0.000 | 0.016 | S | 2.4 | 2.3 |
| 0071 | 1997 TK27 | 2457722.38400 | 1:47 | 16.8 | 19.41 | -0.06±0.03 | -0.09±0.04 | ≥0.36 | ≥107 | 0.000 | 0.042 | 0.959 | 0.000 | C | 2.4 | 3.6 |
| 0072 | 1997 UT20 | 2457724.28943 | 1:08 | 14.8 | 19.17 | 0.02±0.10 | -0.05±0.05 | ≥0.16 | ≥68 | 0.263 | 0.085 | 0.652 | 0.000 | C | 2.5 | 8.7 |
| 0073 | 1998 BG2 | 2457692.38892 | 1:20 | 13.8 | 16.69 | 0.08±0.02 | 0.10±0.03 | ≥0.10 | ≥80 | 0.985 | 0.010 | 0.000 | 0.006 | S | 2.2 | 1.2 |
| 0074 | 1998 CH1 | 2457692.44645 | 1:00 | 14.6 | 18.79 | 0.03±0.04 | 0.03±0.05 | ≥0.08 | ≥60 | 0.400 | 0.324 | 0.269 | 0.007 | - | 2.9 | 1.9 |
| 0075 | 1998 CY4 | 2457724.38316 | 1:08 | 14.4 | 18.88 | 0.13±0.02 | 0.10±0.03 | 0.20±0.04 | 137±25 | 1.000 | 0.000 | 0.000 | 0.000 | S | 2.9 | 1.2 |
| 0076 | 1998 FB6 | 2457722.38400 | 1:47 | 15.7 | 19.40 | 0.05±0.03 | 0.03±0.03 | ≥0.18 | ≥107 | 0.688 | 0.122 | 0.188 | 0.001 | S | 2.2 | 3.9 |
| 0077 | 1998 FG69 | 2457800.27020 | 1:03 | 12.7 | 18.29 | 0.07±0.03 | 0.15±0.04 | ≥0.06 | ≥64 | 0.738 | 0.028 | 0.000 | 0.234 | S | 3.0 | 10.2 |
| 0078 | 1998 FJ97 | 2457692.38892 | 1:20 | 14.3 | 18.83 | 0.02±0.02 | 0.05±0.03 | 0.21±0.04 | 158±37 | 0.409 | 0.439 | 0.152 | 0.000 | - | 3.0 | 8.9 |
| 0079 | 1998 FM12 | 2457722.38400 | 1:47 | 14.3 | 18.26 | 0.01±0.02 | 0.06±0.03 | ≥0.19 | ≥107 | 0.175 | 0.746 | 0.075 | 0.003 | X | 2.6 | 4.0 |
| 0080 | 1998 FR41 | 2457802.39311 | 0:45 | 14.5 | 18.94 | 0.07±0.03 | 0.09±0.04 | ≥0.06 | ≥46 | 0.888 | 0.086 | 0.003 | 0.024 | S | 2.5 | 11.5 |
| 0081 | 1998 FT121 | 2457690.33738 | 1:05 | 14.9 | 18.39 | 0.02±0.02 | 0.01±0.03 | ≥0.12 | ≥66 | 0.230 | 0.190 | 0.580 | 0.000 | C | 2.3 | 5.0 |
| 0082 | 1998 FV48 | 2457802.42659 | 1:45 | 15.4 | 18.45 | -0.01±0.02 | 0.03±0.04 | 0.19±0.04 | 170±36 | 0.038 | 0.570 | 0.389 | 0.003 | X | 2.4 | 7.1 |
| 0083 | 1998 HA142 | 2457691.46127 | 2:00 | 16.3 | 19.20 | 0.09±0.03 | -0.01±0.06 | ≥0.14 | ≥121 | 0.644 | 0.007 | 0.348 | 0.002 | S | 2.3 | 6.6 |
| 0084 | 1998 HC122 | 2457728.49022 | 0:57 | 12.9 | 18.10 | 0.08±0.05 | 0.20±0.06 | ≥0.05 | ≥58 | 0.592 | 0.019 | 0.000 | 0.389 | S | 3.0 | 11.3 |
| 0085 | 1998 HD145 | 2457690.42183 | 2:19 | 15.2 | 19.80 | 0.07±0.05 | 0.09±0.05 | ≥0.35 | ≥139 | 0.714 | 0.182 | 0.024 | 0.081 | S | 2.6 | 8.1 |
| 0086 | 1998 HF149 | 2457724.38316 | 1:08 | 15.5 | 19.79 | -0.02±0.04 | -0.08±0.04 | ≥0.25 | ≥69 | 0.011 | 0.055 | 0.934 | 0.000 | C | 2.6 | 11.7 |
| 0087 | 1998 HH100 | 2457802.50174 | 0:55 | 13.2 | 18.25 | -0.03±0.03 | -0.03±0.04 | ≥0.14 | ≥55 | 0.008 | 0.231 | 0.761 | 0.000 | C | 2.5 | 5.9 |
| 0088 | 1998 HO113 | 2457691.46127 | 2:00 | 15.2 | 20.54 | 0.01±0.07 | -0.02±0.10 | ≥0.42 | ≥121 | 0.202 | 0.217 | 0.543 | 0.037 | C | 3.0 | 6.0 |
| 0089 | 1998 HR6 | 2457724.39253 | 0:55 | 13.7 | 18.16 | 0.06±0.03 | 0.12±0.04 | ≥0.15 | ≥55 | 0.781 | 0.095 | 0.000 | 0.123 | S | 3.1 | 9.4 |
| 0090 | 1998 HS10 | 2457692.38892 | 1:20 | 16.5 | 19.56 | -0.04±0.02 | -0.03±0.03 | ≥0.15 | ≥80 | 0.001 | 0.334 | 0.665 | 0.000 | C | 2.3 | 4.0 |

*Table 1 continued on next page*

**Table 1** *(continued)*

| No.[a] | Object | Obs. Start (JD) | Obs. Duration (h:mm) | H[b] (mag) | V (mag) | V-R[c] (mag) | V-I[c] (mag) | Amplitude[d] (mag) | Rot. Period[d] (min) | S-type | X-type | C-type | D-type | Tax. | a[b] (au) | i[b] (deg) |
|---|---|---|---|---|---|---|---|---|---|---|---|---|---|---|---|---|
| | | | | | | | | | | (probability) | | | | | | |
| 0091 | 1998 HT4 | 2457724.28943 | 1:08 | 14.5 | 18.48 | 0.12±0.02 | 0.20±0.03 | 0.15±0.04 | 135±34 | 0.972 | 0.000 | 0.000 | 0.028 | S | 2.3 | 5.9 |
| 0092 | 1998 KK47 | 2457691.46127 | 2:00 | 15.1 | 18.88 | 0.01±0.03 | 0.08±0.04 | ≥0.10 | ≥121 | 0.185 | 0.727 | 0.045 | 0.043 | X | 2.3 | 7.5 |
| 0093 | 1998 KM19 | 2457802.42659 | 1:21 | 16.2 | 19.76 | -0.02±0.03 | 0.01±0.04 | ≥0.21 | ≥82 | 0.055 | 0.499 | 0.446 | 0.000 | - | 2.2 | 6.2 |
| 0094 | 1998 KO15 | 2457724.34161 | 0:57 | 14.7 | 18.25 | 0.09±0.04 | 0.21±0.05 | ≥0.10 | ≥58 | 0.616 | 0.004 | 0.000 | 0.380 | S | 2.5 | 13.6 |
| 0095 | 1998 KS54 | 2457690.42183 | 2:21 | 14.7 | 18.72 | 0.10±0.03 | 0.13±0.03 | ≥0.15 | ≥142 | 0.963 | 0.009 | 0.000 | 0.028 | S | 2.6 | 7.6 |
| 0096 | 1998 KT28 | 2457692.38892 | 1:20 | 13.2 | 17.03 | 0.05±0.02 | 0.10±0.03 | ≥0.06 | ≥80 | 0.748 | 0.222 | 0.000 | 0.030 | S | 3.1 | 9.5 |
| 0097 | 1998 OR5 | 2457800.44661 | 0:54 | 15.7 | 19.28 | -0.01±0.14 | 0.04±0.22 | ≥0.37 | ≥54 | 0.239 | 0.174 | 0.381 | 0.206 | - | 2.5 | 14.1 |
| 0098 | 1998 QC7 | 2457768.50255 | 2:00 | 13.3 | 18.62 | 0.16±0.10 | 0.18±0.11 | ≥0.30 | ≥120 | 0.817 | 0.034 | 0.014 | 0.135 | S | 3.0 | 11.0 |
| 0099 | 1998 QE107 | 2457800.44661 | 0:54 | 15.4 | 19.72 | 0.13±0.10 | 0.00±0.14 | ≥0.23 | ≥54 | 0.642 | 0.050 | 0.268 | 0.040 | S | 2.6 | 11.5 |
| 0100 | 1998 QN62 | 2457691.37970 | 1:54 | 14.8 | 18.25 | 0.06±0.02 | 0.05±0.02 | ≥0.39 | ≥114 | 0.943 | 0.046 | 0.011 | 0.000 | S | 2.3 | 6.6 |
| 0101 | 1998 QO26 | 2457724.28943 | 1:08 | 13.3 | 18.32 | -0.08±0.02 | -0.07±0.03 | ≥0.08 | ≥68 | 0.000 | 0.054 | 0.946 | 0.000 | C | 3.2 | 4.4 |
| 0102 | 1998 QP33 | 2457691.54720 | 1:27 | 14.5 | 17.99 | 0.12±0.03 | 0.17±0.04 | ≥0.11 | ≥87 | 0.977 | 0.000 | 0.000 | 0.023 | S | 2.7 | 15.9 |
| 0103 | 1998 QT23 | 2457802.50174 | 0:55 | 13.0 | 18.34 | 0.03±0.03 | 0.08±0.04 | ≥0.07 | ≥55 | 0.461 | 0.448 | 0.029 | 0.061 | - | 3.0 | 10.7 |
| 0104 | 1998 SA63 | 2457730.42038 | 1:30 | 14.3 | 19.26 | -0.02±0.04 | -0.00±0.05 | ≥0.39 | ≥90 | 0.062 | 0.437 | 0.499 | 0.001 | - | 3.2 | 7.8 |
| 0105 | 1998 SG142 | 2457691.46127 | 2:00 | 15.3 | 19.18 | 0.09±0.03 | 0.10±0.04 | 0.55±0.07 | 231±51 | 0.944 | 0.040 | 0.001 | 0.016 | S | 2.3 | 5.6 |
| 0106 | 1998 SU37 | 2457724.28943 | 1:08 | 14.5 | 18.58 | 0.14±0.02 | 0.08±0.03 | ≥0.17 | ≥68 | 1.000 | 0.000 | 0.000 | 0.000 | S | 2.3 | 6.9 |
| 0107 | 1998 SW81 | 2457691.37970 | 1:54 | 16.8 | 18.64 | 0.08±0.02 | 0.10±0.03 | ≥0.07 | ≥114 | 0.975 | 0.020 | 0.000 | 0.005 | S | 2.3 | 2.3 |
| 0108 | 1998 SY21 | 2457690.33738 | 1:49 | 15.5 | 17.85 | 0.07±0.02 | 0.14±0.03 | 0.38±0.04 | 219±54 | 0.909 | 0.007 | 0.000 | 0.083 | S | 2.3 | 6.0 |
| 0109 | 1998 TJ36 | 2457801.38674 | 1:43 | 15.2 | 20.23 | 0.02±0.18 | -0.15±0.23 | ≥1.40 | ≥104 | 0.221 | 0.091 | 0.633 | 0.055 | C | 2.6 | 15.5 |
| 0110 | 1998 TR12 | 2457692.44645 | 1:00 | 17.4 | 19.40 | -0.04±0.04 | 0.08±0.06 | ≥0.08 | ≥60 | 0.048 | 0.776 | 0.081 | 0.094 | X | 2.7 | 9.5 |
| 0111 | 1998 TU8 | 2457692.44645 | 1:00 | 16.4 | 19.63 | 0.01±0.05 | 0.05±0.06 | ≥0.14 | ≥60 | 0.331 | 0.414 | 0.215 | 0.040 | - | 2.7 | 1.2 |
| 0112 | 1998 UT9 | 2457724.34310 | 0:55 | 15.9 | 19.77 | 0.15±0.06 | 0.17±0.08 | ≥0.26 | ≥55 | 0.922 | 0.006 | 0.002 | 0.069 | S | 2.7 | 1.6 |
| 0113 | 1998 UY19 | 2457798.41862 | 1:06 | 12.1 | 17.04 | 0.11±0.03 | 0.24±0.05 | ≥0.03 | ≥67 | 0.803 | 0.000 | 0.000 | 0.197 | S | 3.1 | 16.9 |
| 0114 | 1998 VG28 | 2457801.38674 | 1:43 | 13.5 | 18.75 | -0.09±0.08 | -0.00±0.05 | ≥0.15 | ≥104 | 0.053 | 0.451 | 0.496 | 0.001 | - | 3.2 | 13.1 |
| 0115 | 1998 VH32 | 2457690.42183 | 2:21 | 16.4 | 18.23 | 0.16±0.02 | 0.22±0.03 | 0.23±0.06 | 187±45 | 0.999 | 0.000 | 0.000 | 0.001 | S | 2.4 | 6.1 |
| 0116 | 1998 VJ42 | 2457724.28943 | 1:08 | 15.9 | 19.62 | 0.12±0.04 | 0.27±0.05 | ≥0.17 | ≥68 | 0.809 | 0.000 | 0.000 | 0.190 | S | 2.4 | 2.9 |
| 0117 | 1998 VK32 | 2457801.38674 | 1:43 | 14.4 | 19.84 | 0.09±0.11 | 0.22±0.11 | ≥0.33 | ≥104 | 0.577 | 0.081 | 0.014 | 0.328 | S | 2.6 | 15.4 |
| 0118 | 1998 VP29 | 2457798.46669 | 0:53 | 14.6 | 19.74 | -0.00±0.22 | 0.06±0.17 | ≥0.51 | ≥53 | 0.353 | 0.209 | 0.298 | 0.140 | - | 3.2 | 5.7 |
| 0119 | 1998 VS6 | 2457802.40513 | 0:28 | 13.9 | 18.17 | 0.10±0.03 | 0.11±0.04 | ≥0.10 | ≥28 | 0.973 | 0.012 | 0.000 | 0.015 | S | 2.7 | 8.3 |
| 0120 | 1998 VT54 | 2457796.44027 | 0:58 | 13.6 | 18.01 | 0.07±0.04 | 0.11±0.05 | ≥0.12 | ≥59 | 0.774 | 0.102 | 0.013 | 0.111 | S | 2.6 | 14.3 |
| 0121 | 1998 VT7 | 2457798.50513 | 1:09 | 16.3 | 19.46 | 0.26±0.12 | 0.23±0.14 | ≥0.35 | ≥69 | 0.952 | 0.004 | 0.004 | 0.039 | S | 2.3 | 5.4 |
| 0122 | 1998 WJ42 | 2457724.38316 | 1:08 | 16.6 | 20.00 | -0.09±0.05 | -0.04±0.05 | 0.59±0.07 | 91±23 | 0.003 | 0.291 | 0.706 | 0.000 | C | 2.7 | 5.7 |
| 0123 | 1998 WN7 | 2457691.46127 | 1:01 | 14.5 | 18.04 | 0.12±0.03 | 0.18±0.04 | ≥0.28 | ≥61 | 0.962 | 0.000 | 0.000 | 0.037 | S | 2.7 | 10.6 |
| 0124 | 1998 WQ27 | 2457691.37970 | 1:54 | 17.2 | 20.84 | 0.11±0.10 | 0.14±0.11 | 0.64±0.17 | 128±33 | 0.691 | 0.085 | 0.057 | 0.167 | S | 2.4 | 2.8 |
| 0125 | 1998 WU10 | 2457724.34608 | 0:51 | 14.2 | 18.73 | 0.11±0.05 | 0.18±0.05 | ≥0.07 | ≥51 | 0.831 | 0.005 | 0.000 | 0.164 | S | 2.4 | 1.9 |
| 0126 | 1998 XJ15 | 2457722.38400 | 1:47 | 15.1 | 18.84 | 0.10±0.02 | 0.17±0.03 | ≥0.20 | ≥107 | 0.943 | 0.000 | 0.000 | 0.057 | S | 2.7 | 7.2 |
| 0127 | 1998 XM74 | 2457690.33738 | 1:05 | 14.8 | 19.17 | 0.17±0.03 | 0.23±0.03 | 0.30±0.04 | 132±37 | 0.999 | 0.000 | 0.000 | 0.001 | S | 2.4 | 3.1 |
| 0128 | 1999 AJ4 | 2457724.28943 | 1:08 | 15.6 | 18.58 | 0.05±0.02 | 0.10±0.03 | ≥0.12 | ≥68 | 0.711 | 0.206 | 0.002 | 0.080 | S | 2.4 | 1.5 |
| 0129 | 1999 AK9 | 2457690.33738 | 1:05 | 16.2 | 19.25 | 0.14±0.02 | 0.18±0.04 | ≥0.12 | ≥66 | 0.996 | 0.000 | 0.000 | 0.004 | S | 2.4 | 2.5 |
| 0130 | 1999 CD63 | 2457800.27020 | 1:07 | 13.3 | 19.06 | 0.06±0.04 | 0.11±0.05 | ≥0.56 | ≥68 | 0.690 | 0.159 | 0.019 | 0.132 | S | 2.8 | 9.3 |
| 0131 | 1999 CG70 | 2457798.50513 | 1:09 | 14.2 | 18.28 | -0.02±0.05 | -0.03±0.06 | ≥0.24 | ≥69 | 0.059 | 0.271 | 0.669 | 0.001 | C | 2.7 | 7.1 |
| 0132 | 1999 CL83 | 2457724.34310 | 0:55 | 14.2 | 17.26 | -0.06±0.04 | 0.01±0.04 | ≥0.11 | ≥55 | 0.003 | 0.657 | 0.339 | 0.000 | X | 2.8 | 8.9 |
| 0133 | 1999 CN62 | 2457728.49022 | 0:57 | 14.9 | 18.21 | 0.02±0.06 | -0.07±0.07 | ≥0.09 | ≥58 | 0.131 | 0.085 | 0.783 | 0.001 | C | 2.4 | 5.4 |
| 0134 | 1999 CO20 | 2457802.50174 | 0:55 | 14.1 | 19.73 | -0.04±0.06 | -0.23±0.08 | ≥0.32 | ≥55 | 0.001 | 0.002 | 0.997 | 0.000 | C | 3.2 | 6.2 |
| 0135 | 1999 CW70 | 2457802.50174 | 0:55 | 14.8 | 19.24 | 0.07±0.05 | 0.10±0.07 | ≥0.21 | ≥55 | 0.729 | 0.105 | 0.042 | 0.124 | S | 2.7 | 14.0 |
| 0136 | 1999 FK6 | 2457724.38316 | 1:08 | 14.8 | 18.76 | 0.09±0.02 | 0.13±0.03 | ≥0.08 | ≥69 | 0.954 | 0.009 | 0.000 | 0.037 | S | 2.9 | 2.1 |

*Table 1 continued on next page*

**Table 1** *(continued)*

| No.[a] | Object | Obs. Start | Obs. Duration | H[b] | V | V-R[c] | V-I[c] | Amplitude[d] | Rot. Period[d] | S-type | X-type | C-type | D-type | Tax. | a[b] | i[b] |
|---|---|---|---|---|---|---|---|---|---|---|---|---|---|---|---|---|
| | | (JD) | (h:mm) | (mag) | (mag) | (mag) | (mag) | (mag) | (min) | (probability) | | | | | (au) | (deg) |
| 0137 | 1999 FP4 | 2457798.50513 | 1:09 | 16.6 | 19.16 | 0.28±0.10 | 0.37±0.08 | ≥0.29 | ≥69 | 0.983 | 0.000 | 0.000 | 0.017 | S | 2.4 | 7.7 |
| 0138 | 1999 GJ10 | 2457724.36139 | 0:29 | 15.0 | 20.31 | 0.14±0.12 | 0.13±0.22 | ≥0.25 | ≥29 | 0.676 | 0.046 | 0.156 | 0.122 | S | 2.9 | 1.1 |
| 0139 | 1999 HS2 | 2457691.37970 | 1:52 | 14.7 | 18.70 | 0.07±0.02 | 0.11±0.03 | 0.43±0.05 | 176±41 | 0.947 | 0.034 | 0.000 | 0.019 | S | 2.4 | 2.2 |
| 0140 | 1999 JG112 | 2457692.44645 | 1:00 | 14.0 | 17.53 | 0.03±0.04 | 0.10±0.05 | ≥0.18 | ≥60 | 0.475 | 0.365 | 0.029 | 0.131 | - | 2.9 | 2.9 |
| 0141 | 1999 PB4 | 2457690.33738 | 1:57 | 14.5 | 18.00 | -0.06±0.02 | -0.06±0.03 | ≥0.28 | ≥118 | 0.000 | 0.067 | 0.933 | 0.000 | C | 3.2 | 3.0 |
| 0142 | 1999 PR4 | 2457722.38400 | 1:47 | 15.7 | 17.64 | 0.05±0.02 | 0.04±0.03 | 0.16±0.04 | 150±44 | 0.899 | 0.068 | 0.033 | 0.000 | S | 2.2 | 3.4 |
| 0143 | 1999 RA105 | 2457724.34310 | 0:29 | 13.2 | 18.45 | -0.09±0.06 | 0.00±0.08 | ≥0.24 | ≥29 | 0.015 | 0.520 | 0.443 | 0.021 | X | 3.1 | 3.9 |
| 0144 | 1999 RC182 | 2457692.44645 | 1:00 | 15.8 | 19.17 | 0.06±0.04 | 0.12±0.05 | ≥0.49 | ≥60 | 0.652 | 0.142 | 0.006 | 0.200 | S | 2.2 | 2.0 |
| 0145 | 1999 RC93 | 2457800.27321 | 0:59 | 14.7 | 20.15 | 0.02±0.10 | 0.19±0.13 | ≥0.29 | ≥59 | 0.331 | 0.165 | 0.047 | 0.457 | - | 3.1 | 9.6 |
| 0146 | 1999 RE190 | 2457802.38260 | 1:00 | 13.9 | 16.47 | -0.00±0.02 | 0.01±0.03 | 0.17±0.04 | 114±27 | 0.110 | 0.382 | 0.508 | 0.000 | C | 2.5 | 7.2 |
| 0147 | 1999 RE190 | 2457803.38332 | 0:58 | 13.9 | 16.43 | -0.05±0.04 | 0.00±0.05 | 0.20±0.04 | 118±18 | 0.011 | 0.570 | 0.419 | 0.001 | X | 2.5 | 7.2 |
| 0148 | 1999 RF73 | 2457724.34161 | 0:57 | 16.5 | 19.03 | 0.06±0.05 | 0.07±0.06 | ≥0.16 | ≥58 | 0.681 | 0.185 | 0.084 | 0.050 | S | 2.2 | 3.1 |
| 0149 | 1999 RH149 | 2457800.27020 | 1:07 | 14.0 | 20.04 | 0.13±0.14 | 0.15±0.16 | ≥0.25 | ≥68 | 0.673 | 0.079 | 0.083 | 0.165 | S | 3.0 | 10.2 |
| 0150 | 1999 RH15 | 2457766.28699 | 2:15 | 13.7 | 18.60 | 0.02±0.09 | 0.15±0.04 | ≥0.22 | ≥135 | 0.390 | 0.285 | 0.001 | 0.324 | - | 2.6 | 13.5 |
| 0151 | 1999 RN63 | 2457722.38400 | 1:47 | 17.1 | 20.02 | -0.01±0.04 | -0.06±0.05 | ≥0.18 | ≥107 | 0.032 | 0.097 | 0.871 | 0.000 | C | 2.2 | 4.7 |
| 0152 | 1999 RO136 | 2457691.40704 | 1:15 | 15.8 | 19.01 | 0.06±0.03 | 0.08±0.04 | ≥0.15 | ≥75 | 0.780 | 0.186 | 0.013 | 0.021 | S | 2.6 | 0.5 |
| 0153 | 1999 RO211 | 2457691.37970 | 1:54 | 13.6 | 18.12 | -0.05±0.02 | -0.06±0.02 | ≥0.11 | ≥114 | 0.000 | 0.063 | 0.937 | 0.000 | C | 3.2 | 27.5 |
| 0154 | 1999 RO34 | 2457800.27020 | 1:03 | 13.0 | 18.53 | 0.05±0.03 | 0.11±0.04 | ≥0.13 | ≥64 | 0.631 | 0.212 | 0.007 | 0.150 | S | 3.1 | 9.7 |
| 0155 | 1999 RO51 | 2457692.44645 | 1:00 | 15.7 | 18.84 | 0.02±0.04 | 0.08±0.05 | ≥0.10 | ≥60 | 0.386 | 0.460 | 0.075 | 0.079 | - | 2.6 | 4.0 |
| 0156 | 1999 RT85 | 2457800.27020 | 1:03 | 13.8 | 19.26 | 0.06±0.05 | 0.03±0.11 | ≥0.12 | ≥64 | 0.482 | 0.111 | 0.326 | 0.081 | - | 3.0 | 10.0 |
| 0157 | 1999 RT90 | 2457730.42038 | 1:30 | 14.6 | 18.99 | 0.06±0.03 | 0.11±0.03 | ≥0.17 | ≥90 | 0.726 | 0.156 | 0.001 | 0.118 | S | 3.1 | 8.3 |
| 0158 | 1999 RU197 | 2457724.28943 | 1:08 | 13.5 | 18.27 | 0.09±0.03 | 0.16±0.03 | 0.21±0.05 | 136±28 | 0.854 | 0.003 | 0.000 | 0.143 | S | 2.6 | 14.0 |
| 0159 | 1999 RV175 | 2457722.38400 | 1:47 | 15.2 | 19.43 | 0.05±0.03 | 0.09±0.04 | ≥0.34 | ≥107 | 0.692 | 0.235 | 0.020 | 0.053 | S | 2.6 | 4.6 |
| 0160 | 1999 SG6 | 2457800.27020 | 1:07 | 13.1 | 18.58 | 0.05±0.04 | 0.12±0.05 | 0.18±0.07 | 136±40 | 0.642 | 0.173 | 0.008 | 0.176 | S | 3.1 | 11.7 |
| 0161 | 1999 SJ26 | 2457724.38316 | 1:08 | 15.0 | 18.23 | 0.07±0.02 | 0.08±0.03 | ≥0.06 | ≥69 | 0.937 | 0.058 | 0.001 | 0.004 | S | 2.6 | 10.1 |
| 0162 | 1999 SV17 | 2457765.28694 | 1:01 | 14.9 | 19.18 | -0.01±0.05 | -0.00±0.07 | ≥0.31 | ≥62 | 0.137 | 0.338 | 0.511 | 0.013 | C | 2.6 | 12.0 |
| 0163 | 1999 TB121 | 2457692.44645 | 1:00 | 14.2 | 18.54 | -0.07±0.04 | -0.05±0.05 | ≥0.10 | ≥60 | 0.001 | 0.247 | 0.752 | 0.000 | C | 3.1 | 1.4 |
| 0164 | 1999 TB29 | 2457692.44971 | 0:55 | 15.5 | 18.37 | 0.09±0.05 | 0.22±0.06 | ≥0.70 | ≥55 | 0.592 | 0.007 | 0.000 | 0.402 | S | 2.6 | 4.1 |
| 0165 | 1999 TB91 | 2457724.38316 | 1:08 | 16.1 | 18.66 | 0.06±0.02 | 0.09±0.03 | ≥0.05 | ≥69 | 0.858 | 0.114 | 0.002 | 0.026 | S | 2.2 | 3.1 |
| 0166 | 1999 TC171 | 2457692.44645 | 1:00 | 14.2 | 18.23 | -0.02±0.04 | 0.04±0.06 | ≥0.04 | ≥60 | 0.101 | 0.656 | 0.215 | 0.028 | X | 3.2 | 2.5 |
| 0167 | 1999 TE119 | 2457692.44645 | 1:00 | 14.6 | 18.53 | -0.06±0.04 | 0.01±0.05 | ≥0.10 | ≥60 | 0.011 | 0.604 | 0.382 | 0.003 | X | 3.1 | 2.2 |
| 0168 | 1999 TE7 | 2457692.38892 | 1:20 | 15.1 | 18.43 | -0.06±0.02 | -0.08±0.03 | ≥0.06 | ≥80 | 0.000 | 0.037 | 0.963 | 0.000 | C | 2.6 | 4.8 |
| 0169 | 1999 TF222 | 2457724.34310 | 0:55 | 14.1 | 18.80 | 0.04±0.04 | 0.08±0.06 | ≥0.31 | ≥55 | 0.544 | 0.289 | 0.072 | 0.095 | S | 3.1 | 8.4 |
| 0170 | 1999 TH63 | 2457724.38316 | 1:08 | 14.1 | 19.03 | -0.10±0.03 | -0.12±0.03 | ≥0.20 | ≥69 | 0.000 | 0.002 | 0.999 | 0.000 | C | 3.2 | 1.9 |
| 0171 | 1999 TK127 | 2457722.38400 | 1:47 | 16.3 | 19.23 | 0.04±0.02 | 0.06±0.03 | ≥0.39 | ≥107 | 0.646 | 0.271 | 0.078 | 0.005 | S | 2.6 | 4.2 |
| 0172 | 1999 TK129 | 2457802.43399 | 1:34 | 15.2 | 19.96 | -0.15±0.04 | -0.13±0.05 | ≥0.17 | ≥95 | 0.000 | 0.004 | 0.996 | 0.000 | C | 2.9 | 8.8 |
| 0173 | 1999 TK139 | 2457724.28943 | 1:08 | 15.1 | 20.26 | -0.06±0.06 | -0.09±0.12 | ≥0.56 | ≥68 | 0.015 | 0.205 | 0.761 | 0.019 | C | 2.6 | 4.3 |
| 0174 | 1999 TM127 | 2457692.38892 | 1:20 | 15.8 | 19.43 | 0.07±0.02 | 0.12±0.04 | ≥0.17 | ≥80 | 0.892 | 0.035 | 0.000 | 0.073 | S | 2.6 | 3.9 |
| 0175 | 1999 TN33 | 2457724.38316 | 1:08 | 15.9 | 20.13 | 0.05±0.06 | 0.15±0.08 | ≥0.27 | ≥69 | 0.491 | 0.138 | 0.031 | 0.341 | - | 2.6 | 3.6 |
| 0176 | 1999 TO144 | 2457692.39719 | 1:08 | 16.5 | 18.98 | 0.04±0.02 | 0.07±0.03 | ≥0.07 | ≥68 | 0.685 | 0.276 | 0.028 | 0.011 | S | 2.2 | 3.0 |
| 0177 | 1999 TO7 | 2457724.34310 | 0:55 | 13.6 | 16.99 | 0.08±0.04 | 0.17±0.04 | ≥0.41 | ≥55 | 0.649 | 0.025 | 0.000 | 0.326 | S | 2.6 | 14.4 |
| 0178 | 1999 TP18 | 2457692.38892 | 1:20 | 14.8 | 18.89 | -0.05±0.02 | -0.06±0.03 | ≥0.13 | ≥80 | 0.000 | 0.113 | 0.887 | 0.000 | C | 3.2 | 8.9 |
| 0179 | 1999 TU197 | 2457801.38966 | 1:39 | 14.5 | 19.73 | 0.18±0.09 | 0.31±0.08 | ≥0.52 | ≥99 | 0.901 | 0.001 | 0.000 | 0.098 | S | 3.0 | 13.3 |
| 0180 | 1999 TX93 | 2457765.29319 | 0:52 | 13.0 | 17.65 | -0.04±0.03 | -0.06±0.05 | ≥0.18 | ≥53 | 0.002 | 0.144 | 0.855 | 0.000 | C | 2.6 | 11.5 |
| 0181 | 1999 TY98 | 2457690.42183 | 2:21 | 16.2 | 19.25 | 0.03±0.04 | -0.00±0.04 | ≥0.19 | ≥142 | 0.374 | 0.134 | 0.491 | 0.001 | - | 2.2 | 7.2 |
| 0182 | 1999 UG13 | 2457800.27020 | 1:03 | 13.7 | 19.55 | 0.07±0.04 | 0.17±0.06 | ≥0.15 | ≥64 | 0.598 | 0.059 | 0.003 | 0.340 | S | 3.1 | 9.5 |

*Table 1 continued on next page*

**Table 1** *(continued)*

| No.[a] | Object | Obs. Start (JD) | Obs. Duration (h:mm) | H[b] (mag) | V (mag) | V-R[c] (mag) | V-I[c] (mag) | Amplitude[d] (mag) | Rot. Period[d] (min) | S-type | X-type | C-type | D-type | Tax. | a[b] (au) | i[b] (deg) |
|---|---|---|---|---|---|---|---|---|---|---|---|---|---|---|---|---|
| | | | | | | | | | | (probability) | | | | | | |
| 0183 | 1999 UG15 | 2457691.46127 | 2:00 | 13.9 | 18.95 | -0.01±0.03 | 0.05±0.04 | ≥0.17 | ≥121 | 0.104 | 0.714 | 0.177 | 0.005 | X | 3.1 | 17.8 |
| 0184 | 1999 UL41 | 2457728.49022 | 0:55 | 14.8 | 19.54 | 0.22±0.13 | -0.02±0.14 | ≥0.22 | ≥55 | 0.804 | 0.018 | 0.166 | 0.012 | S | 3.1 | 9.1 |
| 0185 | 1999 UO42 | 2457802.38260 | 1:00 | 13.1 | 18.08 | 0.03±0.02 | 0.08±0.03 | ≥0.04 | ≥61 | 0.414 | 0.551 | 0.022 | 0.012 | X | 3.0 | 9.6 |
| 0186 | 1999 UQ12 | 2457691.37970 | 0:59 | 15.7 | 17.86 | -0.01±0.02 | -0.02±0.03 | ≥0.06 | ≥60 | 0.012 | 0.240 | 0.749 | 0.000 | C | 2.3 | 6.3 |
| 0187 | 1999 UQ52 | 2457730.42038 | 1:30 | 16.2 | 19.14 | 0.07±0.03 | 0.09±0.04 | ≥0.14 | ≥90 | 0.842 | 0.129 | 0.005 | 0.024 | S | 2.6 | 8.2 |
| 0188 | 1999 US51 | 2457692.44645 | 1:00 | 15.3 | 19.17 | 0.03±0.05 | 0.04±0.09 | ≥0.19 | ≥60 | 0.373 | 0.205 | 0.322 | 0.100 | - | 2.6 | 5.5 |
| 0189 | 1999 VA64 | 2457690.33738 | 1:57 | 13.8 | 19.02 | -0.10±0.07 | -0.08±0.04 | 0.76±0.08 | 235±41 | 0.006 | 0.055 | 0.940 | 0.000 | C | 3.2 | 7.6 |
| 0190 | 1999 VC163 | 2457691.37970 | 1:54 | 16.8 | 19.02 | 0.07±0.03 | 0.08±0.03 | ≥0.21 | ≥114 | 0.892 | 0.101 | 0.003 | 0.004 | S | 2.3 | 5.2 |
| 0191 | 1999 VF19 | 2457724.34310 | 0:55 | 15.5 | 18.98 | 0.12±0.04 | 0.10±0.06 | ≥0.13 | ≥55 | 0.961 | 0.013 | 0.004 | 0.023 | S | 2.6 | 3.4 |
| 0192 | 1999 VF199 | 2457800.27020 | 1:07 | 14.5 | 19.94 | -0.12±0.17 | -0.15±0.10 | ≥0.33 | ≥68 | 0.049 | 0.043 | 0.908 | 0.000 | C | 3.1 | 11.3 |
| 0193 | 1999 VG3 | 2457724.38316 | 1:08 | 14.0 | 18.01 | -0.01±0.02 | -0.07±0.03 | ≥0.10 | ≥69 | 0.001 | 0.015 | 0.984 | 0.000 | C | 3.2 | 1.9 |
| 0194 | 1999 VH126 | 2457691.37970 | 1:52 | 15.7 | 19.19 | 0.06±0.03 | 0.08±0.03 | ≥0.17 | ≥113 | 0.800 | 0.177 | 0.009 | 0.014 | S | 2.6 | 1.1 |
| 0195 | 1999 VH168 | 2457724.28943 | 1:08 | 16.2 | 19.89 | 0.05±0.05 | -0.03±0.06 | ≥0.24 | ≥68 | 0.365 | 0.083 | 0.550 | 0.002 | C | 2.3 | 5.9 |
| 0196 | 1999 VJ94 | 2457724.34310 | 0:55 | 15.5 | 19.32 | 0.17±0.05 | 0.18±0.06 | ≥0.23 | ≥55 | 0.977 | 0.001 | 0.000 | 0.021 | S | 2.6 | 2.3 |
| 0197 | 1999 VK96 | 2457724.34161 | 0:57 | 15.2 | 19.88 | -0.04±0.07 | -0.12±0.11 | ≥0.18 | ≥58 | 0.039 | 0.113 | 0.844 | 0.004 | C | 3.2 | 0.6 |
| 0198 | 1999 VL208 | 2457766.28699 | 1:41 | 13.7 | 18.99 | 0.06±0.05 | 0.07±0.06 | ≥0.57 | ≥101 | 0.648 | 0.198 | 0.089 | 0.066 | S | 3.1 | 18.0 |
| 0199 | 1999 VL82 | 2457692.38892 | 1:20 | 16.4 | 18.79 | 0.05±0.02 | 0.07±0.03 | ≥0.09 | ≥80 | 0.776 | 0.203 | 0.015 | 0.006 | S | 2.3 | 1.8 |
| 0200 | 1999 VN66 | 2457692.38892 | 1:20 | 14.7 | 18.43 | -0.00±0.02 | 0.03±0.03 | ≥0.06 | ≥80 | 0.084 | 0.538 | 0.379 | 0.000 | X | 3.2 | 3.7 |
| 0201 | 1999 VO130 | 2457692.38892 | 1:20 | 15.4 | 18.96 | -0.05±0.02 | -0.08±0.03 | ≥0.12 | ≥80 | 0.000 | 0.034 | 0.966 | 0.000 | C | 3.2 | 6.4 |
| 0202 | 1999 VP105 | 2457692.44645 | 1:00 | 16.3 | 19.94 | 0.09±0.06 | 0.06±0.07 | ≥0.27 | ≥60 | 0.802 | 0.080 | 0.080 | 0.038 | S | 2.3 | 3.7 |
| 0203 | 1999 VP165 | 2457724.28943 | 1:08 | 15.2 | 19.43 | -0.01±0.03 | 0.04±0.06 | ≥0.24 | ≥68 | 0.080 | 0.568 | 0.300 | 0.052 | X | 2.6 | 11.6 |
| 0204 | 1999 VR198 | 2457690.48064 | 0:52 | 16.0 | 19.49 | 0.05±0.07 | 0.08±0.07 | ≥0.22 | ≥53 | 0.557 | 0.238 | 0.089 | 0.117 | S | 2.3 | 8.8 |
| 0205 | 1999 VS53 | 2457692.44645 | 1:00 | 13.6 | 18.40 | -0.10±0.04 | -0.04±0.05 | ≥0.08 | ≥60 | 0.000 | 0.293 | 0.707 | 0.000 | C | 3.2 | 17.0 |
| 0206 | 1999 VX46 | 2457724.34310 | 0:55 | 15.5 | 18.53 | 0.02±0.04 | 0.07±0.05 | ≥0.12 | ≥55 | 0.369 | 0.494 | 0.082 | 0.055 | - | 2.3 | 10.0 |
| 0207 | 1999 WC18 | 2457724.34161 | 0:57 | 14.8 | 20.26 | 0.26±0.10 | 0.14±0.10 | ≥0.30 | ≥58 | 0.981 | 0.002 | 0.005 | 0.012 | S | 3.2 | 13.7 |
| 0208 | 1999 XG74 | 2457724.34608 | 0:51 | 15.1 | 18.50 | -0.09±0.05 | -0.01±0.05 | ≥0.08 | ≥51 | 0.003 | 0.490 | 0.506 | 0.001 | C | 2.6 | 3.2 |
| 0209 | 1999 XJ131 | 2457802.39604 | 0:41 | 15.1 | 19.43 | -0.07±0.04 | -0.11±0.06 | ≥0.14 | ≥41 | 0.001 | 0.059 | 0.941 | 0.000 | C | 3.1 | 14.8 |
| 0210 | 1999 XL64 | 2457692.44645 | 0:40 | 13.9 | 18.32 | -0.08±0.05 | -0.10±0.05 | ≥0.09 | ≥40 | 0.001 | 0.052 | 0.947 | 0.000 | C | 3.2 | 0.4 |
| 0211 | 1999 XO5 | 2457692.44645 | 1:00 | 16.8 | 18.95 | 0.07±0.04 | 0.06±0.05 | ≥0.08 | ≥60 | 0.753 | 0.130 | 0.090 | 0.027 | S | 2.3 | 2.0 |
| 0212 | 1999 XS4 | 2457692.44645 | 1:00 | 14.9 | 18.70 | -0.07±0.04 | -0.06±0.04 | ≥0.10 | ≥60 | 0.001 | 0.144 | 0.855 | 0.000 | C | 3.2 | 5.0 |
| 0213 | 1999 XZ20 | 2457690.33738 | 1:57 | 15.9 | 17.44 | 0.10±0.01 | 0.15±0.02 | ≥0.83 | ≥118 | 0.994 | 0.000 | 0.000 | 0.006 | S | 2.3 | 7.5 |
| 0214 | 2000 AC165 | 2457724.28943 | 1:08 | 12.7 | 18.08 | -0.06±0.02 | -0.06±0.03 | ≥0.06 | ≥68 | 0.000 | 0.086 | 0.914 | 0.000 | C | 3.2 | 14.4 |
| 0215 | 2000 AC62 | 2457730.42646 | 1:21 | 14.3 | 18.36 | 0.00±0.03 | 0.01±0.03 | ≥0.23 | ≥81 | 0.168 | 0.349 | 0.482 | 0.000 | - | 3.2 | 11.8 |
| 0216 | 2000 AF82 | 2457798.51082 | 1:01 | 14.7 | 20.05 | 0.24±0.27 | 0.54±0.36 | ≥0.89 | ≥61 | 0.788 | 0.021 | 0.031 | 0.160 | S | 3.1 | 7.7 |
| 0217 | 2000 AG38 | 2457724.34310 | 0:55 | 16.3 | 20.11 | 0.14±0.08 | 0.17±0.13 | ≥0.30 | ≥55 | 0.823 | 0.024 | 0.034 | 0.120 | S | 2.3 | 4.7 |
| 0218 | 2000 AK211 | 2457798.46669 | 0:53 | 15.1 | 19.52 | 0.10±0.12 | 0.19±0.20 | ≥0.31 | ≥53 | 0.587 | 0.071 | 0.111 | 0.231 | S | 2.6 | 11.8 |
| 0219 | 2000 AK62 | 2457796.48277 | 0:51 | 13.0 | 17.58 | 0.07±0.07 | 0.02±0.07 | ≥0.13 | ≥52 | 0.581 | 0.139 | 0.263 | 0.016 | S | 3.1 | 12.0 |
| 0220 | 2000 AM66 | 2457691.46127 | 2:00 | 15.2 | 18.53 | 0.11±0.03 | 0.04±0.03 | ≥0.17 | ≥121 | 0.992 | 0.003 | 0.005 | 0.000 | S | 2.3 | 7.5 |
| 0221 | 2000 AN170 | 2457728.49022 | 0:57 | 13.7 | 18.21 | -0.05±0.04 | -0.09±0.07 | ≥0.16 | ≥58 | 0.004 | 0.132 | 0.863 | 0.001 | C | 3.2 | 23.3 |
| 0222 | 2000 AQ75 | 2457768.50255 | 2:00 | 14.8 | 18.57 | 0.15±0.08 | 0.12±0.09 | ≥0.40 | ≥120 | 0.873 | 0.031 | 0.026 | 0.070 | S | 2.2 | 3.8 |
| 0223 | 2000 AV178 | 2457800.27020 | 1:07 | 14.7 | 19.70 | 0.05±0.15 | -0.02±0.12 | ≥0.18 | ≥68 | 0.381 | 0.158 | 0.420 | 0.041 | - | 2.7 | 13.4 |
| 0224 | 2000 AZ65 | 2457722.38979 | 1:38 | 15.1 | 18.75 | 0.13±0.02 | 0.07±0.03 | ≥0.12 | ≥99 | 1.000 | 0.000 | 0.000 | 0.000 | S | 2.3 | 7.1 |
| 0225 | 2000 BQ30 | 2457800.44661 | 0:54 | 14.9 | 20.32 | 0.19±0.18 | 0.38±0.25 | ≥0.32 | ≥54 | 0.791 | 0.024 | 0.028 | 0.157 | S | 3.1 | 21.6 |
| 0226 | 2000 BR28 | 2457724.28943 | 1:08 | 16.1 | 20.05 | -0.01±0.06 | 0.08±0.08 | ≥0.29 | ≥68 | 0.189 | 0.468 | 0.146 | 0.198 | - | 2.3 | 2.2 |
| 0227 | 2000 CB70 | 2457728.49613 | 0:49 | 15.3 | 19.42 | 0.12±0.07 | 0.18±0.11 | ≥0.57 | ≥49 | 0.780 | 0.033 | 0.021 | 0.166 | S | 2.3 | 5.6 |
| 0228 | 2000 CM51 | 2457798.46669 | 0:53 | 13.9 | 17.68 | -0.08±0.03 | -0.11±0.04 | ≥0.04 | ≥53 | 0.000 | 0.016 | 0.984 | 0.000 | C | 3.2 | 4.4 |

*Table 1 continued on next page*

**Table 1** *(continued)*

| No.[a] | Object | Obs. Start (JD) | Obs. Duration (h:mm) | $H^b$ (mag) | V (mag) | V-R[c] (mag) | V-I[c] (mag) | Amplitude[d] (mag) | Rot. Period[d] (min) | S-type | X-type (probability) | C-type | D-type | Tax. | a[b] (au) | i[b] (deg) |
|---|---|---|---|---|---|---|---|---|---|---|---|---|---|---|---|---|
| 0229 | 2000 CN16 | 2457796.48277 | 0:51 | 16.2 | 19.05 | -0.05±0.19 | 0.15±0.25 | ≥0.55 | ≥52 | 0.267 | 0.183 | 0.248 | 0.302 | - | 2.2 | 3.6 |
| 0230 | 2000 CO79 | 2457692.38892 | 1:20 | 16.8 | 19.66 | 0.05±0.03 | 0.06±0.04 | ≥0.75 | ≥80 | 0.662 | 0.246 | 0.070 | 0.021 | S | 2.4 | 1.6 |
| 0231 | 2000 CR97 | 2457692.38892 | 1:20 | 15.6 | 19.24 | 0.05±0.02 | 0.08±0.03 | ≥0.17 | ≥80 | 0.691 | 0.276 | 0.014 | 0.019 | S | 2.4 | 0.7 |
| 0232 | 2000 CW114 | 2457800.27321 | 1:01 | 15.5 | 20.26 | 0.22±0.08 | 0.31±0.11 | ≥0.31 | ≥61 | 0.969 | 0.001 | 0.000 | 0.031 | S | 2.7 | 15.3 |
| 0233 | 2000 DG25 | 2457728.49022 | 0:57 | 14.9 | 19.66 | 0.23±0.07 | 0.29±0.08 | ≥0.24 | ≥58 | 0.980 | 0.000 | 0.000 | 0.020 | S | 2.7 | 6.3 |
| 0234 | 2000 DK103 | 2457690.33738 | 1:20 | 16.8 | 19.76 | 0.10±0.03 | 0.10±0.05 | ≥0.14 | ≥81 | 0.968 | 0.011 | 0.001 | 0.019 | S | 2.3 | 5.6 |
| 0235 | 2000 DK48 | 2457728.49022 | 0:57 | 16.9 | 19.89 | 0.09±0.09 | 0.27±0.09 | ≥0.34 | ≥58 | 0.568 | 0.026 | 0.001 | 0.404 | S | 2.3 | 5.4 |
| 0236 | 2000 DN3 | 2457691.37970 | 1:54 | 13.8 | 17.91 | 0.06±0.02 | 0.10±0.02 | 0.20±0.04 | 224±51 | 0.857 | 0.136 | 0.000 | 0.007 | S | 2.8 | 10.0 |
| 0237 | 2000 DS59 | 2457722.38400 | 1:47 | 16.1 | 20.25 | 0.06±0.04 | 0.10±0.07 | ≥0.20 | ≥107 | 0.677 | 0.130 | 0.043 | 0.150 | S | 2.7 | 5.2 |
| 0238 | 2000 ED23 | 2457692.38892 | 1:20 | 17.3 | 20.35 | 0.09±0.04 | 0.12±0.05 | ≥0.18 | ≥80 | 0.833 | 0.064 | 0.005 | 0.099 | S | 2.4 | 2.0 |
| 0239 | 2000 ED92 | 2457798.46669 | 0:53 | 13.3 | 17.67 | -0.02±0.03 | -0.06±0.04 | ≥0.07 | ≥53 | 0.007 | 0.125 | 0.869 | 0.000 | C | 3.2 | 3.9 |
| 0240 | 2000 EE128 | 2457692.44645 | 1:00 | 14.8 | 18.39 | 0.08±0.04 | 0.27±0.05 | ≥0.14 | ≥60 | 0.538 | 0.000 | 0.000 | 0.462 | S | 2.8 | 3.5 |
| 0241 | 2000 EQ134 | 2457692.38892 | 1:20 | 16.0 | 19.70 | 0.10±0.03 | 0.10±0.03 | ≥0.18 | ≥80 | 0.984 | 0.009 | 0.000 | 0.007 | S | 2.4 | 3.9 |
| 0242 | 2000 ER111 | 2457690.33738 | 1:17 | 13.0 | 16.20 | -0.06±0.02 | -0.03±0.02 | ≥0.05 | ≥78 | 0.000 | 0.345 | 0.655 | 0.000 | C | 2.8 | 18.4 |
| 0243 | 2000 ER180 | 2457802.50174 | 0:55 | 16.1 | 19.93 | 0.02±0.08 | -0.08±0.15 | ≥0.81 | ≥55 | 0.176 | 0.129 | 0.649 | 0.046 | C | 2.2 | 8.4 |
| 0244 | 2000 ES55 | 2457692.44645 | 1:00 | 15.2 | 19.12 | 0.06±0.04 | 0.12±0.05 | ≥0.05 | ≥60 | 0.688 | 0.123 | 0.011 | 0.177 | S | 2.8 | 2.0 |
| 0245 | 2000 ET145 | 2457724.38316 | 1:08 | 16.3 | 19.98 | -0.10±0.05 | -0.04±0.07 | ≥0.12 | ≥69 | 0.001 | 0.312 | 0.685 | 0.001 | C | 2.7 | 0.8 |
| 0246 | 2000 EV32 | 2457690.43605 | 1:58 | 15.7 | 19.20 | 0.08±0.03 | -0.04±0.04 | ≥0.20 | ≥119 | 0.455 | 0.003 | 0.542 | 0.000 | C | 2.4 | 7.4 |
| 0247 | 2000 EY120 | 2457690.42183 | 2:21 | 15.8 | 18.71 | 0.12±0.03 | 0.09±0.03 | ≥0.17 | ≥142 | 0.994 | 0.004 | 0.000 | 0.002 | S | 2.3 | 7.1 |
| 0248 | 2000 FA41 | 2457724.38316 | 1:08 | 14.6 | 18.57 | -0.09±0.02 | -0.04±0.03 | ≥0.56 | ≥69 | 0.000 | 0.216 | 0.784 | 0.000 | C | 2.7 | 5.4 |
| 0249 | 2000 FB31 | 2457724.34310 | 0:55 | 14.5 | 19.53 | 0.21±0.05 | 0.18±0.06 | ≥0.42 | ≥55 | 0.995 | 0.000 | 0.000 | 0.004 | S | 2.8 | 9.8 |
| 0250 | 2000 FX63 | 2457692.38892 | 1:20 | 15.9 | 20.16 | 0.06±0.03 | 0.07±0.04 | ≥0.12 | ≥80 | 0.792 | 0.151 | 0.034 | 0.023 | S | 2.8 | 9.7 |
| 0251 | 2000 FY72 | 2457690.33738 | 1:57 | 15.0 | 18.90 | 0.00±0.12 | 0.04±0.08 | ≥0.21 | ≥118 | 0.344 | 0.356 | 0.253 | 0.047 | - | 2.8 | 11.1 |
| 0252 | 2000 FZ19 | 2457691.54720 | 0:52 | 14.4 | 18.14 | 0.13±0.04 | 0.13±0.10 | ≥0.15 | ≥52 | 0.944 | 0.003 | 0.017 | 0.036 | S | 2.7 | 13.3 |
| 0253 | 2000 GD54 | 2457724.38316 | 1:08 | 15.7 | 19.81 | 0.07±0.03 | 0.02±0.04 | ≥0.29 | ≥69 | 0.794 | 0.050 | 0.155 | 0.001 | S | 2.8 | 2.0 |
| 0254 | 2000 GG108 | 2457722.38400 | 1:47 | 14.4 | 18.53 | -0.05±0.02 | -0.04±0.03 | ≥0.04 | ≥107 | 0.000 | 0.265 | 0.735 | 0.000 | C | 2.8 | 6.5 |
| 0255 | 2000 GG117 | 2457692.38892 | 1:20 | 17.5 | 20.64 | 0.09±0.07 | 0.09±0.08 | ≥0.30 | ≥80 | 0.713 | 0.104 | 0.084 | 0.099 | S | 2.3 | 2.2 |
| 0256 | 2000 GG32 | 2457692.38892 | 1:20 | 15.8 | 19.36 | -0.04±0.02 | -0.04±0.03 | ≥0.30 | ≥80 | 0.000 | 0.178 | 0.822 | 0.000 | C | 2.8 | 0.6 |
| 0257 | 2000 GG6 | 2457798.51082 | 0:59 | 13.5 | 18.29 | -0.18±0.10 | -0.10±0.06 | ≥0.38 | ≥59 | 0.003 | 0.033 | 0.965 | 0.000 | C | 2.7 | 11.7 |
| 0258 | 2000 GL5 | 2457691.37970 | 1:52 | 16.1 | 19.51 | 0.07±0.03 | 0.04±0.04 | ≥0.19 | ≥113 | 0.860 | 0.071 | 0.068 | 0.001 | S | 2.3 | 5.8 |
| 0259 | 2000 GM52 | 2457691.37970 | 1:54 | 16.8 | 20.25 | 0.07±0.05 | 0.15±0.06 | ≥0.40 | ≥114 | 0.615 | 0.084 | 0.003 | 0.298 | S | 2.3 | 5.2 |
| 0260 | 2000 GO144 | 2457692.44645 | 1:00 | 17.6 | 19.82 | 0.02±0.05 | 0.02±0.06 | ≥0.17 | ≥60 | 0.341 | 0.262 | 0.387 | 0.010 | - | 2.1 | 2.1 |
| 0261 | 2000 GX118 | 2457722.38400 | 1:47 | 18.0 | 20.23 | 0.08±0.04 | 0.07±0.05 | ≥0.26 | ≥107 | 0.847 | 0.080 | 0.045 | 0.028 | S | 2.3 | 4.5 |
| 0262 | 2000 GY156 | 2457728.50413 | 0:26 | 15.7 | 20.15 | -0.13±0.15 | -0.03±0.13 | ≥0.52 | ≥27 | 0.091 | 0.281 | 0.576 | 0.052 | C | 2.7 | 6.4 |
| 0263 | 2000 GZ103 | 2457692.38892 | 1:20 | 16.0 | 19.66 | 0.07±0.03 | 0.10±0.03 | ≥0.17 | ≥80 | 0.875 | 0.080 | 0.001 | 0.045 | S | 2.4 | 3.2 |
| 0264 | 2000 HB8 | 2457691.37970 | 1:10 | 16.1 | 19.59 | 0.04±0.03 | -0.01±0.04 | ≥0.20 | ≥70 | 0.410 | 0.061 | 0.529 | 0.000 | C | 2.4 | 6.4 |
| 0265 | 2000 HN50 | 2457690.34053 | 1:53 | 15.7 | 18.63 | 0.05±0.03 | 0.06±0.03 | 0.42±0.04 | 189±42 | 0.763 | 0.206 | 0.030 | 0.001 | S | 2.4 | 2.4 |
| 0266 | 2000 HV16 | 2457690.33738 | 1:57 | 16.4 | 19.47 | 0.05±0.03 | 0.04±0.03 | ≥0.28 | ≥118 | 0.750 | 0.140 | 0.110 | 0.000 | S | 2.4 | 2.0 |
| 0267 | 2000 HV86 | 2457722.38400 | 1:47 | 16.1 | 19.67 | 0.08±0.03 | 0.17±0.04 | 0.30±0.06 | 191±38 | 0.732 | 0.003 | 0.000 | 0.265 | S | 2.3 | 6.7 |
| 0268 | 2000 HX47 | 2457722.38685 | 1:43 | 17.0 | 21.13 | -0.06±0.12 | 0.05±0.12 | ≥0.40 | ≥103 | 0.150 | 0.385 | 0.316 | 0.149 | - | 2.4 | 6.6 |
| 0269 | 2000 JE14 | 2457690.38513 | 0:48 | 15.5 | 19.09 | 0.07±0.04 | 0.05±0.07 | ≥0.10 | ≥49 | 0.706 | 0.097 | 0.151 | 0.046 | S | 2.4 | 3.5 |
| 0270 | 2000 JE23 | 2457796.48277 | 0:51 | 14.9 | 17.79 | -0.02±0.07 | 0.01±0.07 | ≥0.15 | ≥52 | 0.189 | 0.366 | 0.426 | 0.019 | - | 2.7 | 5.5 |
| 0271 | 2000 JJ15 | 2457724.38316 | 1:08 | 16.6 | 20.47 | 0.02±0.08 | -0.03±0.07 | ≥0.27 | ≥69 | 0.289 | 0.168 | 0.539 | 0.004 | C | 2.4 | 3.8 |
| 0272 | 2000 JP12 | 2457724.28943 | 1:08 | 15.3 | 18.50 | -0.06±0.02 | 0.03±0.03 | ≥0.10 | ≥68 | 0.000 | 0.918 | 0.082 | 0.000 | X | 2.4 | 3.9 |
| 0273 | 2000 JP17 | 2457691.46127 | 2:00 | 14.5 | 17.61 | 0.11±0.03 | 0.20±0.03 | 0.27±0.07 | 88±12 | 0.879 | 0.000 | 0.000 | 0.121 | S | 2.4 | 6.2 |
| 0274 | 2000 JV13 | 2457692.38892 | 1:20 | 14.6 | 18.09 | 0.04±0.02 | 0.10±0.03 | 0.17±0.04 | 160±29 | 0.577 | 0.350 | 0.002 | 0.071 | S | 2.8 | 4.6 |

*Table 1 continued on next page*

**Table 1** *(continued)*

| No.[a] | Object | Obs. Start | Obs. Duration | H[b] | V | V-R[c] | V-I[c] | Amplitude[d] | Rot. Period[d] | S-type | X-type | C-type | D-type | Tax. | a[b] | i[b] |
|---|---|---|---|---|---|---|---|---|---|---|---|---|---|---|---|---|
| | | (JD) | (h:mm) | (mag) | (mag) | (mag) | (mag) | (mag) | (min) | (probability) | | | | | (au) | (deg) |
| 0275 | 2000 JX58 | 2457722.38400 | 1:47 | 15.1 | 19.67 | -0.11±0.03 | -0.15±0.04 | ≥0.23 | ≥107 | 0.000 | 0.000 | 1.000 | 0.000 | C | 2.8 | 8.9 |
| 0276 | 2000 KA58 | 2457722.38400 | 1:47 | 16.0 | 19.59 | 0.08±0.03 | 0.16±0.04 | ≥0.18 | ≥107 | 0.790 | 0.010 | 0.000 | 0.200 | S | 2.4 | 9.0 |
| 0277 | 2000 KP45 | 2457692.39994 | 1:04 | 15.5 | 19.12 | -0.06±0.02 | -0.06±0.03 | ≥0.06 | ≥64 | 0.000 | 0.119 | 0.881 | 0.000 | C | 2.9 | 6.3 |
| 0278 | 2000 KT38 | 2457724.38316 | 1:08 | 16.1 | 18.78 | 0.05±0.03 | -0.04±0.04 | ≥0.09 | ≥69 | 0.245 | 0.006 | 0.749 | 0.000 | C | 2.4 | 5.6 |
| 0279 | 2000 KU48 | 2457722.38400 | 1:47 | 15.9 | 19.64 | -0.06±0.03 | -0.06±0.04 | ≥0.22 | ≥107 | 0.000 | 0.149 | 0.851 | 0.000 | C | 2.4 | 4.0 |
| 0280 | 2000 LU2 | 2457802.42659 | 0:37 | 16.4 | 20.10 | -0.08±0.06 | -0.11±0.06 | ≥0.08 | ≥37 | 0.002 | 0.057 | 0.941 | 0.000 | C | 2.3 | 5.1 |
| 0281 | 2000 OO37 | 2457802.38260 | 1:00 | 15.7 | 20.15 | -0.12±0.07 | -0.04±0.07 | ≥0.19 | ≥61 | 0.008 | 0.287 | 0.702 | 0.003 | C | 2.4 | 9.1 |
| 0282 | 2000 PD21 | 2457724.38316 | 1:08 | 15.5 | 19.41 | 0.03±0.03 | 0.04±0.04 | ≥0.17 | ≥69 | 0.521 | 0.287 | 0.191 | 0.001 | S | 2.9 | 1.9 |
| 0283 | 2000 PQ7 | 2457728.49022 | 0:57 | 13.4 | 17.29 | 0.04±0.04 | 0.08±0.06 | ≥0.10 | ≥58 | 0.544 | 0.279 | 0.077 | 0.100 | S | 3.0 | 14.8 |
| 0284 | 2000 QA45 | 2457728.50413 | 0:37 | 15.4 | 19.50 | -0.04±0.08 | -0.01±0.08 | ≥0.16 | ≥38 | 0.110 | 0.343 | 0.535 | 0.013 | C | 2.4 | 6.0 |
| 0285 | 2000 QB53 | 2457692.38892 | 1:20 | 16.5 | 18.75 | -0.03±0.02 | -0.02±0.03 | ≥0.06 | ≥80 | 0.000 | 0.365 | 0.635 | 0.000 | C | 2.5 | 1.5 |
| 0286 | 2000 QE23 | 2457690.42183 | 2:21 | 14.7 | 19.20 | 0.07±0.04 | 0.06±0.04 | ≥0.19 | ≥142 | 0.839 | 0.123 | 0.031 | 0.007 | S | 3.0 | 10.2 |
| 0287 | 2000 QL64 | 2457802.50174 | 0:55 | 15.5 | 19.83 | 0.15±0.08 | 0.12±0.10 | ≥0.46 | ≥55 | 0.884 | 0.027 | 0.024 | 0.065 | S | 2.4 | 9.2 |
| 0288 | 2000 QM183 | 2457800.27642 | 0:54 | 14.8 | 20.23 | 0.02±0.08 | 0.06±0.11 | ≥0.77 | ≥55 | 0.325 | 0.260 | 0.252 | 0.163 | - | 2.9 | 10.3 |
| 0289 | 2000 QT65 | 2457802.50174 | 0:55 | 14.7 | 19.34 | -0.03±0.05 | -0.11±0.07 | ≥0.13 | ≥55 | 0.008 | 0.053 | 0.939 | 0.000 | C | 2.3 | 5.9 |
| 0290 | 2000 QY160 | 2457692.45850 | 0:42 | 14.3 | 18.28 | 0.05±0.04 | 0.11±0.06 | ≥0.17 | ≥43 | 0.563 | 0.220 | 0.025 | 0.192 | S | 3.0 | 8.9 |
| 0291 | 2000 QZ227 | 2457728.49321 | 0:33 | 15.6 | 19.71 | 0.28±0.10 | 0.21±0.14 | ≥0.28 | ≥34 | 0.979 | 0.002 | 0.004 | 0.015 | S | 2.4 | 6.4 |
| 0292 | 2000 RA69 | 2457724.38316 | 1:08 | 15.4 | 18.35 | -0.06±0.02 | -0.11±0.03 | ≥0.04 | ≥69 | 0.000 | 0.004 | 0.996 | 0.000 | C | 2.5 | 1.2 |
| 0293 | 2000 RK53 | 2457724.31706 | 0:28 | 14.7 | 19.07 | 0.03±0.05 | 0.07±0.06 | ≥0.08 | ≥28 | 0.456 | 0.335 | 0.126 | 0.083 | - | 3.1 | 12.3 |
| 0294 | 2000 SC122 | 2457690.33738 | 1:57 | 14.3 | 18.57 | 0.06±0.02 | 0.12±0.03 | 0.35±0.05 | 235±46 | 0.827 | 0.058 | 0.000 | 0.115 | S | 2.5 | 16.0 |
| 0295 | 2000 SD120 | 2457722.38400 | 1:47 | 16.9 | 19.13 | 0.06±0.03 | 0.09±0.03 | ≥0.12 | ≥107 | 0.852 | 0.121 | 0.002 | 0.026 | S | 2.2 | 3.0 |
| 0296 | 2000 SD49 | 2457690.33738 | 1:57 | 15.0 | 19.46 | 0.05±0.04 | 0.05±0.05 | ≥0.31 | ≥118 | 0.674 | 0.189 | 0.116 | 0.021 | S | 3.0 | 10.2 |
| 0297 | 2000 SE144 | 2457724.34161 | 0:57 | 14.7 | 19.04 | 0.16±0.04 | 0.14±0.06 | ≥0.24 | ≥58 | 0.993 | 0.001 | 0.000 | 0.007 | S | 3.0 | 2.6 |
| 0298 | 2000 SE343 | 2457690.33738 | 1:57 | 15.8 | 19.54 | 0.03±0.02 | 0.11±0.03 | 0.49±0.06 | 235±51 | 0.495 | 0.341 | 0.003 | 0.162 | - | 3.1 | 10.7 |
| 0299 | 2000 SG191 | 2457724.28943 | 1:08 | 15.5 | 20.37 | -0.01±0.08 | 0.00±0.11 | ≥0.60 | ≥68 | 0.200 | 0.282 | 0.443 | 0.076 | - | 3.0 | 10.1 |
| 0300 | 2000 SJ50 | 2457691.37970 | 1:28 | 15.0 | 19.32 | -0.06±0.03 | -0.05±0.04 | ≥0.67 | ≥89 | 0.000 | 0.196 | 0.804 | 0.000 | C | 3.0 | 7.2 |
| 0301 | 2000 SL119 | 2457728.49022 | 0:57 | 15.8 | 20.04 | 0.04±0.09 | 0.10±0.09 | ≥0.26 | ≥58 | 0.464 | 0.248 | 0.101 | 0.186 | - | 2.5 | 5.6 |
| 0302 | 2000 SM224 | 2457724.28943 | 1:08 | 15.3 | 20.38 | 0.09±0.09 | 0.08±0.13 | 1.10±0.14 | 136±27 | 0.599 | 0.105 | 0.162 | 0.135 | S | 3.1 | 2.9 |
| 0303 | 2000 SO66 | 2457692.38892 | 1:20 | 16.7 | 19.53 | 0.14±0.02 | 0.20±0.03 | ≥0.13 | ≥80 | 0.997 | 0.000 | 0.000 | 0.003 | S | 2.2 | 3.7 |
| 0304 | 2000 SQ143 | 2457724.38316 | 1:08 | 14.6 | 19.30 | 0.01±0.03 | 0.04±0.05 | 0.25±0.04 | 98±26 | 0.237 | 0.453 | 0.298 | 0.012 | - | 3.0 | 8.9 |
| 0305 | 2000 SR97 | 2457724.38316 | 1:08 | 15.9 | 19.99 | 0.17±0.04 | 0.12±0.05 | ≥0.20 | ≥69 | 0.997 | 0.000 | 0.000 | 0.002 | S | 2.5 | 5.8 |
| 0306 | 2000 SS268 | 2457730.42646 | 1:21 | 15.1 | 18.66 | 0.01±0.03 | 0.03±0.04 | ≥0.12 | ≥81 | 0.262 | 0.398 | 0.339 | 0.001 | - | 3.0 | 7.7 |
| 0307 | 2000 SU1 | 2457800.44661 | 0:54 | 14.1 | 18.57 | 0.08±0.05 | 0.23±0.07 | ≥0.14 | ≥54 | 0.524 | 0.017 | 0.000 | 0.458 | S | 2.4 | 11.7 |
| 0308 | 2000 SY212 | 2457722.38400 | 1:47 | 14.3 | 19.09 | 0.06±0.02 | 0.08±0.03 | ≥0.23 | ≥107 | 0.868 | 0.118 | 0.006 | 0.008 | S | 3.0 | 9.9 |
| 0309 | 2000 SY220 | 2457765.29937 | 0:43 | 13.2 | 18.57 | 0.01±0.05 | 0.07±0.06 | ≥0.42 | ≥44 | 0.294 | 0.466 | 0.152 | 0.087 | - | 3.0 | 11.4 |
| 0310 | 2000 SZ109 | 2457728.49022 | 0:57 | 15.3 | 19.59 | 0.17±0.06 | 0.08±0.08 | 0.31±0.11 | 72±16 | 0.963 | 0.007 | 0.016 | 0.015 | S | 2.4 | 5.7 |
| 0311 | 2000 TO39 | 2457690.33738 | 1:57 | 15.1 | 19.15 | 0.03±0.02 | 0.12±0.03 | ≥0.11 | ≥118 | 0.372 | 0.447 | 0.001 | 0.180 | - | 3.0 | 10.4 |
| 0312 | 2000 UB17 | 2457724.34161 | 0:57 | 17.5 | 19.69 | 0.18±0.10 | 0.24±0.12 | ≥0.36 | ≥58 | 0.890 | 0.007 | 0.004 | 0.098 | S | 2.2 | 4.2 |
| 0313 | 2000 UD77 | 2457724.28943 | 0:41 | 16.9 | 19.73 | 0.04±0.05 | 0.00±0.11 | ≥0.18 | ≥42 | 0.341 | 0.143 | 0.449 | 0.067 | - | 2.5 | 10.1 |
| 0314 | 2000 UH20 | 2457692.38892 | 1:20 | 15.6 | 18.81 | -0.05±0.02 | -0.07±0.03 | ≥0.05 | ≥80 | 0.000 | 0.049 | 0.951 | 0.000 | C | 3.1 | 1.1 |
| 0315 | 2000 UK43 | 2457692.38892 | 1:20 | 15.9 | 19.37 | 0.10±0.02 | 0.15±0.03 | ≥0.14 | ≥80 | 0.957 | 0.001 | 0.000 | 0.041 | S | 2.5 | 5.3 |
| 0316 | 2000 UM15 | 2457724.34904 | 0:46 | 14.8 | 19.44 | 0.09±0.08 | 0.17±0.10 | ≥0.31 | ≥47 | 0.650 | 0.080 | 0.026 | 0.244 | S | 3.1 | 10.2 |
| 0317 | 2000 UM3 | 2457800.44661 | 0:54 | 15.0 | 18.20 | -0.03±0.05 | 0.02±0.06 | ≥0.13 | ≥54 | 0.075 | 0.574 | 0.332 | 0.019 | X | 2.4 | 11.6 |
| 0318 | 2000 UM3 | 2457801.38674 | 1:43 | 15.0 | 18.14 | -0.02±0.04 | 0.06±0.04 | ≥0.24 | ≥104 | 0.093 | 0.785 | 0.099 | 0.024 | X | 2.4 | 11.6 |
| 0319 | 2000 UN25 | 2457802.50174 | 0:55 | 15.1 | 18.59 | 0.09±0.03 | 0.14±0.04 | ≥0.20 | ≥55 | 0.908 | 0.015 | 0.000 | 0.076 | S | 2.4 | 5.9 |
| 0320 | 2000 UO39 | 2457690.33738 | 1:26 | 14.6 | 18.87 | -0.06±0.02 | -0.06±0.03 | ≥0.33 | ≥87 | 0.000 | 0.127 | 0.873 | 0.000 | C | 3.1 | 2.3 |

*Table 1 continued on next page*

**Table 1** *(continued)*

| No.[a] | Object | Obs. Start (JD) | Obs. Duration (h:mm) | H[b] (mag) | V (mag) | V-R[c] (mag) | V-I[c] (mag) | Amplitude[d] (mag) | Rot. Period[d] (min) | S-type (probability) | X-type | C-type | D-type | Tax. | a[b] (au) | i[b] (deg) |
|---|---|---|---|---|---|---|---|---|---|---|---|---|---|---|---|---|
| 0321 | 2000 UO87 | 2457691.46127 | 1:56 | 14.9 | 19.36 | 0.05±0.04 | 0.03±0.04 | ≥0.35 | ≥116 | 0.610 | 0.183 | 0.203 | 0.003 | S | 3.1 | 9.0 |
| 0322 | 2000 UR44 | 2457724.34608 | 0:51 | 14.1 | 18.64 | 0.04±0.05 | 0.16±0.06 | ≥0.14 | ≥51 | 0.368 | 0.162 | 0.003 | 0.467 | - | 3.1 | 13.7 |
| 0323 | 2000 UU104 | 2457692.44645 | 1:00 | 14.4 | 17.77 | 0.04±0.04 | 0.15±0.06 | ≥0.02 | ≥60 | 0.381 | 0.159 | 0.004 | 0.456 | - | 3.1 | 17.7 |
| 0324 | 2000 UU76 | 2457692.44645 | 1:00 | 15.4 | 19.35 | -0.08±0.04 | 0.00±0.06 | ≥0.07 | ≥60 | 0.006 | 0.559 | 0.432 | 0.002 | X | 3.0 | 1.6 |
| 0325 | 2000 VD5 | 2457724.28943 | 1:08 | 15.0 | 19.97 | -0.02±0.05 | -0.06±0.08 | ≥0.20 | ≥68 | 0.053 | 0.181 | 0.762 | 0.004 | C | 3.1 | 5.1 |
| 0326 | 2000 VK1 | 2457692.38892 | 1:20 | 15.2 | 17.27 | 0.08±0.02 | 0.13±0.03 | 0.28±0.03 | 160±34 | 0.958 | 0.004 | 0.000 | 0.038 | S | 2.6 | 4.5 |
| 0327 | 2000 VK24 | 2457724.38626 | 1:04 | 15.9 | 18.66 | 0.11±0.02 | 0.10±0.03 | ≥0.10 | ≥64 | 1.000 | 0.000 | 0.000 | 0.000 | S | 2.5 | 0.9 |
| 0328 | 2000 VL58 | 2457724.38316 | 1:08 | 14.9 | 18.48 | 0.01±0.02 | 0.11±0.03 | ≥0.07 | ≥69 | 0.167 | 0.702 | 0.004 | 0.127 | X | 3.1 | 16.4 |
| 0329 | 2000 VP11 | 2457724.34310 | 0:55 | 15.0 | 18.75 | -0.05±0.04 | -0.05±0.05 | ≥0.12 | ≥55 | 0.005 | 0.221 | 0.774 | 0.000 | C | 3.1 | 1.6 |
| 0330 | 2000 VW10 | 2457691.37970 | 1:54 | 14.6 | 18.79 | -0.06±0.02 | -0.10±0.03 | ≥0.21 | ≥114 | 0.000 | 0.005 | 0.995 | 0.000 | C | 3.1 | 2.1 |
| 0331 | 2000 VW6 | 2457691.38271 | 1:50 | 16.8 | 20.16 | 0.08±0.05 | 0.13±0.05 | ≥0.22 | ≥110 | 0.724 | 0.092 | 0.005 | 0.179 | S | 2.5 | 3.1 |
| 0332 | 2000 WA147 | 2457800.27020 | 1:07 | 14.2 | 19.64 | -0.02±0.13 | 0.05±0.16 | ≥0.55 | ≥68 | 0.229 | 0.241 | 0.343 | 0.187 | - | 3.0 | 9.8 |
| 0333 | 2000 WA60 | 2457691.46127 | 2:00 | 15.8 | 19.21 | -0.02±0.04 | 0.02±0.04 | ≥0.16 | ≥121 | 0.065 | 0.575 | 0.359 | 0.001 | X | 2.5 | 7.6 |
| 0334 | 2000 WD1 | 2457724.28943 | 1:08 | 17.2 | 20.44 | 0.15±0.06 | 0.05±0.10 | ≥0.36 | ≥68 | 0.876 | 0.009 | 0.098 | 0.018 | S | 2.2 | 2.5 |
| 0335 | 2000 WD182 | 2457687.38022 | 0:58 | 14.8 | 19.11 | 0.07±0.04 | 0.23±0.03 | ≥0.27 | ≥58 | 0.346 | 0.000 | 0.000 | 0.654 | D | 3.1 | 16.8 |
| 0336 | 2000 WM165 | 2457730.42038 | 1:30 | 15.4 | 19.32 | 0.10±0.04 | 0.18±0.04 | ≥0.17 | ≥90 | 0.823 | 0.004 | 0.000 | 0.173 | S | 3.0 | 9.7 |
| 0337 | 2000 WN59 | 2457690.33738 | 1:57 | 16.4 | 17.82 | 0.05±0.02 | 0.03±0.02 | ≥0.16 | ≥118 | 0.900 | 0.033 | 0.067 | 0.000 | S | 2.2 | 1.8 |
| 0338 | 2000 WR127 | 2457798.50513 | 1:07 | 15.0 | 19.55 | 0.32±0.12 | 0.42±0.13 | ≥0.72 | ≥67 | 0.985 | 0.000 | 0.000 | 0.015 | S | 2.4 | 6.7 |
| 0339 | 2000 WR45 | 2457724.34310 | 0:55 | 16.2 | 19.64 | 0.10±0.06 | 0.05±0.07 | ≥0.26 | ≥55 | 0.791 | 0.075 | 0.109 | 0.025 | S | 3.1 | 19.2 |
| 0340 | 2000 WT79 | 2457724.28943 | 1:08 | 16.3 | 20.07 | 0.17±0.06 | 0.26±0.07 | ≥0.17 | ≥68 | 0.925 | 0.000 | 0.000 | 0.075 | S | 2.5 | 6.4 |
| 0341 | 2000 WX112 | 2457724.34310 | 0:55 | 15.5 | 18.99 | 0.08±0.04 | 0.04±0.05 | ≥0.12 | ≥55 | 0.827 | 0.054 | 0.110 | 0.010 | S | 2.5 | 11.6 |
| 0342 | 2000 WY18 | 2457690.42183 | 2:21 | 13.7 | 19.17 | 0.16±0.04 | 0.24±0.04 | 0.67±0.08 | 284±64 | 0.969 | 0.000 | 0.000 | 0.031 | S | 3.1 | 18.7 |
| 0343 | 2000 XH43 | 2457691.37970 | 0:33 | 14.9 | 17.50 | 0.05±0.03 | 0.10±0.04 | ≥0.09 | ≥33 | 0.742 | 0.165 | 0.008 | 0.085 | S | 2.6 | 15.6 |
| 0344 | 2000 XK13 | 2457802.50174 | 0:55 | 14.7 | 19.83 | 0.06±0.06 | 0.06±0.10 | ≥0.29 | ≥55 | 0.525 | 0.162 | 0.204 | 0.109 | S | 2.4 | 13.3 |
| 0345 | 2000 YF140 | 2457692.44645 | 1:00 | 14.7 | 19.07 | -0.01±0.05 | 0.07±0.06 | ≥0.15 | ≥60 | 0.200 | 0.604 | 0.103 | 0.093 | X | 2.6 | 13.8 |
| 0346 | 2000 YH124 | 2457724.38316 | 1:08 | 14.8 | 19.21 | -0.03±0.03 | -0.03±0.04 | ≥0.11 | ≥69 | 0.015 | 0.319 | 0.666 | 0.000 | C | 3.1 | 8.2 |
| 0347 | 2000 YJ135 | 2457724.34310 | 0:55 | 15.2 | 18.34 | -0.02±0.04 | 0.03±0.05 | ≥0.16 | ≥55 | 0.108 | 0.570 | 0.317 | 0.005 | X | 2.5 | 25.9 |
| 0348 | 2000 YK27 | 2457687.38022 | 3:03 | 14.7 | 19.18 | -0.11±0.02 | -0.11±0.02 | 0.18±0.04 | 233±59 | 0.000 | 0.000 | 1.000 | 0.000 | C | 3.2 | 15.6 |
| 0349 | 2000 YM136 | 2457690.33738 | 1:57 | 13.9 | 18.23 | 0.06±0.02 | 0.21±0.03 | ≥0.18 | ≥118 | 0.213 | 0.000 | 0.000 | 0.787 | D | 3.1 | 19.4 |
| 0350 | 2000 YT117 | 2457802.42659 | 1:45 | 15.4 | 19.66 | -0.03±0.03 | 0.00±0.04 | ≥0.20 | ≥106 | 0.022 | 0.513 | 0.465 | 0.000 | X | 3.0 | 10.4 |
| 0351 | 2000 YV105 | 2457800.27020 | 1:07 | 13.8 | 19.23 | 0.09±0.04 | 0.15±0.05 | ≥0.10 | ≥68 | 0.825 | 0.024 | 0.001 | 0.151 | S | 2.6 | 14.2 |
| 0352 | 2001 BB46 | 2457800.27020 | 1:07 | 14.5 | 19.14 | 0.03±0.04 | 0.09±0.09 | ≥0.61 | ≥68 | 0.381 | 0.218 | 0.173 | 0.229 | - | 2.6 | 10.8 |
| 0353 | 2001 BG30 | 2457691.46127 | 2:00 | 14.7 | 19.90 | -0.05±0.04 | -0.02±0.05 | ≥0.29 | ≥121 | 0.016 | 0.425 | 0.559 | 0.001 | C | 3.2 | 6.3 |
| 0354 | 2001 BH27 | 2457728.49022 | 0:57 | 14.8 | 19.45 | 0.12±0.07 | 0.24±0.08 | ≥0.15 | ≥58 | 0.731 | 0.014 | 0.001 | 0.254 | S | 3.1 | 10.5 |
| 0355 | 2001 BL30 | 2457728.49022 | 0:57 | 13.7 | 19.27 | 0.01±0.07 | 0.01±0.08 | ≥0.14 | ≥58 | 0.263 | 0.308 | 0.398 | 0.031 | - | 3.1 | 6.6 |
| 0356 | 2001 BN58 | 2457798.46669 | 0:53 | 14.1 | 19.08 | 0.14±0.08 | 0.22±0.09 | ≥0.35 | ≥53 | 0.805 | 0.015 | 0.001 | 0.178 | S | 3.0 | 8.9 |
| 0357 | 2001 BV65 | 2457692.44645 | 1:00 | 16.2 | 18.73 | 0.04±0.04 | 0.07±0.04 | ≥0.11 | ≥60 | 0.520 | 0.350 | 0.095 | 0.035 | S | 2.2 | 5.0 |
| 0358 | 2001 CK30 | 2457692.44645 | 1:00 | 15.7 | 20.27 | -0.02±0.13 | 0.13±0.09 | ≥0.45 | ≥60 | 0.287 | 0.370 | 0.076 | 0.267 | - | 2.6 | 1.1 |
| 0359 | 2001 DB96 | 2457691.42249 | 0:52 | 16.1 | 20.03 | 0.02±0.09 | 0.06±0.09 | ≥0.28 | ≥53 | 0.378 | 0.325 | 0.185 | 0.112 | - | 2.6 | 3.5 |
| 0360 | 2001 DH92 | 2457796.44027 | 0:58 | 13.6 | 18.65 | 0.06±0.07 | 0.06±0.08 | ≥0.14 | ≥59 | 0.580 | 0.190 | 0.149 | 0.081 | S | 3.1 | 14.0 |
| 0361 | 2001 DK36 | 2457690.38513 | 0:48 | 15.7 | 19.76 | 0.16±0.08 | 0.13±0.12 | ≥0.22 | ≥49 | 0.884 | 0.020 | 0.038 | 0.059 | S | 2.6 | 3.0 |
| 0362 | 2001 DL27 | 2457802.50174 | 0:55 | 15.1 | 20.08 | -0.03±0.07 | -0.05±0.15 | ≥0.24 | ≥55 | 0.080 | 0.233 | 0.613 | 0.074 | C | 3.0 | 6.0 |
| 0363 | 2001 DL37 | 2457691.55067 | 1:22 | 15.3 | 19.75 | -0.02±0.15 | 0.06±0.07 | ≥0.18 | ≥82 | 0.337 | 0.437 | 0.174 | 0.052 | - | 2.6 | 15.1 |
| 0364 | 2001 DU20 | 2457687.38617 | 2:54 | 14.4 | 18.42 | 0.06±0.01 | 0.08±0.02 | 0.43±0.03 | 350±85 | 0.898 | 0.102 | 0.000 | 0.000 | S | 2.7 | 15.4 |
| 0365 | 2001 DU68 | 2457691.49049 | 1:18 | 14.2 | 18.97 | -0.02±0.04 | -0.00±0.05 | ≥0.25 | ≥79 | 0.068 | 0.411 | 0.520 | 0.001 | C | 3.2 | 7.4 |
| 0366 | 2001 DV19 | 2457722.38400 | 1:47 | 15.8 | 19.42 | 0.04±0.02 | 0.06±0.04 | ≥0.10 | ≥107 | 0.580 | 0.332 | 0.079 | 0.009 | S | 2.6 | 8.6 |

*Table 1 continued on next page*

**Table 1** *(continued)*

| No.[a] | Object | Obs. Start | Obs. Duration | H[b] | V | V-R[c] | V-I[c] | Amplitude[d] | Rot. Period[d] | S-type | X-type | C-type | D-type | Tax. | a[b] | i[b] |
|---|---|---|---|---|---|---|---|---|---|---|---|---|---|---|---|---|
| | | (JD) | (h:mm) | (mag) | (mag) | (mag) | (mag) | (mag) | (min) | (probability) | | | | | (au) | (deg) |
| 0367 | 2001 DV88 | 2457802.39311 | 0:45 | 14.6 | 19.15 | -0.04±0.03 | -0.06±0.04 | ≥0.17 | ≥46 | 0.001 | 0.107 | 0.892 | 0.000 | C | 3.1 | 22.1 |
| 0368 | 2001 EM19 | 2457798.46669 | 0:53 | 14.9 | 18.85 | 0.20±0.07 | 0.30±0.09 | ≥0.52 | ≥53 | 0.952 | 0.000 | 0.000 | 0.047 | S | 2.5 | 12.9 |
| 0369 | 2001 EQ8 | 2457690.33738 | 1:57 | 15.7 | 20.03 | 0.03±0.04 | 0.10±0.05 | ≥0.19 | ≥118 | 0.448 | 0.366 | 0.021 | 0.165 | - | 2.7 | 2.9 |
| 0370 | 2001 ET23 | 2457691.37970 | 1:54 | 16.2 | 20.52 | -0.08±0.10 | -0.13±0.09 | ≥0.35 | ≥114 | 0.023 | 0.057 | 0.919 | 0.000 | C | 2.7 | 1.8 |
| 0371 | 2001 EU14 | 2457730.42038 | 1:30 | 14.6 | 19.56 | 0.06±0.04 | 0.10±0.04 | ≥0.28 | ≥90 | 0.704 | 0.200 | 0.017 | 0.079 | S | 3.2 | 12.1 |
| 0372 | 2001 EX7 | 2457730.42927 | 1:17 | 15.9 | 20.54 | 0.11±0.07 | 0.04±0.08 | ≥0.30 | ≥77 | 0.770 | 0.061 | 0.145 | 0.024 | S | 2.6 | 8.5 |
| 0373 | 2001 FB88 | 2457722.38979 | 1:38 | 14.0 | 17.02 | -0.04±0.02 | -0.03±0.03 | ≥0.07 | ≥99 | 0.000 | 0.297 | 0.703 | 0.000 | C | 2.7 | 7.7 |
| 0374 | 2001 FG45 | 2457722.38400 | 1:47 | 16.2 | 19.72 | 0.05±0.03 | 0.06±0.04 | ≥0.17 | ≥107 | 0.662 | 0.232 | 0.095 | 0.011 | S | 2.3 | 4.0 |
| 0375 | 2001 FH75 | 2457692.38892 | 1:20 | 16.5 | 18.67 | 0.07±0.02 | 0.09±0.03 | ≥0.10 | ≥80 | 0.949 | 0.044 | 0.000 | 0.007 | S | 2.2 | 7.6 |
| 0376 | 2001 FP22 | 2457798.50513 | 1:09 | 13.4 | 18.19 | -0.06±0.04 | -0.10±0.05 | ≥0.15 | ≥69 | 0.002 | 0.045 | 0.953 | 0.000 | C | 3.2 | 13.5 |
| 0377 | 2001 FT2 | 2457692.38892 | 1:20 | 15.9 | 19.70 | 0.05±0.03 | 0.08±0.04 | ≥0.11 | ≥80 | 0.714 | 0.241 | 0.018 | 0.027 | S | 2.6 | 3.8 |
| 0378 | 2001 FV134 | 2457801.38674 | 1:43 | 13.1 | 18.21 | 0.03±0.04 | 0.14±0.04 | ≥0.08 | ≥104 | 0.391 | 0.243 | 0.002 | 0.365 | - | 3.1 | 12.2 |
| 0379 | 2001 FV46 | 2457800.27321 | 0:59 | 14.0 | 19.63 | -0.10±0.13 | -0.02±0.11 | ≥0.71 | ≥59 | 0.100 | 0.309 | 0.556 | 0.035 | C | 3.2 | 9.2 |
| 0380 | 2001 GM9 | 2457722.38400 | 1:47 | 14.4 | 18.89 | -0.10±0.02 | -0.13±0.03 | ≥0.13 | ≥107 | 0.000 | 0.001 | 1.000 | 0.000 | C | 2.7 | 12.9 |
| 0381 | 2001 HH35 | 2457796.44027 | 0:25 | 14.5 | 19.06 | 0.15±0.12 | 0.04±0.21 | ≥1.12 | ≥26 | 0.647 | 0.037 | 0.245 | 0.071 | S | 2.6 | 10.7 |
| 0382 | 2001 HN28 | 2457728.49022 | 0:57 | 14.4 | 18.95 | 0.09±0.07 | -0.01±0.08 | ≥0.14 | ≥58 | 0.576 | 0.076 | 0.336 | 0.012 | S | 2.7 | 8.0 |
| 0383 | 2001 HU8 | 2457801.39929 | 1:25 | 12.8 | 18.60 | 0.03±0.13 | 0.13±0.11 | ≥0.23 | ≥86 | 0.426 | 0.243 | 0.085 | 0.246 | - | 3.1 | 15.2 |
| 0384 | 2001 HZ12 | 2457691.37970 | 1:54 | 16.4 | 19.82 | 0.09±0.04 | 0.08±0.04 | ≥0.25 | ≥114 | 0.913 | 0.066 | 0.006 | 0.015 | S | 2.3 | 7.5 |
| 0385 | 2001 HZ41 | 2457691.46127 | 1:58 | 15.9 | 20.71 | 0.04±0.07 | 0.12±0.09 | ≥0.27 | ≥119 | 0.449 | 0.214 | 0.079 | 0.259 | - | 2.7 | 9.2 |
| 0386 | 2001 HZ45 | 2457730.42038 | 1:30 | 15.5 | 19.75 | 0.07±0.04 | 0.08±0.05 | ≥0.14 | ≥90 | 0.832 | 0.111 | 0.027 | 0.029 | S | 2.6 | 8.3 |
| 0387 | 2001 KA14 | 2457690.44264 | 1:49 | 14.9 | 19.57 | 0.04±0.04 | 0.01±0.05 | ≥0.17 | ≥109 | 0.440 | 0.185 | 0.373 | 0.002 | - | 2.7 | 14.5 |
| 0388 | 2001 KE19 | 2457800.44661 | 0:54 | 14.0 | 19.59 | 0.06±0.09 | 0.06±0.10 | ≥0.32 | ≥54 | 0.496 | 0.189 | 0.200 | 0.115 | - | 3.2 | 15.4 |
| 0389 | 2001 KJ43 | 2457691.47191 | 1:43 | 16.7 | 20.43 | 0.10±0.07 | 0.19±0.08 | ≥0.28 | ≥103 | 0.699 | 0.045 | 0.004 | 0.252 | S | 2.3 | 6.3 |
| 0390 | 2001 KL58 | 2457796.29214 | 1:07 | 15.0 | 19.85 | 0.17±0.10 | 0.17±0.10 | ≥0.29 | ≥68 | 0.854 | 0.030 | 0.013 | 0.104 | S | 2.6 | 22.6 |
| 0391 | 2001 KS16 | 2457800.44661 | 0:54 | 14.7 | 19.48 | -0.04±0.11 | 0.10±0.12 | ≥0.34 | ≥54 | 0.195 | 0.366 | 0.192 | 0.247 | - | 3.2 | 13.7 |
| 0392 | 2001 MW7 | 2457724.39556 | 0:50 | 15.7 | 19.39 | 0.11±0.04 | 0.16±0.04 | ≥0.34 | ≥51 | 0.883 | 0.011 | 0.000 | 0.105 | S | 2.3 | 4.3 |
| 0393 | 2001 NX19 | 2457802.38260 | 1:00 | 14.4 | 18.71 | 0.07±0.03 | 0.08±0.04 | ≥0.05 | ≥61 | 0.882 | 0.097 | 0.006 | 0.014 | S | 2.7 | 13.8 |
| 0394 | 2001 OF45 | 2457796.44027 | 0:58 | 14.5 | 18.64 | 0.14±0.07 | 0.16±0.08 | ≥0.25 | ≥59 | 0.867 | 0.022 | 0.005 | 0.106 | S | 2.6 | 13.5 |
| 0395 | 2001 OM29 | 2457730.42038 | 1:30 | 15.2 | 19.45 | 0.11±0.04 | 0.21±0.04 | ≥0.18 | ≥90 | 0.777 | 0.000 | 0.000 | 0.222 | S | 2.8 | 13.9 |
| 0396 | 2001 ON53 | 2457724.34310 | 0:55 | 16.4 | 19.68 | 0.21±0.08 | -0.01±0.12 | ≥0.18 | ≥55 | 0.874 | 0.003 | 0.120 | 0.003 | S | 2.4 | 7.4 |
| 0397 | 2001 PE31 | 2457800.44661 | 0:54 | 16.1 | 20.18 | 0.05±0.16 | 0.32±0.16 | ≥0.51 | ≥54 | 0.488 | 0.068 | 0.013 | 0.431 | - | 2.7 | 12.9 |
| 0398 | 2001 PV42 | 2457802.42659 | 1:45 | 15.6 | 19.83 | 0.05±0.03 | 0.07±0.03 | ≥1.07 | ≥106 | 0.757 | 0.210 | 0.021 | 0.012 | S | 2.7 | 10.4 |
| 0399 | 2001 QH246 | 2457724.28943 | 1:05 | 16.7 | 19.77 | 0.11±0.05 | 0.09±0.06 | ≥0.26 | ≥66 | 0.934 | 0.027 | 0.014 | 0.025 | S | 2.4 | 2.8 |
| 0400 | 2001 QJ296 | 2457691.37970 | 1:54 | 14.4 | 18.22 | 0.07±0.02 | 0.11±0.02 | 0.68±0.05 | 229±54 | 0.922 | 0.052 | 0.000 | 0.025 | S | 2.9 | 1.7 |
| 0401 | 2001 QK328 | 2457691.46127 | 2:00 | 18.0 | 20.75 | 0.13±0.10 | 0.14±0.15 | ≥0.62 | ≥121 | 0.733 | 0.054 | 0.077 | 0.136 | S | 2.4 | 6.4 |
| 0402 | 2001 QN128 | 2457691.37970 | 1:54 | 16.2 | 19.92 | -0.02±0.03 | 0.01±0.04 | ≥0.23 | ≥114 | 0.060 | 0.518 | 0.421 | 0.001 | X | 2.4 | 7.6 |
| 0403 | 2001 QT105 | 2457692.38892 | 1:20 | 15.2 | 17.46 | 0.02±0.02 | -0.04±0.03 | 0.06±0.03 | 160±36 | 0.022 | 0.002 | 0.976 | 0.000 | C | 2.4 | 6.2 |
| 0404 | 2001 QT236 | 2457724.34310 | 0:55 | 16.1 | 20.22 | 0.29±0.09 | 0.26±0.16 | ≥0.35 | ≥55 | 0.993 | 0.000 | 0.003 | 0.004 | S | 2.4 | 1.2 |
| 0405 | 2001 QT54 | 2457724.38316 | 1:08 | 16.0 | 18.54 | 0.10±0.02 | 0.07±0.03 | ≥0.16 | ≥69 | 0.996 | 0.003 | 0.000 | 0.001 | S | 2.4 | 3.2 |
| 0406 | 2001 QW258 | 2457724.38316 | 1:08 | 15.0 | 17.74 | 0.06±0.02 | -0.05±0.03 | ≥0.36 | ≥69 | 0.208 | 0.000 | 0.792 | 0.000 | C | 2.4 | 5.7 |
| 0407 | 2001 QW292 | 2457722.38400 | 1:47 | 16.3 | 20.22 | -0.02±0.05 | -0.06±0.06 | ≥1.02 | ≥107 | 0.044 | 0.154 | 0.802 | 0.000 | C | 2.9 | 14.2 |
| 0408 | 2001 QZ288 | 2457724.34310 | 0:55 | 16.4 | 18.95 | -0.09±0.05 | -0.06±0.06 | ≥0.18 | ≥55 | 0.003 | 0.202 | 0.795 | 0.000 | C | 2.4 | 2.3 |
| 0409 | 2001 RA13 | 2457724.28943 | 1:05 | 16.6 | 19.20 | 0.23±0.03 | 0.45±0.04 | ≥0.14 | ≥66 | 1.000 | 0.000 | 0.000 | 0.000 | S | 2.4 | 2.1 |
| 0410 | 2001 RA89 | 2457692.38892 | 1:20 | 14.8 | 18.08 | 0.09±0.02 | 0.14±0.03 | ≥0.19 | ≥80 | 0.969 | 0.000 | 0.000 | 0.030 | S | 2.4 | 3.0 |
| 0411 | 2001 RD115 | 2457690.38513 | 0:48 | 16.6 | 19.89 | 0.14±0.07 | 0.19±0.09 | ≥0.44 | ≥49 | 0.860 | 0.016 | 0.003 | 0.121 | S | 2.4 | 2.9 |
| 0412 | 2001 RJ32 | 2457798.46669 | 0:53 | 15.3 | 18.56 | -0.06±0.06 | -0.06±0.07 | ≥0.11 | ≥53 | 0.018 | 0.253 | 0.727 | 0.002 | C | 2.7 | 13.5 |

*Table 1 continued on next page*

**Table 1** *(continued)*

| No.[a] | Object | Obs. Start (JD) | Obs. Duration (h:mm) | H[b] (mag) | V (mag) | V-R[c] (mag) | V-I[c] (mag) | Amplitude[d] (mag) | Rot. Period[d] (min) | S-type (probability) | X-type | C-type | D-type | Tax. | a[b] (au) | i[b] (deg) |
|---|---|---|---|---|---|---|---|---|---|---|---|---|---|---|---|---|
| 0413 | 2001 RL28 | 2457722.38400 | 1:47 | 16.2 | 18.16 | 0.06±0.02 | 0.07±0.02 | 0.08±0.06 | 122±35 | 0.881 | 0.116 | 0.002 | 0.000 | S | 2.4 | 3.7 |
| 0414 | 2001 RR6 | 2457692.38892 | 1:20 | 15.3 | 19.20 | 0.04±0.02 | 0.02±0.03 | ≥0.06 | ≥80 | 0.581 | 0.122 | 0.297 | 0.000 | S | 2.9 | 1.6 |
| 0415 | 2001 RT85 | 2457692.38892 | 1:20 | 16.5 | 18.49 | -0.08±0.02 | -0.14±0.03 | ≥0.11 | ≥80 | 0.000 | 0.000 | 1.000 | 0.000 | C | 2.4 | 2.3 |
| 0416 | 2001 RW109 | 2457690.33738 | 1:57 | 17.5 | 20.99 | 0.10±0.11 | 0.14±0.10 | ≥0.72 | ≥118 | 0.646 | 0.119 | 0.040 | 0.195 | S | 2.4 | 3.1 |
| 0417 | 2001 RW61 | 2457690.33738 | 1:05 | 15.7 | 20.58 | 0.11±0.07 | 0.11±0.08 | ≥0.37 | ≥66 | 0.788 | 0.073 | 0.030 | 0.108 | S | 2.9 | 2.9 |
| 0418 | 2001 SA36 | 2457690.33738 | 1:57 | 14.7 | 18.99 | 0.04±0.02 | 0.08±0.03 | ≥0.13 | ≥118 | 0.568 | 0.421 | 0.004 | 0.007 | S | 2.9 | 3.0 |
| 0419 | 2001 SB251 | 2457690.33738 | 1:57 | 16.0 | 19.91 | 0.03±0.03 | 0.09±0.04 | ≥0.18 | ≥118 | 0.415 | 0.446 | 0.029 | 0.110 | - | 2.9 | 3.2 |
| 0420 | 2001 SB320 | 2457692.44645 | 1:00 | 16.5 | 19.76 | 0.03±0.06 | -0.14±0.12 | ≥0.18 | ≥60 | 0.091 | 0.052 | 0.850 | 0.007 | C | 2.4 | 2.1 |
| 0421 | 2001 SC294 | 2457724.34161 | 0:57 | 16.5 | 20.38 | 0.20±0.13 | 0.05±0.12 | ≥0.85 | ≥58 | 0.827 | 0.039 | 0.099 | 0.035 | S | 2.9 | 2.0 |
| 0422 | 2001 SC49 | 2457802.50174 | 0:46 | 15.8 | 20.19 | 0.13±0.10 | -0.04±0.23 | ≥0.22 | ≥47 | 0.518 | 0.029 | 0.401 | 0.052 | S | 2.3 | 8.7 |
| 0423 | 2001 SF76 | 2457724.38316 | 1:08 | 15.2 | 19.73 | 0.09±0.04 | 0.02±0.05 | ≥0.13 | ≥69 | 0.800 | 0.039 | 0.158 | 0.003 | S | 2.8 | 4.5 |
| 0424 | 2001 SG136 | 2457802.38260 | 1:00 | 16.0 | 19.33 | -0.10±0.05 | -0.09±0.09 | ≥0.21 | ≥61 | 0.001 | 0.201 | 0.796 | 0.003 | C | 2.8 | 10.2 |
| 0425 | 2001 SJ32 | 2457724.34904 | 0:46 | 15.8 | 19.50 | 0.06±0.08 | 0.12±0.10 | ≥0.16 | ≥47 | 0.508 | 0.166 | 0.090 | 0.237 | S | 2.9 | 1.5 |
| 0426 | 2001 SL211 | 2457691.37970 | 1:54 | 17.5 | 19.82 | 0.04±0.03 | 0.06±0.04 | ≥0.15 | ≥114 | 0.562 | 0.326 | 0.095 | 0.017 | S | 2.4 | 0.6 |
| 0427 | 2001 SL256 | 2457691.37970 | 1:54 | 17.4 | 19.71 | 0.04±0.03 | 0.03±0.04 | ≥0.48 | ≥114 | 0.525 | 0.204 | 0.269 | 0.001 | S | 2.4 | 1.3 |
| 0428 | 2001 SM20 | 2457690.33738 | 1:57 | 14.7 | 18.78 | 0.05±0.02 | 0.07±0.03 | ≥0.13 | ≥118 | 0.745 | 0.246 | 0.006 | 0.003 | S | 2.9 | 2.6 |
| 0429 | 2001 SO266 | 2457724.28943 | 1:08 | 15.1 | 20.03 | 0.16±0.05 | 0.19±0.07 | ≥0.40 | ≥68 | 0.970 | 0.001 | 0.000 | 0.029 | S | 2.9 | 1.2 |
| 0430 | 2001 SO278 | 2457798.46669 | 0:53 | 15.9 | 19.00 | 0.19±0.08 | 0.04±0.10 | ≥0.19 | ≥53 | 0.912 | 0.011 | 0.068 | 0.010 | S | 2.3 | 4.9 |
| 0431 | 2001 SS232 | 2457691.37970 | 1:54 | 15.1 | 19.68 | -0.02±0.04 | 0.02±0.05 | 0.88±0.09 | 229±50 | 0.112 | 0.518 | 0.365 | 0.005 | X | 2.9 | 7.5 |
| 0432 | 2001 ST175 | 2457690.33738 | 1:57 | 15.7 | 19.55 | 0.09±0.03 | 0.15±0.04 | ≥0.33 | ≥118 | 0.854 | 0.010 | 0.000 | 0.136 | S | 2.4 | 3.2 |
| 0433 | 2001 SU69 | 2457728.49022 | 0:57 | 14.8 | 18.75 | 0.11±0.05 | 0.10±0.06 | ≥0.15 | ≥58 | 0.897 | 0.039 | 0.013 | 0.050 | S | 2.4 | 7.1 |
| 0434 | 2001 TD148 | 2457730.42038 | 1:30 | 16.3 | 19.33 | 0.08±0.04 | 0.06±0.05 | ≥0.16 | ≥90 | 0.861 | 0.078 | 0.048 | 0.013 | S | 2.4 | 6.9 |
| 0435 | 2001 TD6 | 2457802.50174 | 0:55 | 15.6 | 20.01 | -0.11±0.07 | -0.16±0.11 | ≥0.23 | ≥55 | 0.003 | 0.076 | 0.919 | 0.001 | C | 2.7 | 5.9 |
| 0436 | 2001 TF126 | 2457724.38316 | 1:08 | 16.5 | 19.56 | 0.10±0.03 | 0.13±0.04 | ≥0.23 | ≥69 | 0.969 | 0.004 | 0.000 | 0.027 | S | 2.4 | 1.8 |
| 0437 | 2001 TG129 | 2457692.44645 | 1:00 | 15.6 | 19.66 | 0.10±0.05 | 0.13±0.06 | ≥0.11 | ≥60 | 0.843 | 0.041 | 0.003 | 0.113 | S | 2.4 | 1.6 |
| 0438 | 2001 TG68 | 2457802.50174 | 0:55 | 15.5 | 19.49 | 0.08±0.06 | 0.02±0.08 | ≥0.29 | ≥55 | 0.647 | 0.099 | 0.234 | 0.021 | S | 2.3 | 7.6 |
| 0439 | 2001 TJ174 | 2457692.38892 | 1:20 | 16.8 | 19.85 | 0.07±0.03 | -0.03±0.05 | ≥0.11 | ≥80 | 0.505 | 0.007 | 0.488 | 0.000 | S | 2.4 | 6.6 |
| 0440 | 2001 TJ255 | 2457722.38400 | 1:47 | 16.5 | 20.54 | 0.09±0.10 | 0.21±0.11 | ≥0.26 | ≥107 | 0.619 | 0.067 | 0.013 | 0.301 | S | 2.9 | 6.2 |
| 0441 | 2001 TL76 | 2457690.33738 | 1:57 | 16.7 | 18.82 | -0.06±0.02 | -0.09±0.03 | ≥0.17 | ≥118 | 0.000 | 0.002 | 0.998 | 0.000 | C | 2.4 | 3.9 |
| 0442 | 2001 TN29 | 2457692.38892 | 1:20 | 16.3 | 19.79 | 0.05±0.03 | 0.03±0.04 | ≥0.30 | ≥80 | 0.657 | 0.151 | 0.191 | 0.001 | S | 2.9 | 5.7 |
| 0443 | 2001 TQ55 | 2457724.28943 | 1:08 | 15.1 | 19.62 | -0.13±0.04 | -0.12±0.07 | ≥0.19 | ≥68 | 0.000 | 0.053 | 0.947 | 0.000 | C | 2.9 | 0.3 |
| 0444 | 2001 TU134 | 2457798.51082 | 1:01 | 15.1 | 18.82 | 0.21±0.07 | 0.26±0.07 | ≥0.14 | ≥61 | 0.978 | 0.000 | 0.000 | 0.022 | S | 2.7 | 8.3 |
| 0445 | 2001 UA49 | 2457724.38316 | 1:08 | 15.2 | 19.06 | 0.10±0.03 | 0.09±0.03 | ≥0.09 | ≥69 | 0.987 | 0.010 | 0.000 | 0.003 | S | 2.9 | 2.8 |
| 0446 | 2001 UA81 | 2457692.45553 | 0:46 | 16.4 | 19.84 | 0.12±0.06 | 0.03±0.07 | ≥0.41 | ≥47 | 0.836 | 0.035 | 0.120 | 0.008 | S | 2.9 | 7.1 |
| 0447 | 2001 UA86 | 2457690.33738 | 1:57 | 16.1 | 20.37 | 0.03±0.06 | 0.05±0.07 | ≥0.39 | ≥118 | 0.406 | 0.313 | 0.220 | 0.061 | - | 2.9 | 3.2 |
| 0448 | 2001 UD131 | 2457692.38892 | 1:20 | 16.7 | 19.78 | -0.03±0.03 | -0.06±0.04 | ≥0.11 | ≥80 | 0.001 | 0.080 | 0.920 | 0.000 | C | 2.4 | 5.5 |
| 0449 | 2001 UD163 | 2457724.34161 | 0:57 | 16.3 | 20.38 | 0.28±0.12 | 0.19±0.13 | ≥0.63 | ≥58 | 0.968 | 0.004 | 0.005 | 0.023 | S | 2.4 | 6.7 |
| 0450 | 2001 UD227 | 2457690.33738 | 1:57 | 16.3 | 20.60 | 0.08±0.07 | 0.04±0.08 | ≥1.02 | ≥118 | 0.645 | 0.126 | 0.188 | 0.041 | S | 2.9 | 11.8 |
| 0451 | 2001 UD46 | 2457692.44645 | 1:00 | 15.7 | 18.93 | 0.04±0.04 | 0.06±0.06 | ≥0.08 | ≥60 | 0.507 | 0.300 | 0.128 | 0.065 | S | 2.4 | 7.9 |
| 0452 | 2001 UK40 | 2457690.33738 | 1:57 | 17.5 | 20.04 | 0.06±0.04 | 0.04±0.05 | ≥0.24 | ≥118 | 0.708 | 0.139 | 0.142 | 0.012 | S | 2.4 | 2.9 |
| 0453 | 2001 UK8 | 2457690.38513 | 0:48 | 17.0 | 19.43 | 0.09±0.04 | 0.15±0.08 | ≥0.36 | ≥49 | 0.765 | 0.031 | 0.015 | 0.190 | S | 2.4 | 3.3 |
| 0454 | 2001 UL108 | 2457691.37970 | 1:54 | 17.5 | 19.98 | 0.01±0.04 | 0.05±0.05 | ≥0.26 | ≥114 | 0.305 | 0.504 | 0.167 | 0.024 | X | 2.4 | 1.3 |
| 0455 | 2001 UQ153 | 2457692.44645 | 1:00 | 16.0 | 18.63 | 0.13±0.04 | 0.10±0.05 | ≥0.05 | ≥60 | 0.987 | 0.005 | 0.001 | 0.007 | S | 2.4 | 5.7 |
| 0456 | 2001 UU138 | 2457691.37970 | 1:54 | 16.5 | 20.39 | 0.08±0.06 | 0.07±0.07 | ≥0.59 | ≥114 | 0.743 | 0.127 | 0.078 | 0.052 | S | 2.4 | 1.3 |
| 0457 | 2001 UV72 | 2457692.44645 | 1:00 | 15.7 | 19.85 | -0.03±0.05 | 0.02±0.06 | ≥0.43 | ≥60 | 0.077 | 0.559 | 0.356 | 0.009 | X | 2.9 | 3.2 |
| 0458 | 2001 UW35 | 2457802.50174 | 0:55 | 14.9 | 19.57 | 0.03±0.05 | -0.00±0.07 | ≥0.15 | ≥55 | 0.351 | 0.177 | 0.460 | 0.012 | - | 2.7 | 10.1 |

*Table 1 continued on next page*

**Table 1** *(continued)*

| No.[a] | Object | Obs. Start (JD) | Obs. Duration (h:mm) | H[b] (mag) | V (mag) | V-R[c] (mag) | V-I[c] (mag) | Amplitude[d] (mag) | Rot. Period[d] (min) | S-type (probability) | X-type | C-type | D-type | Tax. | a[b] (au) | i[b] (deg) |
|---|---|---|---|---|---|---|---|---|---|---|---|---|---|---|---|---|
| 0459 | 2001 UY117 | 2457692.38892 | 1:20 | 15.6 | 19.11 | 0.04±0.02 | 0.02±0.03 | ≥0.17 | ≥80 | 0.682 | 0.093 | 0.225 | 0.000 | S | 2.9 | 2.5 |
| 0460 | 2001 VH18 | 2457798.47394 | 0:42 | 15.2 | 19.50 | 0.14±0.12 | 0.21±0.12 | ≥0.37 | ≥43 | 0.731 | 0.049 | 0.014 | 0.206 | S | 2.8 | 4.6 |
| 0461 | 2001 VR90 | 2457687.38022 | 3:03 | 14.9 | 18.19 | -0.09±0.01 | -0.05±0.02 | ≥0.13 | ≥183 | 0.000 | 0.060 | 0.940 | 0.000 | C | 2.9 | 15.0 |
| 0462 | 2001 VT103 | 2457724.28943 | 1:08 | 14.9 | 19.79 | -0.01±0.04 | -0.07±0.06 | ≥0.24 | ≥68 | 0.034 | 0.108 | 0.857 | 0.000 | C | 3.0 | 10.2 |
| 0463 | 2001 VZ23 | 2457692.38892 | 1:20 | 15.0 | 18.97 | 0.02±0.02 | 0.05±0.03 | ≥0.21 | ≥80 | 0.325 | 0.541 | 0.131 | 0.003 | X | 3.0 | 5.6 |
| 0464 | 2001 WQ83 | 2457724.39853 | 0:46 | 15.7 | 20.04 | 0.03±0.07 | 0.10±0.09 | ≥0.31 | ≥47 | 0.415 | 0.248 | 0.114 | 0.223 | - | 2.9 | 3.5 |
| 0465 | 2001 XA185 | 2457724.38316 | 1:06 | 15.6 | 19.91 | -0.05±0.07 | -0.15±0.06 | ≥0.15 | ≥66 | 0.004 | 0.011 | 0.985 | 0.000 | C | 3.0 | 0.9 |
| 0466 | 2001 XA195 | 2457691.37970 | 1:54 | 16.5 | 18.72 | 0.03±0.02 | 0.07±0.02 | ≥0.14 | ≥114 | 0.440 | 0.541 | 0.018 | 0.001 | X | 2.5 | 2.0 |
| 0467 | 2001 XD194 | 2457692.44645 | 1:00 | 15.2 | 20.01 | -0.02±0.06 | 0.06±0.07 | ≥0.20 | ≥60 | 0.141 | 0.604 | 0.185 | 0.069 | X | 3.1 | 1.7 |
| 0468 | 2001 XE251 | 2457690.33738 | 1:57 | 15.6 | 18.92 | -0.04±0.02 | -0.02±0.02 | ≥0.12 | ≥118 | 0.000 | 0.395 | 0.605 | 0.000 | C | 3.0 | 2.8 |
| 0469 | 2001 XG260 | 2457690.33738 | 1:57 | 17.2 | 20.29 | 0.05±0.05 | 0.02±0.06 | ≥0.37 | ≥118 | 0.545 | 0.151 | 0.290 | 0.013 | S | 2.4 | 2.6 |
| 0470 | 2001 XJ228 | 2457692.44645 | 1:00 | 17.2 | 19.37 | 0.06±0.04 | 0.18±0.06 | ≥0.11 | ≥60 | 0.485 | 0.069 | 0.001 | 0.445 | - | 2.4 | 3.9 |
| 0471 | 2001 XL240 | 2457691.37970 | 1:54 | 16.4 | 19.15 | 0.08±0.03 | 0.15±0.03 | ≥0.20 | ≥114 | 0.807 | 0.012 | 0.000 | 0.181 | S | 2.1 | 2.7 |
| 0472 | 2001 XO80 | 2457730.42038 | 1:30 | 15.8 | 18.92 | 0.09±0.03 | 0.03±0.04 | ≥0.09 | ≥90 | 0.948 | 0.013 | 0.038 | 0.000 | S | 2.4 | 7.3 |
| 0473 | 2001 XR167 | 2457724.28943 | 1:08 | 16.8 | 20.26 | 0.09±0.08 | -0.04±0.12 | ≥0.46 | ≥68 | 0.499 | 0.060 | 0.417 | 0.024 | - | 2.4 | 1.2 |
| 0474 | 2001 XS151 | 2457690.33738 | 1:57 | 15.4 | 19.41 | 0.05±0.03 | 0.08±0.04 | ≥0.23 | ≥118 | 0.719 | 0.237 | 0.015 | 0.030 | S | 3.0 | 3.0 |
| 0475 | 2001 XW51 | 2457724.38316 | 1:08 | 15.9 | 18.89 | 0.04±0.03 | 0.07±0.04 | ≥0.10 | ≥69 | 0.658 | 0.283 | 0.032 | 0.027 | S | 2.4 | 1.3 |
| 0476 | 2001 XZ191 | 2457802.50174 | 0:55 | 16.0 | 19.74 | -0.03±0.06 | 0.02±0.08 | ≥0.32 | ≥55 | 0.112 | 0.458 | 0.390 | 0.040 | - | 2.4 | 5.5 |
| 0477 | 2001 YH117 | 2457802.50174 | 0:55 | 16.4 | 20.04 | 0.03±0.09 | -0.06±0.11 | ≥0.19 | ≥55 | 0.252 | 0.127 | 0.606 | 0.015 | C | 2.4 | 5.0 |
| 0478 | 2001 YL13 | 2457802.50174 | 0:55 | 16.3 | 19.81 | 0.08±0.08 | -0.07±0.10 | ≥0.14 | ≥55 | 0.385 | 0.053 | 0.555 | 0.006 | C | 2.3 | 5.4 |
| 0479 | 2001 YU121 | 2457692.44645 | 1:00 | 14.7 | 18.97 | 0.01±0.04 | 0.07±0.05 | ≥0.11 | ≥60 | 0.296 | 0.509 | 0.106 | 0.089 | X | 3.0 | 13.0 |
| 0480 | 2001 YU25 | 2457692.38892 | 1:20 | 17.8 | 20.02 | 0.06±0.03 | 0.09±0.04 | ≥0.13 | ≥80 | 0.763 | 0.167 | 0.016 | 0.053 | S | 2.4 | 2.0 |
| 0481 | 2002 AG175 | 2457690.33738 | 1:57 | 15.5 | 20.13 | -0.01±0.05 | -0.05±0.05 | ≥0.23 | ≥118 | 0.052 | 0.148 | 0.800 | 0.000 | C | 3.1 | 2.3 |
| 0482 | 2002 AG203 | 2457730.42038 | 1:30 | 17.0 | 20.03 | 0.15±0.05 | 0.10±0.05 | ≥0.56 | ≥90 | 0.991 | 0.003 | 0.001 | 0.006 | S | 2.4 | 7.1 |
| 0483 | 2002 AH145 | 2457722.38400 | 1:47 | 17.5 | 19.84 | -0.05±0.04 | -0.05±0.05 | ≥0.47 | ≥107 | 0.002 | 0.251 | 0.747 | 0.000 | C | 2.4 | 3.2 |
| 0484 | 2002 AQ164 | 2457690.33738 | 1:57 | 16.3 | 18.92 | 0.05±0.02 | 0.07±0.03 | ≥0.11 | ≥118 | 0.786 | 0.205 | 0.007 | 0.002 | S | 2.5 | 3.1 |
| 0485 | 2002 AR114 | 2457691.37970 | 1:54 | 14.9 | 18.79 | -0.03±0.04 | -0.01±0.05 | ≥0.09 | ≥114 | 0.034 | 0.411 | 0.555 | 0.001 | C | 3.1 | 1.5 |
| 0486 | 2002 AV9 | 2457728.49022 | 0:57 | 14.5 | 18.94 | 0.08±0.06 | 0.10±0.08 | ≥0.36 | ≥58 | 0.699 | 0.102 | 0.065 | 0.134 | S | 3.0 | 12.4 |
| 0487 | 2002 CA248 | 2457690.33738 | 1:57 | 14.8 | 19.11 | -0.07±0.02 | -0.09±0.03 | ≥0.13 | ≥118 | 0.000 | 0.017 | 0.983 | 0.000 | C | 3.1 | 2.7 |
| 0488 | 2002 CA292 | 2457691.37970 | 1:54 | 15.8 | 20.53 | -0.05±0.08 | -0.23±0.10 | ≥0.27 | ≥114 | 0.006 | 0.017 | 0.977 | 0.000 | C | 3.1 | 1.4 |
| 0489 | 2002 CL215 | 2457692.44645 | 1:00 | 16.5 | 20.00 | 0.06±0.06 | 0.16±0.07 | ≥0.17 | ≥60 | 0.499 | 0.118 | 0.005 | 0.377 | - | 2.5 | 1.2 |
| 0490 | 2002 CM265 | 2457724.38316 | 1:08 | 14.4 | 19.57 | 0.05±0.03 | 0.11±0.04 | ≥0.14 | ≥69 | 0.635 | 0.189 | 0.008 | 0.168 | S | 3.1 | 9.9 |
| 0491 | 2002 CN110 | 2457692.39079 | 1:15 | 15.4 | 20.23 | -0.10±0.05 | -0.13±0.05 | ≥0.44 | ≥75 | 0.000 | 0.008 | 0.992 | 0.000 | C | 3.2 | 2.0 |
| 0492 | 2002 CR75 | 2457692.38892 | 1:20 | 16.2 | 19.57 | 0.03±0.03 | 0.03±0.03 | ≥0.14 | ≥80 | 0.547 | 0.210 | 0.242 | 0.001 | S | 2.5 | 2.4 |
| 0493 | 2002 CS194 | 2457690.33738 | 1:57 | 16.4 | 20.25 | 0.08±0.04 | 0.12±0.06 | ≥0.47 | ≥118 | 0.764 | 0.079 | 0.009 | 0.148 | S | 2.5 | 2.9 |
| 0494 | 2002 ED135 | 2457722.38400 | 1:45 | 15.9 | 20.66 | 0.10±0.06 | 0.07±0.07 | ≥0.27 | ≥105 | 0.790 | 0.090 | 0.072 | 0.048 | S | 2.6 | 7.2 |
| 0495 | 2002 FL16 | 2457687.38022 | 3:03 | 14.5 | 19.37 | -0.06±0.02 | -0.10±0.02 | ≥0.31 | ≥183 | 0.000 | 0.001 | 0.999 | 0.000 | C | 3.2 | 21.6 |
| 0496 | 2002 GC80 | 2457690.43605 | 1:58 | 14.7 | 19.39 | 0.08±0.04 | 0.14±0.05 | ≥0.41 | ≥119 | 0.754 | 0.062 | 0.002 | 0.182 | S | 3.1 | 9.5 |
| 0497 | 2002 GG | 2457691.37970 | 1:54 | 17.4 | 20.67 | 0.07±0.08 | -0.15±0.12 | ≥0.32 | ≥114 | 0.200 | 0.026 | 0.771 | 0.003 | C | 2.2 | 7.1 |
| 0498 | 2002 GK9 | 2457690.33738 | 1:57 | 15.7 | 19.98 | 0.04±0.04 | 0.14±0.04 | ≥0.22 | ≥118 | 0.454 | 0.215 | 0.002 | 0.329 | - | 2.6 | 16.6 |
| 0499 | 2002 GS81 | 2457800.27020 | 1:03 | 14.8 | 20.04 | 0.11±0.06 | 0.07±0.09 | ≥0.17 | ≥64 | 0.825 | 0.047 | 0.086 | 0.042 | S | 3.1 | 10.4 |
| 0500 | 2002 GV150 | 2457722.38400 | 1:47 | 17.1 | 19.91 | 0.05±0.03 | 0.08±0.04 | ≥0.22 | ≥107 | 0.644 | 0.267 | 0.033 | 0.056 | S | 2.2 | 3.6 |
| 0501 | 2002 HV7 | 2457722.38400 | 1:47 | 15.8 | 19.36 | -0.06±0.02 | -0.12±0.04 | ≥0.14 | ≥107 | 0.000 | 0.004 | 0.996 | 0.000 | C | 3.0 | 5.0 |
| 0502 | 2002 JJ72 | 2457722.38400 | 1:47 | 16.2 | 19.06 | 0.07±0.02 | 0.11±0.03 | ≥0.93 | ≥107 | 0.936 | 0.026 | 0.000 | 0.037 | S | 2.2 | 3.9 |
| 0503 | 2002 JP83 | 2457691.37970 | 1:54 | 16.3 | 18.27 | 0.05±0.02 | 0.06±0.02 | ≥0.07 | ≥114 | 0.852 | 0.142 | 0.005 | 0.000 | S | 2.2 | 4.3 |
| 0504 | 2002 JQ1 | 2457724.38316 | 1:08 | 15.8 | 19.38 | 0.08±0.03 | 0.04±0.04 | ≥0.09 | ≥69 | 0.904 | 0.045 | 0.050 | 0.001 | S | 2.2 | 1.4 |

*Table 1 continued on next page*

**Table 1** *(continued)*

| No.[a] | Object | Obs. Start (JD) | Obs. Duration (h:mm) | H[b] (mag) | V (mag) | V-R[c] (mag) | V-I[c] (mag) | Amplitude[d] (mag) | Rot. Period[d] (min) | S-type | X-type (probability) | C-type | D-type | Tax. | a[b] (au) | i[b] (deg) |
|---|---|---|---|---|---|---|---|---|---|---|---|---|---|---|---|---|
| 0505 | 2002 JS28 | 2457692.44645 | 1:00 | 15.4 | 19.79 | 0.08±0.04 | 0.15±0.06 | ≥0.49 | ≥60 | 0.733 | 0.047 | 0.003 | 0.217 | S | 2.6 | 3.9 |
| 0506 | 2002 JT48 | 2457724.34161 | 0:57 | 15.6 | 19.94 | 0.21±0.08 | 0.07±0.10 | ≥0.25 | ≥58 | 0.965 | 0.005 | 0.021 | 0.009 | S | 2.6 | 8.7 |
| 0507 | 2002 JU53 | 2457692.38892 | 1:20 | 16.3 | 19.19 | 0.12±0.02 | 0.19±0.03 | ≥0.12 | ≥80 | 0.975 | 0.000 | 0.000 | 0.025 | S | 2.2 | 3.0 |
| 0508 | 2002 JV75 | 2457722.38400 | 1:47 | 15.5 | 20.08 | 0.01±0.04 | 0.15±0.05 | ≥0.17 | ≥107 | 0.184 | 0.327 | 0.004 | 0.484 | - | 3.1 | 9.9 |
| 0509 | 2002 JX63 | 2457724.38316 | 1:08 | 15.8 | 19.91 | 0.05±0.04 | 0.05±0.05 | ≥0.19 | ≥69 | 0.637 | 0.193 | 0.147 | 0.024 | S | 2.6 | 6.8 |
| 0510 | 2002 LB5 | 2457796.44027 | 0:58 | 14.8 | 19.30 | -0.05±0.12 | -0.03±0.15 | ≥0.25 | ≥59 | 0.146 | 0.232 | 0.539 | 0.083 | C | 3.0 | 12.2 |
| 0511 | 2002 LO19 | 2457796.44027 | 0:58 | 14.9 | 19.48 | 0.00±0.26 | -0.05±0.25 | ≥0.54 | ≥59 | 0.300 | 0.119 | 0.476 | 0.104 | - | 3.0 | 11.9 |
| 0512 | 2002 LW6 | 2457802.39311 | 0:45 | 14.8 | 19.68 | 0.01±0.04 | 0.02±0.06 | ≥0.14 | ≥46 | 0.245 | 0.370 | 0.365 | 0.021 | - | 3.1 | 9.6 |
| 0513 | 2002 NV20 | 2457690.38513 | 0:48 | 16.4 | 19.39 | 0.09±0.05 | 0.06±0.07 | ≥0.16 | ≥49 | 0.819 | 0.064 | 0.083 | 0.034 | S | 2.3 | 5.8 |
| 0514 | 2002 NV34 | 2457724.28943 | 1:08 | 15.8 | 20.45 | 0.15±0.07 | 0.22±0.12 | ≥0.69 | ≥68 | 0.868 | 0.011 | 0.007 | 0.113 | S | 2.7 | 6.3 |
| 0515 | 2002 NW46 | 2457692.38892 | 1:20 | 16.1 | 19.54 | -0.03±0.03 | -0.04±0.04 | ≥0.12 | ≥80 | 0.003 | 0.146 | 0.851 | 0.000 | C | 2.7 | 4.5 |
| 0516 | 2002 PK11 | 2457692.38892 | 1:20 | 17.2 | 20.45 | -0.03±0.04 | 0.02±0.07 | ≥0.19 | ≥80 | 0.064 | 0.547 | 0.360 | 0.028 | X | 2.7 | 2.6 |
| 0517 | 2002 PL116 | 2457722.38400 | 1:47 | 16.4 | 19.31 | 0.04±0.03 | 0.06±0.04 | 0.29±0.06 | 202±39 | 0.611 | 0.311 | 0.067 | 0.011 | S | 2.7 | 12.9 |
| 0518 | 2002 PU102 | 2457691.54720 | 1:27 | 15.4 | 19.30 | 0.09±0.04 | 0.07±0.05 | ≥0.13 | ≥87 | 0.898 | 0.046 | 0.034 | 0.021 | S | 2.7 | 15.1 |
| 0519 | 2002 PY105 | 2457691.37970 | 1:50 | 17.8 | 20.75 | 0.13±0.08 | 0.05±0.12 | ≥0.45 | ≥111 | 0.741 | 0.045 | 0.159 | 0.055 | S | 2.3 | 6.5 |
| 0520 | 2002 QJ30 | 2457691.46127 | 2:00 | 16.1 | 19.57 | 0.06±0.04 | 0.11±0.05 | 0.60±0.09 | 241±43 | 0.644 | 0.176 | 0.012 | 0.168 | S | 2.7 | 6.1 |
| 0521 | 2002 QK91 | 2457691.46127 | 2:00 | 17.9 | 20.77 | 0.08±0.10 | 0.03±0.17 | ≥0.35 | ≥121 | 0.484 | 0.096 | 0.313 | 0.107 | - | 2.3 | 7.0 |
| 0522 | 2002 QQ116 | 2457691.37970 | 0:59 | 17.7 | 19.79 | -0.00±0.04 | 0.00±0.07 | ≥0.16 | ≥60 | 0.142 | 0.353 | 0.488 | 0.017 | - | 2.3 | 3.0 |
| 0523 | 2002 RA85 | 2457691.37970 | 1:54 | 15.1 | 17.97 | -0.07±0.02 | -0.09±0.02 | 0.31±0.05 | 229±47 | 0.000 | 0.002 | 0.998 | 0.000 | C | 2.8 | 2.9 |
| 0524 | 2002 RB251 | 2457691.46127 | 2:00 | 16.5 | 20.49 | 0.10±0.06 | 0.14±0.08 | ≥0.37 | ≥121 | 0.778 | 0.061 | 0.017 | 0.144 | S | 2.7 | 6.5 |
| 0525 | 2002 RE94 | 2457690.42183 | 2:21 | 15.2 | 19.14 | 0.05±0.08 | 0.07±0.04 | ≥0.28 | ≥142 | 0.596 | 0.351 | 0.041 | 0.013 | S | 2.7 | 9.3 |
| 0526 | 2002 RM6 | 2457692.44645 | 1:00 | 17.5 | 18.64 | 0.01±0.04 | 0.08±0.05 | ≥0.09 | ≥60 | 0.319 | 0.498 | 0.082 | 0.100 | - | 2.3 | 1.0 |
| 0527 | 2002 RX41 | 2457690.33738 | 1:57 | 17.1 | 19.61 | 0.03±0.03 | 0.10±0.04 | ≥0.18 | ≥118 | 0.410 | 0.444 | 0.015 | 0.130 | - | 2.3 | 2.6 |
| 0528 | 2002 RZ59 | 2457722.38400 | 1:47 | 16.6 | 19.91 | 0.06±0.03 | -0.19±0.05 | ≥0.38 | ≥107 | 0.002 | 0.000 | 0.998 | 0.000 | C | 2.2 | 6.8 |
| 0529 | 2002 SB26 | 2457690.33738 | 1:57 | 15.3 | 18.70 | 0.04±0.02 | 0.13±0.02 | ≥0.14 | ≥118 | 0.610 | 0.114 | 0.000 | 0.275 | S | 2.8 | 4.2 |
| 0530 | 2002 SZ38 | 2457690.33738 | 1:05 | 16.0 | 19.80 | -0.02±0.04 | -0.06±0.05 | ≥0.11 | ≥66 | 0.010 | 0.101 | 0.890 | 0.000 | C | 2.8 | 5.0 |
| 0531 | 2002 TC205 | 2457691.46127 | 2:00 | 17.0 | 20.23 | 0.01±0.06 | -0.05±0.07 | ≥0.31 | ≥121 | 0.161 | 0.143 | 0.694 | 0.002 | C | 2.3 | 7.4 |
| 0532 | 2002 TE17 | 2457692.38892 | 1:20 | 17.1 | 19.38 | 0.00±0.02 | 0.01±0.03 | ≥0.09 | ≥80 | 0.136 | 0.301 | 0.563 | 0.000 | C | 2.3 | 4.5 |
| 0533 | 2002 TG20 | 2457692.44645 | 1:00 | 17.6 | 19.14 | 0.04±0.04 | 0.04±0.05 | ≥0.10 | ≥60 | 0.508 | 0.264 | 0.213 | 0.015 | S | 2.3 | 3.9 |
| 0534 | 2002 TJ259 | 2457722.38400 | 1:47 | 16.4 | 18.88 | -0.01±0.03 | 0.00±0.03 | 0.68±0.04 | 214±47 | 0.095 | 0.349 | 0.555 | 0.000 | C | 2.3 | 3.7 |
| 0535 | 2002 TK19 | 2457691.42249 | 0:52 | 15.3 | 19.33 | -0.00±0.04 | 0.03±0.06 | ≥0.12 | ≥53 | 0.154 | 0.486 | 0.341 | 0.019 | - | 2.8 | 5.2 |
| 0536 | 2002 TM286 | 2457691.37970 | 0:59 | 15.6 | 19.19 | 0.12±0.03 | 0.17±0.04 | ≥0.15 | ≥60 | 0.965 | 0.000 | 0.000 | 0.035 | S | 2.8 | 9.4 |
| 0537 | 2002 TO175 | 2457796.44027 | 0:58 | 14.7 | 18.42 | 0.02±0.08 | 0.10±0.09 | ≥0.09 | ≥59 | 0.358 | 0.319 | 0.100 | 0.222 | - | 2.7 | 12.7 |
| 0538 | 2002 TQ7 | 2457690.33738 | 1:55 | 15.4 | 20.00 | 0.06±0.04 | 0.04±0.05 | ≥0.22 | ≥116 | 0.781 | 0.107 | 0.103 | 0.008 | S | 2.8 | 4.8 |
| 0539 | 2002 TT52 | 2457691.37970 | 1:54 | 16.2 | 18.14 | 0.12±0.02 | 0.14±0.03 | ≥0.15 | ≥114 | 1.000 | 0.000 | 0.000 | 0.001 | S | 2.3 | 0.8 |
| 0540 | 2002 TW30 | 2457691.37970 | 0:59 | 15.7 | 19.37 | -0.08±0.03 | -0.10±0.05 | ≥0.24 | ≥60 | 0.000 | 0.043 | 0.957 | 0.000 | C | 2.8 | 5.4 |
| 0541 | 2002 TZ241 | 2457724.34161 | 0:57 | 16.4 | 20.18 | 0.09±0.10 | -0.10±0.13 | ≥0.33 | ≥58 | 0.339 | 0.049 | 0.600 | 0.011 | C | 2.3 | 5.3 |
| 0542 | 2002 US27 | 2457692.38892 | 1:20 | 18.1 | 19.31 | 0.04±0.02 | 0.03±0.03 | ≥0.22 | ≥80 | 0.698 | 0.093 | 0.210 | 0.000 | S | 2.3 | 2.3 |
| 0543 | 2002 VH18 | 2457690.33738 | 1:57 | 15.8 | 19.66 | 0.15±0.03 | 0.21±0.04 | ≥0.19 | ≥118 | 0.980 | 0.000 | 0.000 | 0.020 | S | 2.8 | 7.0 |
| 0544 | 2002 VH67 | 2457724.34904 | 0:44 | 16.5 | 20.27 | 0.39±0.09 | 0.04±0.12 | ≥0.22 | ≥45 | 0.997 | 0.000 | 0.003 | 0.000 | S | 2.3 | 6.7 |
| 0545 | 2002 VM11 | 2457724.34904 | 0:46 | 17.3 | 19.55 | -0.08±0.06 | -0.12±0.08 | ≥0.25 | ≥47 | 0.003 | 0.077 | 0.920 | 0.000 | C | 2.3 | 2.8 |
| 0546 | 2002 VZ133 | 2457691.46127 | 2:00 | 16.2 | 19.40 | 0.11±0.03 | 0.11±0.04 | ≥0.15 | ≥121 | 0.984 | 0.007 | 0.000 | 0.009 | S | 2.3 | 6.2 |
| 0547 | 2002 XD94 | 2457722.38400 | 1:47 | 16.2 | 19.92 | 0.09±0.03 | 0.12±0.04 | ≥0.20 | ≥107 | 0.903 | 0.033 | 0.001 | 0.062 | S | 2.8 | 5.3 |
| 0548 | 2002 XS21 | 2457802.40209 | 0:32 | 14.9 | 19.24 | 0.08±0.04 | 0.16±0.05 | ≥0.09 | ≥33 | 0.695 | 0.034 | 0.000 | 0.270 | S | 2.7 | 13.5 |
| 0549 | 2002 XW90 | 2457690.42183 | 2:21 | 16.4 | 18.94 | 0.12±0.04 | 0.08±0.05 | ≥0.37 | ≥142 | 0.984 | 0.008 | 0.004 | 0.004 | S | 2.3 | 8.0 |
| 0550 | 2002 XZ80 | 2457692.44645 | 1:00 | 16.7 | 17.94 | -0.08±0.04 | -0.06±0.04 | ≥0.08 | ≥60 | 0.000 | 0.159 | 0.840 | 0.000 | C | 2.3 | 5.9 |

*Table 1 continued on next page*

**Table 1** *(continued)*

| No.[a] | Object | Obs. Start (JD) | Obs. Duration (h:mm) | H[b] (mag) | V (mag) | V-R[c] (mag) | V-I[c] (mag) | Amplitude[d] (mag) | Rot. Period[d] (min) | S-type (probability) | X-type (probability) | C-type (probability) | D-type (probability) | Tax. | a[b] (au) | i[b] (deg) |
|---|---|---|---|---|---|---|---|---|---|---|---|---|---|---|---|---|
| 0551 | 2002 YQ29 | 2457690.33738 | 1:57 | 16.9 | 20.14 | 0.04±0.05 | 0.15±0.06 | ≥0.30 | ≥118 | 0.375 | 0.177 | 0.005 | 0.442 | - | 2.4 | 3.3 |
| 0552 | 2003 AR14 | 2457724.38316 | 1:08 | 16.5 | 19.42 | -0.10±0.04 | -0.13±0.04 | ≥0.12 | ≥69 | 0.000 | 0.001 | 0.999 | 0.000 | C | 2.4 | 7.4 |
| 0553 | 2003 AT36 | 2457692.44645 | 1:00 | 17.5 | 19.52 | 0.03±0.04 | 0.03±0.05 | ≥0.13 | ≥60 | 0.396 | 0.307 | 0.281 | 0.016 | - | 2.3 | 3.8 |
| 0554 | 2003 BC51 | 2457692.44645 | 1:00 | 16.9 | 19.74 | 0.06±0.05 | 0.11±0.06 | ≥0.43 | ≥60 | 0.625 | 0.171 | 0.028 | 0.177 | S | 2.4 | 1.5 |
| 0555 | 2003 BX88 | 2457690.33738 | 1:57 | 14.9 | 19.90 | 0.04±0.04 | 0.09±0.05 | ≥0.27 | ≥118 | 0.531 | 0.339 | 0.039 | 0.090 | S | 3.0 | 12.8 |
| 0556 | 2003 DO7 | 2457690.33738 | 1:57 | 16.7 | 20.69 | 0.11±0.06 | 0.12±0.08 | ≥0.45 | ≥118 | 0.814 | 0.057 | 0.022 | 0.107 | S | 2.4 | 3.5 |
| 0557 | 2003 DV3 | 2457724.38316 | 1:08 | 17.4 | 19.59 | 0.06±0.03 | 0.03±0.04 | 0.19±0.06 | 108±26 | 0.769 | 0.095 | 0.134 | 0.002 | S | 2.4 | 1.3 |
| 0558 | 2003 EQ54 | 2457692.44645 | 1:00 | 17.0 | 20.36 | 0.01±0.07 | 0.06±0.08 | ≥0.15 | ≥60 | 0.290 | 0.378 | 0.206 | 0.126 | - | 2.4 | 1.0 |
| 0559 | 2003 FH45 | 2457722.38400 | 1:47 | 16.7 | 20.66 | 0.06±0.07 | 0.01±0.09 | ≥0.39 | ≥107 | 0.472 | 0.141 | 0.351 | 0.035 | - | 2.4 | 5.1 |
| 0560 | 2003 FK131 | 2457691.37970 | 1:54 | 16.3 | 19.71 | -0.03±0.03 | -0.05±0.04 | ≥0.28 | ≥114 | 0.005 | 0.136 | 0.859 | 0.000 | C | 2.4 | 4.7 |
| 0561 | 2003 FO127 | 2457691.41976 | 0:56 | 16.3 | 19.73 | 0.08±0.05 | 0.17±0.06 | ≥0.24 | ≥57 | 0.635 | 0.041 | 0.001 | 0.323 | S | 2.5 | 4.0 |
| 0562 | 2003 FP74 | 2457691.46127 | 0:57 | 15.4 | 19.66 | 0.08±0.06 | 0.11±0.07 | ≥0.23 | ≥57 | 0.735 | 0.103 | 0.025 | 0.137 | S | 2.5 | 7.3 |
| 0563 | 2003 FW104 | 2457730.42038 | 1:30 | 17.3 | 20.22 | 0.18±0.05 | 0.09±0.07 | ≥0.29 | ≥90 | 0.991 | 0.001 | 0.005 | 0.003 | S | 2.4 | 6.5 |
| 0564 | 2003 FX104 | 2457690.38513 | 0:48 | 16.5 | 20.14 | 0.09±0.09 | 0.14±0.11 | ≥0.22 | ≥49 | 0.645 | 0.098 | 0.057 | 0.199 | S | 2.4 | 3.3 |
| 0565 | 2003 FY41 | 2457691.38271 | 1:50 | 16.5 | 20.82 | 0.08±0.10 | 0.05±0.10 | ≥0.70 | ≥110 | 0.564 | 0.159 | 0.198 | 0.079 | S | 2.5 | 2.9 |
| 0566 | 2003 GA23 | 2457724.35824 | 0:33 | 14.6 | 19.85 | 0.17±0.10 | -0.04±0.11 | ≥0.40 | ≥34 | 0.710 | 0.021 | 0.262 | 0.006 | S | 3.1 | 6.6 |
| 0567 | 2003 HC45 | 2457692.38892 | 1:20 | 16.2 | 19.46 | 0.04±0.02 | 0.04±0.03 | ≥0.25 | ≥80 | 0.663 | 0.200 | 0.136 | 0.001 | S | 2.5 | 1.6 |
| 0568 | 2003 HN47 | 2457692.38892 | 1:20 | 15.4 | 20.46 | 0.06±0.04 | 0.08±0.05 | ≥0.86 | ≥80 | 0.701 | 0.191 | 0.058 | 0.050 | S | 3.1 | 3.4 |
| 0569 | 2003 HN51 | 2457690.42183 | 2:19 | 15.0 | 19.69 | 0.08±0.04 | 0.22±0.04 | 0.76±0.10 | 279±49 | 0.560 | 0.003 | 0.000 | 0.438 | S | 3.0 | 9.9 |
| 0570 | 2003 HS55 | 2457692.44645 | 1:00 | 16.0 | 20.01 | 0.03±0.06 | 0.06±0.08 | ≥0.12 | ≥60 | 0.381 | 0.282 | 0.219 | 0.118 | - | 2.5 | 1.9 |
| 0571 | 2003 HT39 | 2457690.33738 | 1:57 | 16.8 | 20.17 | 0.01±0.05 | 0.03±0.07 | ≥0.20 | ≥118 | 0.268 | 0.384 | 0.308 | 0.040 | - | 2.5 | 2.9 |
| 0572 | 2003 HW13 | 2457722.38400 | 1:47 | 15.2 | 20.03 | -0.01±0.04 | 0.06±0.05 | ≥0.28 | ≥107 | 0.128 | 0.693 | 0.151 | 0.028 | X | 3.0 | 11.9 |
| 0573 | 2003 HY57 | 2457691.46411 | 1:56 | 15.5 | 20.29 | 0.00±0.06 | 0.09±0.07 | 0.81±0.13 | 214±49 | 0.269 | 0.436 | 0.114 | 0.181 | - | 3.1 | 11.0 |
| 0574 | 2003 JC5 | 2457690.33738 | 1:57 | 17.1 | 20.61 | 0.05±0.06 | 0.06±0.08 | ≥0.37 | ≥118 | 0.563 | 0.198 | 0.158 | 0.081 | S | 2.6 | 4.2 |
| 0575 | 2003 JX10 | 2457690.33738 | 1:01 | 15.7 | 20.48 | -0.04±0.07 | 0.02±0.08 | ≥0.26 | ≥62 | 0.098 | 0.488 | 0.377 | 0.037 | - | 3.2 | 16.6 |
| 0576 | 2003 KL32 | 2457691.46127 | 2:00 | 15.3 | 19.69 | -0.04±0.04 | -0.13±0.05 | ≥0.21 | ≥121 | 0.000 | 0.010 | 0.990 | 0.000 | C | 3.1 | 13.3 |
| 0577 | 2003 MW5 | 2457691.54720 | 0:52 | 16.1 | 19.65 | 0.00±0.06 | -0.00±0.12 | ≥0.13 | ≥52 | 0.154 | 0.260 | 0.486 | 0.100 | - | 2.6 | 15.0 |
| 0578 | 2003 NF13 | 2457690.42183 | 1:37 | 17.2 | 19.72 | 0.09±0.04 | 0.11±0.06 | ≥0.23 | ≥97 | 0.846 | 0.062 | 0.012 | 0.079 | S | 2.6 | 6.6 |
| 0579 | 2003 QE96 | 2457691.54720 | 1:27 | 16.1 | 19.51 | 0.04±0.04 | 0.06±0.06 | ≥0.13 | ≥87 | 0.582 | 0.245 | 0.117 | 0.056 | S | 2.6 | 14.4 |
| 0580 | 2003 QF115 | 2457691.38580 | 1:45 | 17.6 | 20.40 | 0.07±0.05 | 0.07±0.08 | ≥0.57 | ≥106 | 0.639 | 0.133 | 0.137 | 0.091 | S | 2.6 | 12.9 |
| 0581 | 2003 QT69 | 2457690.33738 | 1:43 | 16.9 | 19.38 | 0.08±0.04 | 0.13±0.04 | ≥1.23 | ≥103 | 0.789 | 0.059 | 0.001 | 0.152 | S | 2.6 | 4.1 |
| 0582 | 2003 SA160 | 2457691.38961 | 1:40 | 15.6 | 19.55 | 0.06±0.03 | 0.01±0.04 | ≥0.12 | ≥100 | 0.703 | 0.045 | 0.252 | 0.000 | S | 2.6 | 14.6 |
| 0583 | 2003 SM70 | 2457691.46127 | 2:00 | 16.8 | 20.04 | 0.13±0.05 | 0.22±0.06 | ≥0.33 | ≥121 | 0.826 | 0.001 | 0.000 | 0.173 | S | 2.6 | 5.6 |
| 0584 | 2003 SO237 | 2457690.33738 | 1:57 | 16.1 | 19.49 | 0.05±0.03 | 0.07±0.04 | ≥0.21 | ≥118 | 0.740 | 0.208 | 0.028 | 0.024 | S | 2.7 | 4.4 |
| 0585 | 2003 SO264 | 2457722.38400 | 1:47 | 16.6 | 19.25 | 0.08±0.02 | 0.07±0.03 | ≥0.14 | ≥107 | 0.955 | 0.038 | 0.004 | 0.003 | S | 2.6 | 5.2 |
| 0586 | 2003 SO9 | 2457722.38685 | 1:43 | 17.3 | 20.11 | 0.00±0.04 | -0.02±0.07 | ≥0.23 | ≥103 | 0.134 | 0.245 | 0.610 | 0.011 | C | 2.6 | 5.6 |
| 0587 | 2003 SQ91 | 2457690.33738 | 1:57 | 17.2 | 20.16 | 0.02±0.07 | 0.07±0.07 | ≥0.68 | ≥118 | 0.395 | 0.387 | 0.111 | 0.107 | - | 2.6 | 11.1 |
| 0588 | 2003 SR165 | 2457690.42183 | 2:19 | 15.9 | 19.14 | -0.00±0.03 | 0.01±0.05 | ≥0.30 | ≥139 | 0.130 | 0.394 | 0.473 | 0.002 | - | 2.6 | 7.3 |
| 0589 | 2003 UD101 | 2457690.33738 | 1:57 | 16.9 | 19.87 | -0.08±0.04 | -0.03±0.05 | 1.00±0.07 | 235±34 | 0.002 | 0.403 | 0.595 | 0.000 | C | 2.7 | 3.0 |
| 0590 | 2003 UD260 | 2457800.44661 | 0:54 | 15.5 | 19.07 | -0.06±0.09 | -0.02±0.17 | ≥0.41 | ≥54 | 0.073 | 0.270 | 0.531 | 0.126 | C | 2.6 | 10.9 |
| 0591 | 2003 UH126 | 2457691.54720 | 1:27 | 15.2 | 18.86 | 0.02±0.03 | 0.08±0.05 | ≥0.11 | ≥87 | 0.300 | 0.522 | 0.074 | 0.103 | X | 2.7 | 14.4 |
| 0592 | 2003 UH90 | 2457724.38316 | 1:08 | 17.3 | 19.89 | -0.03±0.05 | -0.01±0.06 | ≥0.26 | ≥69 | 0.057 | 0.424 | 0.517 | 0.002 | C | 2.6 | 5.9 |
| 0593 | 2003 UJ245 | 2457724.38316 | 1:08 | 16.2 | 19.61 | 0.04±0.03 | 0.08±0.04 | ≥0.10 | ≥69 | 0.606 | 0.324 | 0.034 | 0.037 | S | 2.6 | 1.6 |
| 0594 | 2003 UN22 | 2457690.33738 | 1:57 | 15.4 | 19.74 | 0.10±0.04 | 0.12±0.04 | ≥1.13 | ≥118 | 0.930 | 0.023 | 0.001 | 0.047 | S | 2.7 | 15.3 |
| 0595 | 2003 UT39 | 2457692.44645 | 1:00 | 16.2 | 19.13 | 0.04±0.04 | 0.17±0.05 | ≥0.19 | ≥60 | 0.383 | 0.093 | 0.000 | 0.524 | D | 2.7 | 4.0 |
| 0596 | 2003 UV64 | 2457690.42183 | 2:21 | 16.7 | 19.25 | 0.07±0.03 | 0.07±0.04 | ≥0.29 | ≥142 | 0.856 | 0.104 | 0.028 | 0.012 | S | 2.6 | 14.2 |

*Table 1 continued on next page*

**Table 1** *(continued)*

| No.[a] | Object | Obs. Start (JD) | Obs. Duration (h:mm) | H[b] (mag) | V (mag) | V-R[c] (mag) | V-I[c] (mag) | Amplitude[d] (mag) | Rot. Period[d] (min) | S-type (probability) | X-type | C-type | D-type | Tax. | a[b] (au) | i[b] (deg) |
|---|---|---|---|---|---|---|---|---|---|---|---|---|---|---|---|---|
| 0597 | 2003 UZ251 | 2457692.38892 | 1:20 | 16.3 | 19.08 | 0.04±0.02 | 0.08±0.03 | ≥0.31 | ≥80 | 0.612 | 0.365 | 0.013 | 0.011 | S | 2.7 | 28.1 |
| 0598 | 2003 UZ3 | 2457692.38892 | 1:20 | 17.5 | 20.71 | 0.06±0.05 | 0.09±0.07 | ≥0.21 | ≥80 | 0.621 | 0.171 | 0.082 | 0.126 | S | 2.6 | 4.5 |
| 0599 | 2003 UZ82 | 2457765.28694 | 1:01 | 15.5 | 19.80 | -0.04±0.07 | -0.01±0.10 | ≥0.22 | ≥62 | 0.084 | 0.353 | 0.530 | 0.033 | C | 2.7 | 11.7 |
| 0600 | 2003 VV5 | 2457692.44645 | 1:00 | 16.8 | 19.82 | 0.10±0.05 | 0.23±0.06 | ≥0.12 | ≥60 | 0.703 | 0.004 | 0.000 | 0.294 | S | 2.7 | 3.8 |
| 0601 | 2003 WQ149 | 2457691.37970 | 1:54 | 16.4 | 19.98 | 0.11±0.04 | 0.22±0.05 | ≥0.13 | ≥114 | 0.823 | 0.000 | 0.000 | 0.176 | S | 2.7 | 5.9 |
| 0602 | 2003 WX101 | 2457692.44645 | 1:00 | 16.1 | 19.66 | 0.05±0.05 | 0.17±0.06 | ≥0.08 | ≥60 | 0.430 | 0.102 | 0.002 | 0.466 | - | 2.7 | 2.3 |
| 0603 | 2003 WY115 | 2457692.38892 | 1:20 | 16.7 | 19.17 | 0.04±0.03 | 0.06±0.03 | ≥0.13 | ≥80 | 0.594 | 0.338 | 0.062 | 0.006 | S | 2.7 | 8.4 |
| 0604 | 2003 XU21 | 2457724.28943 | 1:08 | 15.5 | 19.37 | 0.07±0.04 | 0.11±0.05 | ≥0.40 | ≥68 | 0.781 | 0.095 | 0.010 | 0.114 | S | 2.7 | 8.5 |
| 0605 | 2003 YC138 | 2457690.42183 | 2:19 | 16.9 | 20.29 | 0.14±0.09 | 0.05±0.07 | ≥0.32 | ≥139 | 0.836 | 0.066 | 0.078 | 0.020 | S | 2.2 | 7.6 |
| 0606 | 2003 YE28 | 2457690.42183 | 2:21 | 16.0 | 19.69 | 0.09±0.04 | 0.13±0.05 | ≥0.25 | ≥142 | 0.842 | 0.043 | 0.001 | 0.114 | S | 2.7 | 14.2 |
| 0607 | 2003 YK61 | 2457802.50174 | 0:55 | 16.0 | 19.66 | 0.03±0.06 | -0.03±0.07 | ≥0.21 | ≥55 | 0.252 | 0.153 | 0.587 | 0.008 | C | 2.6 | 5.3 |
| 0608 | 2003 YL137 | 2457691.47338 | 1:43 | 16.3 | 20.09 | 0.07±0.04 | 0.19±0.06 | ≥0.25 | ≥103 | 0.529 | 0.037 | 0.001 | 0.433 | S | 2.7 | 7.6 |
| 0609 | 2003 YL26 | 2457802.39604 | 0:41 | 16.8 | 19.84 | 0.01±0.05 | -0.02±0.06 | ≥0.12 | ≥41 | 0.174 | 0.192 | 0.629 | 0.004 | C | 2.6 | 11.9 |
| 0610 | 2003 YQ12 | 2457802.50174 | 0:55 | 17.1 | 20.30 | 0.07±0.11 | 0.02±0.12 | ≥0.29 | ≥55 | 0.490 | 0.156 | 0.284 | 0.069 | - | 2.6 | 5.4 |
| 0611 | 2003 YY83 | 2457691.37970 | 1:54 | 16.3 | 19.90 | 0.08±0.04 | 0.14±0.05 | ≥0.24 | ≥114 | 0.817 | 0.037 | 0.000 | 0.146 | S | 2.2 | 3.8 |
| 0612 | 2004 BA26 | 2457690.33738 | 1:57 | 16.5 | 19.71 | 0.08±0.04 | 0.09±0.04 | ≥0.51 | ≥118 | 0.864 | 0.102 | 0.005 | 0.029 | S | 2.8 | 9.3 |
| 0613 | 2004 BM46 | 2457724.34608 | 0:51 | 15.7 | 20.17 | 0.28±0.12 | 0.20±0.17 | ≥0.29 | ≥51 | 0.953 | 0.005 | 0.017 | 0.025 | S | 2.8 | 3.4 |
| 0614 | 2004 BU96 | 2457730.42927 | 1:17 | 16.1 | 20.28 | 0.18±0.07 | 0.19±0.07 | ≥0.22 | ≥77 | 0.955 | 0.003 | 0.000 | 0.042 | S | 2.7 | 8.5 |
| 0615 | 2004 BV83 | 2457798.46669 | 0:53 | 15.0 | 18.28 | 0.09±0.05 | 0.16±0.07 | ≥0.11 | ≥53 | 0.723 | 0.049 | 0.006 | 0.222 | S | 2.7 | 13.0 |
| 0616 | 2004 BY117 | 2457691.46127 | 2:00 | 15.8 | 19.25 | -0.02±0.03 | 0.04±0.04 | ≥0.52 | ≥121 | 0.048 | 0.741 | 0.210 | 0.002 | X | 2.7 | 7.1 |
| 0617 | 2004 BY42 | 2457730.42038 | 1:30 | 16.9 | 19.46 | 0.07±0.04 | -0.02±0.04 | ≥0.30 | ≥90 | 0.552 | 0.015 | 0.433 | 0.000 | S | 2.2 | 6.5 |
| 0618 | 2004 BZ1 | 2457722.38400 | 1:47 | 15.9 | 19.03 | -0.06±0.02 | -0.08±0.03 | ≥0.09 | ≥107 | 0.000 | 0.042 | 0.958 | 0.000 | C | 2.7 | 6.4 |
| 0619 | 2004 CB76 | 2457691.46127 | 2:00 | 16.3 | 20.75 | 0.13±0.09 | 0.04±0.13 | ≥0.41 | ≥121 | 0.726 | 0.052 | 0.171 | 0.051 | S | 2.8 | 5.9 |
| 0620 | 2004 CW57 | 2457722.38400 | 1:47 | 16.2 | 19.27 | 0.02±0.02 | 0.06±0.03 | ≥0.20 | ≥107 | 0.373 | 0.499 | 0.123 | 0.005 | - | 2.7 | 9.0 |
| 0621 | 2004 CZ26 | 2457722.38400 | 1:47 | 16.2 | 19.70 | 0.07±0.03 | 0.08±0.04 | ≥0.16 | ≥107 | 0.897 | 0.083 | 0.009 | 0.011 | S | 2.8 | 4.9 |
| 0622 | 2004 EA50 | 2457691.37970 | 1:54 | 16.7 | 20.00 | -0.05±0.04 | -0.06±0.05 | ≥0.24 | ≥114 | 0.005 | 0.167 | 0.828 | 0.000 | C | 2.3 | 2.3 |
| 0623 | 2004 EC22 | 2457692.39079 | 1:17 | 16.0 | 20.01 | 0.04±0.03 | 0.06±0.04 | ≥0.27 | ≥77 | 0.599 | 0.308 | 0.075 | 0.018 | S | 2.9 | 1.9 |
| 0624 | 2004 ER79 | 2457801.38674 | 1:43 | 15.4 | 19.03 | 0.14±0.05 | 0.16±0.06 | ≥0.13 | ≥104 | 0.931 | 0.007 | 0.000 | 0.062 | S | 2.7 | 13.7 |
| 0625 | 2004 EV42 | 2457692.38892 | 1:20 | 16.8 | 20.30 | 0.07±0.04 | 0.10±0.05 | ≥1.08 | ≥80 | 0.800 | 0.119 | 0.012 | 0.069 | S | 2.4 | 3.2 |
| 0626 | 2004 FB118 | 2457690.33738 | 1:57 | 15.8 | 20.38 | 0.08±0.05 | 0.06±0.07 | ≥0.39 | ≥118 | 0.715 | 0.107 | 0.120 | 0.058 | S | 2.9 | 3.2 |
| 0627 | 2004 FB142 | 2457692.38892 | 1:02 | 16.6 | 19.10 | 0.08±0.03 | -0.10±0.03 | ≥0.66 | ≥63 | 0.043 | 0.000 | 0.957 | 0.000 | C | 2.3 | 4.5 |
| 0628 | 2004 FR43 | 2457724.34310 | 0:55 | 16.4 | 19.91 | 0.21±0.07 | 0.13±0.13 | ≥0.49 | ≥55 | 0.968 | 0.002 | 0.018 | 0.012 | S | 2.3 | 3.9 |
| 0629 | 2004 GA8 | 2457692.38892 | 1:20 | 17.3 | 20.38 | -0.08±0.05 | -0.04±0.05 | ≥0.17 | ≥80 | 0.004 | 0.308 | 0.689 | 0.000 | C | 2.3 | 4.9 |
| 0630 | 2004 GD61 | 2457690.33738 | 1:26 | 15.3 | 19.97 | -0.05±0.04 | -0.00±0.05 | ≥0.23 | ≥87 | 0.009 | 0.542 | 0.448 | 0.001 | X | 3.0 | 3.1 |
| 0631 | 2004 GY78 | 2457692.44645 | 1:00 | 17.5 | 20.24 | 0.07±0.06 | 0.16±0.07 | ≥0.16 | ≥60 | 0.587 | 0.093 | 0.009 | 0.311 | S | 2.3 | 2.1 |
| 0632 | 2004 JV32 | 2457722.38400 | 1:47 | 17.0 | 20.67 | 0.06±0.06 | -0.00±0.08 | ≥0.32 | ≥107 | 0.463 | 0.122 | 0.392 | 0.023 | - | 2.3 | 4.8 |
| 0633 | 2004 KA10 | 2457798.41862 | 1:06 | 15.1 | 19.16 | 0.09±0.07 | 0.20±0.09 | ≥0.24 | ≥67 | 0.604 | 0.053 | 0.006 | 0.337 | S | 2.7 | 16.4 |
| 0634 | 2004 KA4 | 2457724.39556 | 0:50 | 15.0 | 19.48 | 0.08±0.04 | 0.06±0.05 | ≥0.36 | ≥51 | 0.854 | 0.079 | 0.057 | 0.011 | S | 3.0 | 12.3 |
| 0635 | 2004 KR | 2457802.39604 | 0:41 | 16.4 | 19.48 | 0.06±0.04 | 0.03±0.04 | ≥0.14 | ≥41 | 0.704 | 0.119 | 0.173 | 0.004 | S | 2.3 | 7.3 |
| 0636 | 2004 LK29 | 2457692.44645 | 1:00 | 16.7 | 19.89 | -0.06±0.04 | -0.10±0.06 | ≥0.14 | ≥60 | 0.002 | 0.078 | 0.920 | 0.000 | C | 2.4 | 3.4 |
| 0637 | 2004 NA4 | 2457730.42038 | 1:30 | 15.1 | 19.98 | -0.00±0.11 | 0.11±0.08 | ≥0.24 | ≥90 | 0.344 | 0.388 | 0.064 | 0.203 | - | 3.1 | 15.4 |
| 0638 | 2004 NB1 | 2457691.46127 | 2:00 | 16.5 | 20.42 | 0.12±0.05 | 0.12±0.07 | ≥0.30 | ≥121 | 0.900 | 0.024 | 0.009 | 0.067 | S | 2.4 | 6.6 |
| 0639 | 2004 NN33 | 2457722.38400 | 1:47 | 15.1 | 19.68 | -0.09±0.03 | -0.09±0.04 | ≥0.15 | ≥107 | 0.000 | 0.050 | 0.950 | 0.000 | C | 3.1 | 17.6 |
| 0640 | 2004 OG13 | 2457730.42038 | 1:30 | 16.4 | 19.51 | 0.16±0.05 | 0.26±0.06 | 0.97±0.08 | 180±36 | 0.950 | 0.000 | 0.000 | 0.050 | S | 2.4 | 6.1 |
| 0641 | 2004 PN30 | 2457691.50542 | 0:55 | 16.6 | 19.45 | 0.05±0.13 | 0.06±0.07 | ≥0.14 | ≥55 | 0.513 | 0.316 | 0.133 | 0.038 | S | 2.4 | 6.2 |
| 0642 | 2004 PR94 | 2457691.46127 | 2:00 | 15.1 | 20.03 | -0.11±0.09 | 0.04±0.07 | ≥0.33 | ≥121 | 0.054 | 0.622 | 0.291 | 0.033 | X | 3.2 | 6.7 |

*Table 1 continued on next page*

**Table 1** *(continued)*

| No.[a] | Object | Obs. Start | Obs. Duration | H[b] | V | V-R[c] | V-I[c] | Amplitude[d] | Rot. Period[d] | S-type | X-type | C-type | D-type | Tax. | a[b] | i[b] |
|---|---|---|---|---|---|---|---|---|---|---|---|---|---|---|---|---|
| | | (JD) | (h:mm) | (mag) | (mag) | (mag) | (mag) | (mag) | (min) | (probability) | | | | | (au) | (deg) |
| 0643 | 2004 PT18 | 2457692.38892 | 1:20 | 17.5 | 20.69 | 0.08±0.06 | 0.15±0.06 | ≥0.37 | ≥80 | 0.665 | 0.082 | 0.006 | 0.247 | S | 2.5 | 2.2 |
| 0644 | 2004 PU111 | 2457691.47191 | 1:45 | 15.5 | 19.97 | -0.01±0.05 | 0.10±0.06 | 0.70±0.10 | 155±34 | 0.188 | 0.557 | 0.040 | 0.215 | X | 3.2 | 16.3 |
| 0645 | 2004 QG8 | 2457692.39380 | 1:12 | 15.0 | 19.20 | -0.02±0.03 | 0.01±0.03 | ≥0.08 | ≥73 | 0.044 | 0.519 | 0.437 | 0.000 | X | 3.2 | 15.6 |
| 0646 | 2004 RL137 | 2457691.37970 | 0:50 | 15.7 | 20.40 | -0.11±0.09 | -0.13±0.13 | ≥0.17 | ≥51 | 0.015 | 0.132 | 0.844 | 0.009 | C | 3.2 | 13.6 |
| 0647 | 2004 RP94 | 2457690.33738 | 1:57 | 15.5 | 19.84 | 0.08±0.04 | 0.11±0.09 | ≥0.30 | ≥118 | 0.724 | 0.055 | 0.063 | 0.158 | S | 3.2 | 12.2 |
| 0648 | 2004 RY59 | 2457691.42249 | 0:52 | 15.7 | 20.22 | 0.06±0.10 | 0.03±0.17 | ≥0.20 | ≥53 | 0.401 | 0.127 | 0.337 | 0.135 | - | 3.2 | 4.1 |
| 0649 | 2004 TC344 | 2457730.42354 | 1:25 | 15.0 | 19.71 | -0.00±0.06 | -0.04±0.08 | ≥0.19 | ≥86 | 0.121 | 0.222 | 0.646 | 0.011 | C | 3.1 | 9.9 |
| 0650 | 2004 TC60 | 2457690.42183 | 2:19 | 15.5 | 20.05 | -0.02±0.08 | 0.01±0.08 | ≥0.18 | ≥139 | 0.175 | 0.366 | 0.427 | 0.032 | - | 3.2 | 15.7 |
| 0651 | 2004 TF333 | 2457692.38892 | 1:20 | 16.4 | 19.51 | 0.02±0.03 | 0.02±0.04 | ≥0.19 | ≥80 | 0.361 | 0.237 | 0.402 | 0.000 | - | 2.5 | 8.0 |
| 0652 | 2004 TG55 | 2457690.33738 | 1:55 | 16.5 | 20.00 | 0.12±0.04 | 0.17±0.04 | ≥0.26 | ≥116 | 0.952 | 0.002 | 0.000 | 0.046 | S | 2.5 | 6.8 |
| 0653 | 2004 TQ2 | 2457796.44609 | 0:50 | 17.1 | 19.30 | -0.14±0.14 | -0.12±0.14 | ≥0.40 | ≥51 | 0.034 | 0.139 | 0.813 | 0.015 | C | 2.4 | 9.5 |
| 0654 | 2004 VN62 | 2457765.29937 | 0:43 | 15.6 | 20.12 | 0.11±0.14 | 0.27±0.18 | ≥0.23 | ≥44 | 0.631 | 0.057 | 0.037 | 0.275 | S | 2.5 | 15.5 |
| 0655 | 2004 VS72 | 2457800.44661 | 0:54 | 15.2 | 19.92 | 0.06±0.16 | 0.26±0.18 | ≥0.43 | ≥54 | 0.494 | 0.098 | 0.042 | 0.366 | - | 3.1 | 11.5 |
| 0656 | 2004 XC65 | 2457800.44661 | 0:54 | 15.3 | 19.99 | 0.00±0.15 | 0.15±0.22 | ≥0.26 | ≥54 | 0.328 | 0.165 | 0.219 | 0.288 | - | 3.1 | 12.1 |
| 0657 | 2004 XE30 | 2457691.37970 | 0:59 | 17.0 | 18.98 | -0.07±0.03 | -0.08±0.04 | ≥0.09 | ≥60 | 0.000 | 0.056 | 0.944 | 0.000 | C | 2.5 | 0.9 |
| 0658 | 2004 XK87 | 2457692.44645 | 1:00 | 16.9 | 19.68 | 0.04±0.04 | 0.14±0.06 | ≥0.11 | ≥60 | 0.452 | 0.199 | 0.008 | 0.340 | - | 2.5 | 2.6 |
| 0659 | 2004 XY24 | 2457687.42207 | 2:03 | 16.0 | 19.24 | 0.14±0.02 | 0.18±0.03 | ≥0.21 | ≥123 | 1.000 | 0.000 | 0.000 | 0.000 | S | 2.6 | 14.3 |
| 0660 | 2004 YT16 | 2457724.38316 | 1:08 | 16.7 | 20.35 | 0.08±0.06 | 0.02±0.06 | ≥0.25 | ≥69 | 0.683 | 0.095 | 0.211 | 0.010 | S | 2.6 | 3.2 |
| 0661 | 2005 AW13 | 2457724.34310 | 0:55 | 16.3 | 19.18 | 0.00±0.04 | 0.02±0.06 | ≥0.18 | ≥55 | 0.213 | 0.401 | 0.370 | 0.016 | - | 2.6 | 3.5 |
| 0662 | 2005 BG25 | 2457724.38316 | 1:08 | 16.1 | 19.50 | 0.15±0.05 | 0.10±0.05 | ≥0.36 | ≥69 | 0.977 | 0.011 | 0.002 | 0.010 | S | 2.6 | 5.9 |
| 0663 | 2005 EB163 | 2457692.38892 | 1:20 | 16.6 | 20.15 | -0.01±0.04 | -0.00±0.05 | ≥0.19 | ≥80 | 0.076 | 0.383 | 0.539 | 0.003 | C | 2.7 | 1.9 |
| 0664 | 2005 EG18 | 2457692.45553 | 0:46 | 16.8 | 19.45 | 0.08±0.05 | 0.19±0.06 | ≥0.10 | ≥47 | 0.615 | 0.029 | 0.001 | 0.355 | S | 2.6 | 1.3 |
| 0665 | 2005 EL167 | 2457692.38892 | 1:20 | 17.9 | 20.55 | -0.06±0.05 | -0.03±0.06 | ≥0.14 | ≥80 | 0.012 | 0.348 | 0.639 | 0.002 | C | 2.7 | 2.0 |
| 0666 | 2005 ER186 | 2457691.38580 | 1:45 | 16.1 | 20.15 | 0.06±0.04 | 0.09±0.06 | ≥0.39 | ≥106 | 0.694 | 0.150 | 0.046 | 0.110 | S | 2.8 | 3.0 |
| 0667 | 2005 EU | 2457691.54720 | 1:27 | 16.0 | 19.52 | 0.05±0.04 | 0.08±0.05 | ≥0.14 | ≥87 | 0.657 | 0.203 | 0.067 | 0.073 | S | 2.6 | 14.8 |
| 0668 | 2005 EY236 | 2457692.38892 | 1:20 | 16.7 | 20.31 | 0.07±0.04 | 0.08±0.05 | ≥0.59 | ≥80 | 0.783 | 0.132 | 0.034 | 0.051 | S | 2.7 | 3.6 |
| 0669 | 2005 GA133 | 2457690.33738 | 0:59 | 16.4 | 20.66 | 0.11±0.09 | 0.18±0.10 | ≥0.14 | ≥60 | 0.720 | 0.061 | 0.015 | 0.204 | S | 2.8 | 5.7 |
| 0670 | 2005 GR68 | 2457730.42038 | 1:26 | 15.5 | 19.83 | 0.16±0.04 | 0.21±0.05 | ≥0.20 | ≥86 | 0.969 | 0.000 | 0.000 | 0.031 | S | 2.7 | 14.3 |
| 0671 | 2005 GT107 | 2457691.37970 | 1:54 | 16.2 | 20.18 | 0.01±0.04 | 0.04±0.07 | ≥0.68 | ≥114 | 0.236 | 0.398 | 0.307 | 0.059 | - | 2.8 | 5.4 |
| 0672 | 2005 GU74 | 2457690.33738 | 1:05 | 16.7 | 19.92 | 0.10±0.04 | 0.18±0.04 | ≥0.21 | ≥66 | 0.796 | 0.006 | 0.000 | 0.199 | S | 2.2 | 4.6 |
| 0673 | 2005 GY22 | 2457692.44645 | 1:00 | 15.5 | 20.18 | 0.08±0.05 | 0.17±0.07 | ≥0.17 | ≥60 | 0.670 | 0.046 | 0.003 | 0.281 | S | 2.7 | 13.8 |
| 0674 | 2005 LH31 | 2457802.39604 | 0:41 | 16.6 | 20.37 | 0.05±0.07 | 0.12±0.08 | ≥0.24 | ≥41 | 0.507 | 0.196 | 0.048 | 0.250 | S | 2.6 | 12.1 |
| 0675 | 2005 MN35 | 2457730.42038 | 1:30 | 17.2 | 19.62 | 0.12±0.03 | 0.13±0.04 | ≥0.11 | ≥90 | 0.985 | 0.002 | 0.000 | 0.014 | S | 2.3 | 9.9 |
| 0676 | 2005 NW94 | 2457692.44645 | 1:00 | 17.6 | 19.21 | 0.04±0.04 | -0.04±0.10 | ≥0.16 | ≥60 | 0.276 | 0.095 | 0.604 | 0.026 | C | 2.3 | 2.3 |
| 0677 | 2005 OT27 | 2457724.38316 | 1:08 | 17.3 | 19.79 | -0.03±0.04 | -0.09±0.06 | ≥0.13 | ≥69 | 0.005 | 0.077 | 0.918 | 0.000 | C | 2.3 | 3.1 |
| 0678 | 2005 QK79 | 2457691.46735 | 1:51 | 15.5 | 20.06 | 0.04±0.05 | 0.09±0.06 | ≥0.28 | ≥112 | 0.483 | 0.306 | 0.057 | 0.154 | - | 3.0 | 9.2 |
| 0679 | 2005 QL18 | 2457690.37553 | 1:02 | 16.2 | 19.74 | -0.10±0.07 | 0.01±0.11 | ≥0.66 | ≥63 | 0.027 | 0.460 | 0.449 | 0.064 | - | 3.1 | 2.1 |
| 0680 | 2005 QM37 | 2457691.46127 | 1:03 | 17.0 | 19.23 | 0.07±0.05 | 0.10±0.05 | ≥0.11 | ≥64 | 0.766 | 0.116 | 0.020 | 0.099 | S | 2.4 | 5.9 |
| 0681 | 2005 QT17 | 2457691.37970 | 1:54 | 17.1 | 19.48 | 0.05±0.03 | 0.04±0.03 | ≥0.39 | ≥114 | 0.743 | 0.127 | 0.130 | 0.000 | S | 2.4 | 3.3 |
| 0682 | 2005 RL21 | 2457724.34310 | 0:53 | 15.7 | 20.20 | 0.18±0.10 | 0.04±0.12 | ≥0.68 | ≥53 | 0.847 | 0.022 | 0.109 | 0.022 | S | 3.0 | 11.5 |
| 0683 | 2005 RZ40 | 2457692.38892 | 1:20 | 17.0 | 20.90 | 0.06±0.07 | 0.11±0.09 | ≥0.69 | ≥80 | 0.542 | 0.180 | 0.060 | 0.219 | S | 2.3 | 3.4 |
| 0684 | 2005 SB23 | 2457692.38892 | 1:18 | 18.0 | 20.61 | 0.07±0.05 | 0.18±0.06 | ≥0.26 | ≥78 | 0.581 | 0.043 | 0.002 | 0.375 | S | 2.4 | 3.0 |
| 0685 | 2005 SB85 | 2457724.34310 | 0:55 | 16.1 | 19.98 | 0.14±0.06 | 0.01±0.13 | ≥0.66 | ≥55 | 0.717 | 0.014 | 0.244 | 0.025 | S | 3.1 | 1.0 |
| 0686 | 2005 SC50 | 2457692.38892 | 1:20 | 17.2 | 20.40 | 0.03±0.05 | 0.02±0.07 | ≥0.20 | ≥80 | 0.411 | 0.237 | 0.330 | 0.022 | - | 2.4 | 2.4 |
| 0687 | 2005 SC97 | 2457690.33738 | 1:57 | 15.8 | 19.60 | -0.06±0.03 | -0.09±0.04 | ≥0.26 | ≥118 | 0.000 | 0.027 | 0.973 | 0.000 | C | 3.1 | 2.8 |
| 0688 | 2005 SE77 | 2457692.38892 | 1:20 | 16.2 | 20.53 | -0.10±0.05 | -0.06±0.06 | ≥0.23 | ≥80 | 0.002 | 0.211 | 0.787 | 0.000 | C | 3.1 | 1.3 |

*Table 1 continued on next page*

**Table 1** *(continued)*

| No.[a] | Object | Obs. Start (JD) | Obs. Duration (h:mm) | H[b] (mag) | V (mag) | V-R[c] (mag) | V-I[c] (mag) | Amplitude[d] (mag) | Rot. Period[d] (min) | S-type (probability) | X-type | C-type | D-type | Tax. | a[b] (au) | i[b] (deg) |
|---|---|---|---|---|---|---|---|---|---|---|---|---|---|---|---|---|
| 0689 | 2005 SF252 | 2457692.38892 | 1:20 | 18.1 | 19.95 | 0.08±0.04 | 0.12±0.05 | ≥0.14 | ≥80 | 0.819 | 0.068 | 0.003 | 0.111 | S | 2.4 | 1.9 |
| 0690 | 2005 SG264 | 2457722.38400 | 1:47 | 17.0 | 19.22 | -0.06±0.03 | -0.08±0.04 | ≥0.13 | ≥107 | 0.000 | 0.034 | 0.966 | 0.000 | C | 2.3 | 2.9 |
| 0691 | 2005 SL55 | 2457691.37970 | 1:54 | 16.2 | 20.01 | -0.11±0.04 | -0.14±0.05 | ≥0.92 | ≥114 | 0.000 | 0.004 | 0.996 | 0.000 | C | 3.1 | 1.2 |
| 0692 | 2005 SN286 | 2457691.46127 | 2:00 | 16.9 | 20.80 | -0.13±0.14 | -0.06±0.14 | ≥0.37 | ≥121 | 0.058 | 0.240 | 0.661 | 0.041 | C | 3.1 | 6.6 |
| 0693 | 2005 SO155 | 2457691.58506 | 0:32 | 15.0 | 19.57 | 0.06±0.06 | 0.14±0.09 | ≥0.14 | ≥33 | 0.536 | 0.136 | 0.043 | 0.285 | S | 3.0 | 16.6 |
| 0694 | 2005 SQ51 | 2457722.38400 | 1:47 | 18.0 | 20.26 | 0.09±0.06 | 0.10±0.08 | ≥0.25 | ≥107 | 0.785 | 0.075 | 0.044 | 0.096 | S | 2.4 | 5.2 |
| 0695 | 2005 SV190 | 2457690.33738 | 1:03 | 15.9 | 20.09 | 0.10±0.04 | 0.11±0.05 | ≥0.22 | ≥64 | 0.889 | 0.037 | 0.005 | 0.068 | S | 3.1 | 2.5 |
| 0696 | 2005 SW49 | 2457691.37970 | 1:54 | 17.5 | 20.73 | 0.16±0.09 | 0.06±0.17 | ≥0.40 | ≥114 | 0.780 | 0.022 | 0.146 | 0.052 | S | 2.4 | 5.9 |
| 0697 | 2005 SY10 | 2457691.37970 | 1:54 | 15.5 | 20.28 | -0.07±0.08 | 0.02±0.06 | ≥0.30 | ≥114 | 0.088 | 0.541 | 0.358 | 0.012 | X | 3.1 | 15.0 |
| 0698 | 2005 SZ23 | 2457691.37970 | 0:59 | 16.2 | 19.55 | 0.00±0.05 | 0.07±0.05 | ≥0.23 | ≥60 | 0.258 | 0.611 | 0.087 | 0.045 | X | 3.1 | 11.0 |
| 0699 | 2005 SZ53 | 2457724.38316 | 1:08 | 17.7 | 20.21 | -0.07±0.06 | -0.00±0.06 | ≥0.47 | ≥69 | 0.037 | 0.484 | 0.472 | 0.007 | - | 2.4 | 1.3 |
| 0700 | 2005 TC133 | 2457802.42659 | 1:45 | 15.5 | 20.14 | 0.06±0.04 | 0.08±0.04 | ≥0.25 | ≥106 | 0.773 | 0.169 | 0.023 | 0.035 | S | 3.0 | 9.9 |
| 0701 | 2005 TD59 | 2457692.38892 | 1:20 | 16.6 | 20.59 | -0.02±0.05 | 0.02±0.06 | ≥0.24 | ≥80 | 0.128 | 0.494 | 0.356 | 0.021 | - | 3.1 | 12.2 |
| 0702 | 2005 TE44 | 2457691.37970 | 1:54 | 16.5 | 20.01 | -0.08±0.04 | -0.16±0.05 | ≥0.51 | ≥114 | 0.000 | 0.001 | 0.999 | 0.000 | C | 3.1 | 1.3 |
| 0703 | 2005 TH63 | 2457691.37970 | 1:54 | 18.0 | 20.59 | -0.13±0.13 | -0.04±0.11 | ≥0.33 | ≥114 | 0.064 | 0.271 | 0.645 | 0.019 | C | 2.4 | 4.8 |
| 0704 | 2005 TJ36 | 2457691.46127 | 2:00 | 16.1 | 20.39 | 0.09±0.06 | 0.08±0.08 | ≥0.19 | ≥121 | 0.733 | 0.113 | 0.083 | 0.071 | S | 3.0 | 8.3 |
| 0705 | 2005 TN79 | 2457724.38316 | 1:08 | 15.7 | 19.45 | 0.00±0.03 | -0.03±0.04 | ≥0.11 | ≥69 | 0.068 | 0.132 | 0.799 | 0.000 | C | 3.0 | 7.7 |
| 0706 | 2005 TP59 | 2457802.42659 | 1:45 | 17.5 | 19.82 | -0.01±0.03 | -0.03±0.03 | ≥0.22 | ≥106 | 0.031 | 0.143 | 0.826 | 0.000 | C | 2.3 | 5.6 |
| 0707 | 2005 TQ27 | 2457690.42183 | 2:19 | 15.7 | 19.97 | 0.03±0.04 | 0.03±0.05 | ≥0.27 | ≥139 | 0.405 | 0.299 | 0.285 | 0.011 | - | 3.1 | 12.3 |
| 0708 | 2005 TT77 | 2457690.33738 | 1:57 | 15.2 | 19.46 | -0.02±0.03 | -0.02±0.04 | ≥0.14 | ≥118 | 0.017 | 0.305 | 0.678 | 0.000 | C | 3.1 | 17.1 |
| 0709 | 2005 TU98 | 2457724.28943 | 1:08 | 17.0 | 19.87 | 0.12±0.04 | 0.14±0.07 | ≥0.29 | ≥68 | 0.926 | 0.009 | 0.003 | 0.062 | S | 2.4 | 4.0 |
| 0710 | 2005 TV25 | 2457692.38892 | 1:20 | 16.0 | 19.70 | 0.03±0.03 | 0.06±0.04 | ≥0.09 | ≥80 | 0.521 | 0.387 | 0.074 | 0.018 | S | 3.1 | 2.0 |
| 0711 | 2005 TW183 | 2457690.42183 | 2:19 | 17.0 | 19.89 | 0.10±0.05 | -0.00±0.06 | ≥0.20 | ≥139 | 0.743 | 0.026 | 0.231 | 0.000 | S | 2.3 | 6.5 |
| 0712 | 2005 UD14 | 2457722.38400 | 1:47 | 16.2 | 20.50 | 0.03±0.05 | 0.01±0.07 | ≥0.26 | ≥107 | 0.346 | 0.245 | 0.384 | 0.025 | - | 3.1 | 10.5 |
| 0713 | 2005 UD2 | 2457690.33738 | 1:57 | 15.3 | 19.33 | -0.03±0.06 | 0.01±0.04 | ≥0.09 | ≥118 | 0.131 | 0.486 | 0.382 | 0.000 | - | 3.1 | 9.2 |
| 0714 | 2005 UD216 | 2457724.38316 | 1:08 | 15.2 | 19.43 | 0.06±0.03 | 0.08±0.04 | ≥0.20 | ≥69 | 0.738 | 0.197 | 0.023 | 0.042 | S | 3.1 | 3.1 |
| 0715 | 2005 UJ8 | 2457724.38316 | 0:40 | 17.0 | 19.45 | -0.01±0.04 | 0.05±0.05 | ≥0.14 | ≥40 | 0.142 | 0.643 | 0.191 | 0.024 | X | 2.4 | 3.5 |
| 0716 | 2005 UN54 | 2457692.38892 | 1:20 | 16.0 | 19.56 | -0.03±0.03 | -0.02±0.04 | ≥0.12 | ≥80 | 0.014 | 0.282 | 0.703 | 0.000 | C | 3.1 | 13.6 |
| 0717 | 2005 UP402 | 2457690.33738 | 1:57 | 17.6 | 20.34 | 0.08±0.06 | 0.13±0.08 | ≥0.38 | ≥118 | 0.672 | 0.094 | 0.023 | 0.211 | S | 2.4 | 3.3 |
| 0718 | 2005 UQ47 | 2457690.33738 | 1:57 | 17.5 | 19.77 | 0.04±0.04 | 0.04±0.06 | ≥0.26 | ≥118 | 0.495 | 0.214 | 0.262 | 0.028 | - | 2.4 | 4.2 |
| 0719 | 2005 UX279 | 2457722.38400 | 1:47 | 16.6 | 20.54 | 0.07±0.06 | 0.00±0.08 | ≥0.37 | ≥107 | 0.570 | 0.081 | 0.331 | 0.018 | S | 3.1 | 9.4 |
| 0720 | 2005 UX346 | 2457690.33738 | 1:57 | 16.1 | 20.10 | -0.11±0.07 | -0.01±0.06 | ≥0.22 | ≥118 | 0.013 | 0.436 | 0.550 | 0.001 | C | 3.2 | 7.1 |
| 0721 | 2005 UY281 | 2457690.33738 | 1:57 | 17.5 | 20.53 | 0.11±0.06 | 0.04±0.11 | ≥0.42 | ≥118 | 0.741 | 0.030 | 0.191 | 0.038 | S | 2.4 | 2.2 |
| 0722 | 2005 UZ156 | 2457690.33738 | 1:57 | 16.3 | 20.43 | -0.06±0.05 | -0.10±0.07 | ≥1.01 | ≥118 | 0.003 | 0.080 | 0.917 | 0.000 | C | 3.1 | 7.7 |
| 0723 | 2005 UZ40 | 2457690.33738 | 1:20 | 16.3 | 20.77 | 0.19±0.07 | 0.22±0.11 | ≥0.30 | ≥81 | 0.951 | 0.003 | 0.003 | 0.043 | S | 3.2 | 2.8 |
| 0724 | 2005 VC98 | 2457690.33738 | 1:55 | 16.4 | 20.53 | -0.02±0.06 | 0.03±0.08 | ≥0.50 | ≥116 | 0.124 | 0.482 | 0.337 | 0.057 | - | 3.2 | 11.6 |
| 0725 | 2005 VE61 | 2457691.46127 | 1:01 | 16.2 | 20.19 | -0.04±0.07 | 0.01±0.08 | ≥0.18 | ≥61 | 0.114 | 0.440 | 0.411 | 0.035 | - | 3.1 | 5.6 |
| 0726 | 2005 VT40 | 2457692.44645 | 1:00 | 16.1 | 20.24 | 0.05±0.06 | 0.06±0.08 | ≥0.19 | ≥60 | 0.529 | 0.218 | 0.177 | 0.077 | S | 3.2 | 9.1 |
| 0727 | 2005 VU10 | 2457722.38400 | 1:47 | 18.2 | 20.34 | 0.07±0.05 | 0.06±0.07 | ≥0.23 | ≥107 | 0.695 | 0.128 | 0.117 | 0.060 | S | 2.4 | 3.3 |
| 0728 | 2005 WB111 | 2457692.38892 | 1:20 | 17.2 | 20.66 | 0.10±0.05 | 0.05±0.07 | ≥0.19 | ≥80 | 0.837 | 0.046 | 0.095 | 0.022 | S | 2.4 | 0.8 |
| 0729 | 2005 WM73 | 2457690.42183 | 2:17 | 15.4 | 19.32 | 0.13±0.04 | 0.19±0.04 | ≥0.47 | ≥137 | 0.907 | 0.001 | 0.000 | 0.092 | S | 3.2 | 19.0 |
| 0730 | 2005 WO105 | 2457724.38316 | 1:08 | 17.4 | 19.17 | 0.08±0.03 | 0.13±0.04 | ≥0.10 | ≥69 | 0.908 | 0.017 | 0.000 | 0.075 | S | 2.4 | 5.2 |
| 0731 | 2005 WP163 | 2457692.39079 | 1:17 | 16.0 | 20.53 | -0.04±0.07 | 0.07±0.09 | ≥0.31 | ≥77 | 0.123 | 0.518 | 0.199 | 0.160 | X | 3.2 | 5.0 |
| 0732 | 2005 WP7 | 2457692.44645 | 1:00 | 17.5 | 20.10 | 0.02±0.05 | 0.05±0.07 | ≥0.54 | ≥60 | 0.351 | 0.351 | 0.231 | 0.067 | - | 2.4 | 2.4 |
| 0733 | 2005 XO71 | 2457692.38892 | 1:20 | 15.7 | 19.61 | -0.11±0.02 | -0.15±0.04 | ≥0.13 | ≥80 | 0.000 | 0.000 | 1.000 | 0.000 | C | 3.2 | 4.7 |
| 0734 | 2005 YG40 | 2457724.39404 | 0:53 | 15.2 | 19.76 | -0.95±0.13 | -0.46±0.05 | ≥0.46 | ≥53 | 0.000 | 0.000 | 1.000 | 0.000 | C | 3.2 | 1.3 |

*Table 1 continued on next page*

**Table 1** *(continued)*

| No.[a] | Object | Obs. Start | Obs. Duration | H[b] | V | V-R[c] | V-I[c] | Amplitude[d] | Rot. Period[d] | S-type | X-type | C-type | D-type | Tax. | a[b] | i[b] |
|---|---|---|---|---|---|---|---|---|---|---|---|---|---|---|---|---|
| | | (JD) | (h:mm) | (mag) | (mag) | (mag) | (mag) | (mag) | (min) | (probability) | | | | | (au) | (deg) |
| 0735 | 2005 YZ48 | 2457722.38400 | 1:47 | 17.6 | 20.66 | 0.08±0.06 | 0.02±0.08 | ≥0.40 | ≥107 | 0.642 | 0.093 | 0.239 | 0.026 | S | 2.4 | 4.9 |
| 0736 | 2006 AK14 | 2457692.38892 | 1:20 | 17.2 | 19.27 | 0.01±0.02 | -0.01±0.03 | ≥0.10 | ≥80 | 0.122 | 0.153 | 0.725 | 0.000 | C | 2.5 | 1.1 |
| 0737 | 2006 AN29 | 2457690.35201 | 1:34 | 17.7 | 20.07 | 0.13±0.05 | 0.21±0.06 | ≥0.42 | ≥95 | 0.866 | 0.003 | 0.000 | 0.131 | S | 2.4 | 3.8 |
| 0738 | 2006 AW69 | 2457691.37970 | 0:59 | 17.1 | 19.42 | 0.11±0.04 | 0.15±0.04 | ≥0.14 | ≥60 | 0.923 | 0.006 | 0.000 | 0.071 | S | 2.5 | 3.2 |
| 0739 | 2006 AX46 | 2457730.42354 | 1:25 | 15.8 | 20.05 | -0.05±0.06 | -0.05±0.08 | ≥0.17 | ≥86 | 0.032 | 0.237 | 0.727 | 0.004 | C | 3.2 | 10.0 |
| 0740 | 2006 BB22 | 2457730.42038 | 1:30 | 16.4 | 19.17 | 0.11±0.03 | 0.12±0.04 | ≥0.19 | ≥90 | 0.965 | 0.009 | 0.000 | 0.025 | S | 2.4 | 7.8 |
| 0741 | 2006 BP161 | 2457800.27020 | 1:03 | 16.1 | 20.03 | -0.05±0.07 | -0.20±0.11 | ≥0.17 | ≥64 | 0.007 | 0.041 | 0.951 | 0.001 | C | 2.4 | 8.6 |
| 0742 | 2006 DG8 | 2457798.46669 | 0:53 | 15.8 | 19.56 | 0.19±0.15 | 0.18±0.28 | ≥0.40 | ≥53 | 0.718 | 0.033 | 0.140 | 0.110 | S | 3.2 | 10.7 |
| 0743 | 2006 DY9 | 2457691.37970 | 1:54 | 16.6 | 19.11 | 0.07±0.03 | 0.12±0.03 | ≥0.12 | ≥114 | 0.842 | 0.075 | 0.000 | 0.083 | S | 2.5 | 6.4 |
| 0744 | 2006 FF10 | 2457800.44661 | 0:54 | 15.2 | 19.30 | 0.02±0.08 | -0.01±0.09 | ≥0.17 | ≥54 | 0.293 | 0.208 | 0.466 | 0.033 | - | 3.2 | 10.5 |
| 0745 | 2006 FV17 | 2457690.33738 | 1:57 | 16.2 | 20.22 | 0.08±0.05 | 0.10±0.06 | ≥0.27 | ≥118 | 0.775 | 0.111 | 0.019 | 0.094 | S | 2.6 | 4.4 |
| 0746 | 2006 HC71 | 2457691.46127 | 2:00 | 16.9 | 20.21 | -0.03±0.04 | 0.09±0.07 | ≥0.23 | ≥121 | 0.093 | 0.646 | 0.095 | 0.166 | X | 2.7 | 8.8 |
| 0747 | 2006 KM26 | 2457687.38022 | 2:02 | 16.0 | 20.05 | 0.10±0.03 | 0.16±0.04 | ≥0.11 | ≥123 | 0.889 | 0.004 | 0.000 | 0.107 | S | 2.7 | 13.7 |
| 0748 | 2006 KT102 | 2457692.38892 | 1:20 | 16.0 | 19.86 | 0.09±0.03 | 0.10±0.04 | ≥0.28 | ≥80 | 0.949 | 0.031 | 0.001 | 0.019 | S | 2.6 | 11.9 |
| 0749 | 2006 RO11 | 2457690.33738 | 1:57 | 17.7 | 19.86 | 0.04±0.04 | 0.03±0.06 | ≥0.14 | ≥118 | 0.444 | 0.225 | 0.305 | 0.026 | - | 2.2 | 4.1 |
| 0750 | 2006 RP26 | 2457690.33738 | 0:57 | 16.7 | 20.48 | 0.04±0.06 | 0.00±0.09 | ≥0.26 | ≥57 | 0.345 | 0.173 | 0.442 | 0.040 | - | 2.9 | 2.8 |
| 0751 | 2006 TB27 | 2457691.37970 | 1:54 | 17.6 | 19.97 | 0.10±0.04 | 0.12±0.04 | ≥0.27 | ≥114 | 0.927 | 0.023 | 0.001 | 0.050 | S | 2.2 | 3.6 |
| 0752 | 2006 TE30 | 2457692.44645 | 1:00 | 17.9 | 19.66 | 0.02±0.04 | -0.00±0.06 | ≥0.19 | ≥60 | 0.258 | 0.228 | 0.509 | 0.004 | C | 2.2 | 3.8 |
| 0753 | 2006 TJ105 | 2457690.33738 | 1:57 | 17.6 | 20.51 | -0.02±0.07 | -0.01±0.09 | ≥0.27 | ≥118 | 0.150 | 0.331 | 0.489 | 0.030 | - | 2.2 | 3.3 |
| 0754 | 2006 TS54 | 2457724.38316 | 1:08 | 16.9 | 18.91 | -0.03±0.02 | -0.05±0.04 | ≥0.15 | ≥69 | 0.001 | 0.103 | 0.895 | 0.000 | C | 2.2 | 6.9 |
| 0755 | 2006 TX77 | 2457692.38892 | 1:20 | 18.2 | 20.31 | 0.06±0.04 | 0.05±0.05 | ≥0.16 | ≥80 | 0.683 | 0.163 | 0.132 | 0.022 | S | 2.2 | 6.3 |
| 0756 | 2006 UC45 | 2457690.33738 | 1:57 | 17.6 | 20.43 | 0.08±0.09 | 0.09±0.10 | ≥0.74 | ≥118 | 0.623 | 0.131 | 0.108 | 0.138 | S | 2.3 | 6.8 |
| 0757 | 2006 UC9 | 2457691.37970 | 1:54 | 15.6 | 20.20 | 0.02±0.05 | 0.04±0.06 | ≥0.23 | ≥114 | 0.327 | 0.416 | 0.231 | 0.026 | - | 3.0 | 8.3 |
| 0758 | 2006 UE168 | 2457690.33738 | 1:57 | 17.5 | 19.48 | 0.04±0.03 | 0.04±0.03 | ≥0.13 | ≥118 | 0.642 | 0.207 | 0.151 | 0.000 | S | 2.2 | 5.0 |
| 0759 | 2006 UF91 | 2457724.28943 | 1:05 | 15.7 | 20.43 | 0.03±0.07 | 0.04±0.09 | ≥0.63 | ≥66 | 0.361 | 0.268 | 0.267 | 0.103 | - | 3.0 | 11.0 |
| 0760 | 2006 UG215 | 2457692.44645 | 1:00 | 17.9 | 19.13 | -0.01±0.04 | 0.01±0.05 | ≥0.06 | ≥60 | 0.096 | 0.440 | 0.462 | 0.002 | - | 2.2 | 8.8 |
| 0761 | 2006 UG68 | 2457690.33738 | 1:55 | 16.2 | 20.08 | 0.02±0.04 | 0.09±0.05 | ≥0.22 | ≥116 | 0.386 | 0.398 | 0.054 | 0.163 | - | 2.9 | 3.2 |
| 0762 | 2006 UV37 | 2457691.37970 | 1:54 | 16.0 | 20.62 | 0.01±0.06 | 0.04±0.10 | ≥0.34 | ≥114 | 0.250 | 0.319 | 0.305 | 0.127 | - | 3.0 | 4.5 |
| 0763 | 2006 UX335 | 2457691.37970 | 1:54 | 16.9 | 20.57 | 0.01±0.07 | 0.14±0.08 | ≥0.45 | ≥114 | 0.304 | 0.304 | 0.036 | 0.356 | - | 3.0 | 1.5 |
| 0764 | 2006 UZ226 | 2457692.38892 | 1:20 | 17.9 | 20.44 | 0.11±0.04 | 0.16±0.05 | ≥0.14 | ≥80 | 0.864 | 0.012 | 0.000 | 0.124 | S | 2.2 | 4.3 |
| 0765 | 2006 VY112 | 2457690.33738 | 1:57 | 15.8 | 20.34 | 0.02±0.05 | 0.04±0.07 | ≥0.54 | ≥118 | 0.377 | 0.313 | 0.248 | 0.062 | - | 3.0 | 9.7 |
| 0766 | 2006 WL94 | 2457690.33738 | 1:57 | 16.5 | 20.64 | 0.01±0.07 | -0.11±0.11 | ≥0.37 | ≥118 | 0.112 | 0.085 | 0.797 | 0.006 | C | 3.0 | 3.0 |
| 0767 | 2007 AN6 | 2457687.38022 | 3:03 | 15.1 | 20.16 | -0.08±0.03 | -0.02±0.04 | ≥0.33 | ≥183 | 0.000 | 0.471 | 0.528 | 0.000 | C | 3.1 | 15.7 |
| 0768 | 2007 BK17 | 2457730.42646 | 1:21 | 15.8 | 20.08 | 0.05±0.05 | 0.08±0.06 | ≥0.23 | ≥81 | 0.588 | 0.256 | 0.077 | 0.079 | S | 3.0 | 10.8 |
| 0769 | 2007 BN18 | 2457690.33738 | 1:05 | 17.5 | 19.97 | 0.07±0.04 | 0.15±0.05 | ≥0.18 | ≥66 | 0.639 | 0.063 | 0.001 | 0.297 | S | 2.3 | 3.5 |
| 0770 | 2007 CT15 | 2457722.38400 | 1:47 | 17.5 | 20.38 | 0.09±0.04 | 0.07±0.06 | ≥0.33 | ≥107 | 0.855 | 0.058 | 0.050 | 0.037 | S | 2.3 | 5.4 |
| 0771 | 2007 DC8 | 2457692.44645 | 1:00 | 15.3 | 20.03 | -0.03±0.06 | -0.04±0.06 | ≥0.14 | ≥60 | 0.053 | 0.229 | 0.717 | 0.001 | C | 3.2 | 6.4 |
| 0772 | 2007 EH38 | 2457690.33738 | 1:57 | 16.6 | 20.35 | -0.02±0.06 | 0.10±0.08 | ≥0.64 | ≥118 | 0.135 | 0.536 | 0.095 | 0.234 | X | 2.4 | 6.2 |
| 0773 | 2007 KO9 | 2457690.38513 | 0:48 | 16.7 | 19.30 | 0.07±0.05 | 0.11±0.10 | ≥0.48 | ≥49 | 0.617 | 0.096 | 0.084 | 0.203 | S | 2.6 | 2.7 |
| 0774 | 2007 LP37 | 2457802.39910 | 0:37 | 15.8 | 20.07 | 0.06±0.07 | 0.07±0.08 | ≥0.15 | ≥37 | 0.603 | 0.164 | 0.142 | 0.090 | S | 3.1 | 9.9 |
| 0775 | 2007 PT43 | 2457690.42183 | 2:19 | 16.9 | 19.86 | -0.00±0.05 | -0.04±0.06 | ≥0.66 | ≥139 | 0.110 | 0.178 | 0.711 | 0.001 | C | 2.7 | 8.1 |
| 0776 | 2007 PZ42 | 2457687.38022 | 2:02 | 16.8 | 20.21 | 0.11±0.04 | 0.17±0.06 | ≥0.26 | ≥123 | 0.860 | 0.008 | 0.000 | 0.131 | S | 2.7 | 13.4 |
| 0777 | 2007 RD280 | 2457722.38400 | 1:47 | 15.9 | 19.69 | 0.11±0.03 | 0.16±0.04 | ≥0.11 | ≥107 | 0.958 | 0.001 | 0.000 | 0.041 | S | 2.6 | 14.3 |
| 0778 | 2007 RG84 | 2457724.28943 | 1:08 | 16.1 | 20.21 | -0.07±0.10 | -0.02±0.10 | ≥1.11 | ≥68 | 0.087 | 0.314 | 0.566 | 0.034 | C | 2.7 | 4.7 |
| 0779 | 2007 RO277 | 2457687.46786 | 0:55 | 16.1 | 20.29 | -0.05±0.05 | -0.03±0.07 | ≥0.17 | ≥55 | 0.029 | 0.368 | 0.599 | 0.004 | C | 2.6 | 15.5 |
| 0780 | 2007 RR173 | 2457722.38400 | 1:47 | 16.6 | 20.55 | 0.17±0.05 | 0.18±0.06 | ≥0.19 | ≥107 | 0.979 | 0.000 | 0.000 | 0.021 | S | 2.6 | 4.8 |

*Table 1 continued on next page*

**Table 1** *(continued)*

| No.[a] | Object | Obs. Start (JD) | Obs. Duration (h:mm) | H[b] (mag) | V (mag) | V-R[c] (mag) | V-I[c] (mag) | Amplitude[d] (mag) | Rot. Period[d] (min) | S-type (probability) | X-type (probability) | C-type (probability) | D-type (probability) | Tax. | a[b] (au) | i[b] (deg) |
|---|---|---|---|---|---|---|---|---|---|---|---|---|---|---|---|---|
| 0781 | 2007 TA406 | 2457692.44645 | 1:00 | 16.5 | 20.08 | 0.03±0.07 | 0.06±0.13 | ≥0.22 | ≥60 | 0.354 | 0.190 | 0.286 | 0.170 | - | 2.7 | 14.3 |
| 0782 | 2007 TF322 | 2457692.44645 | 1:00 | 16.9 | 19.61 | 0.06±0.04 | 0.05±0.05 | ≥0.11 | ≥60 | 0.685 | 0.172 | 0.117 | 0.026 | S | 2.7 | 6.5 |
| 0783 | 2007 TH54 | 2457691.37970 | 1:54 | 15.7 | 19.16 | 0.06±0.02 | 0.10±0.03 | ≥0.09 | ≥114 | 0.833 | 0.123 | 0.000 | 0.044 | S | 2.7 | 16.0 |
| 0784 | 2007 TJ169 | 2457722.38400 | 1:47 | 16.9 | 19.80 | -0.08±0.03 | -0.15±0.05 | ≥0.25 | ≥107 | 0.000 | 0.002 | 0.998 | 0.000 | C | 2.7 | 9.0 |
| 0785 | 2007 TK238 | 2457690.33738 | 1:55 | 16.0 | 19.62 | -0.05±0.03 | -0.05±0.04 | ≥0.14 | ≥116 | 0.001 | 0.175 | 0.825 | 0.000 | C | 2.8 | 5.8 |
| 0786 | 2007 TL95 | 2457724.34310 | 0:55 | 15.6 | 19.74 | 0.13±0.06 | 0.18±0.09 | ≥0.34 | ≥55 | 0.850 | 0.016 | 0.004 | 0.130 | S | 2.7 | 13.3 |
| 0787 | 2007 TX330 | 2457802.39604 | 0:41 | 17.6 | 20.04 | -0.00±0.06 | 0.08±0.06 | ≥0.33 | ≥41 | 0.269 | 0.503 | 0.107 | 0.121 | X | 2.6 | 6.5 |
| 0788 | 2007 UY139 | 2457691.47191 | 1:45 | 16.0 | 20.50 | 0.10±0.08 | 0.12±0.09 | ≥0.27 | ≥105 | 0.737 | 0.086 | 0.035 | 0.142 | S | 2.8 | 12.4 |
| 0789 | 2007 VC199 | 2457692.39380 | 1:12 | 17.0 | 20.59 | 0.08±0.05 | 0.09±0.06 | ≥0.19 | ≥73 | 0.813 | 0.098 | 0.032 | 0.057 | S | 2.7 | 9.1 |
| 0790 | 2007 VD310 | 2457690.33738 | 1:55 | 17.6 | 18.89 | 0.08±0.02 | 0.12±0.03 | ≥0.19 | ≥116 | 0.965 | 0.009 | 0.000 | 0.026 | S | 2.1 | 4.8 |
| 0791 | 2007 VO142 | 2457692.38892 | 1:20 | 17.0 | 20.43 | -0.07±0.04 | 0.01±0.06 | ≥0.23 | ≥80 | 0.008 | 0.629 | 0.358 | 0.006 | X | 2.8 | 4.1 |
| 0792 | 2007 VR142 | 2457691.37970 | 1:54 | 16.7 | 20.21 | 0.01±0.05 | -0.06±0.06 | ≥0.62 | ≥114 | 0.106 | 0.086 | 0.808 | 0.000 | C | 2.8 | 4.1 |
| 0793 | 2007 VS270 | 2457690.33738 | 1:57 | 16.7 | 20.55 | 0.01±0.05 | -0.01±0.07 | ≥0.27 | ≥118 | 0.200 | 0.239 | 0.552 | 0.010 | C | 2.8 | 5.3 |
| 0794 | 2007 VY315 | 2457730.42354 | 1:25 | 15.4 | 20.09 | 0.12±0.07 | 0.22±0.07 | ≥0.46 | ≥86 | 0.739 | 0.016 | 0.001 | 0.244 | S | 2.7 | 14.6 |
| 0795 | 2008 CU75 | 2457722.38400 | 1:47 | 16.1 | 20.28 | 0.07±0.05 | 0.09±0.06 | ≥0.27 | ≥107 | 0.719 | 0.151 | 0.050 | 0.080 | S | 2.9 | 7.0 |
| 0796 | 2008 EM167 | 2457691.46127 | 2:00 | 15.2 | 20.51 | -0.05±0.09 | 0.05±0.08 | ≥0.49 | ≥121 | 0.133 | 0.531 | 0.252 | 0.084 | X | 3.2 | 10.6 |
| 0797 | 2008 EO19 | 2457692.44645 | 1:00 | 16.5 | 19.71 | -0.06±0.04 | 0.05±0.05 | ≥0.12 | ≥60 | 0.018 | 0.819 | 0.135 | 0.028 | X | 3.0 | 2.9 |
| 0798 | 2008 EO78 | 2457722.38400 | 1:47 | 17.3 | 20.71 | 0.14±0.06 | -0.00±0.09 | ≥0.32 | ≥107 | 0.810 | 0.014 | 0.171 | 0.005 | S | 2.2 | 6.5 |
| 0799 | 2008 EX92 | 2457691.37970 | 1:54 | 15.4 | 20.48 | -0.06±0.07 | -0.07±0.08 | ≥0.45 | ≥114 | 0.022 | 0.201 | 0.775 | 0.002 | C | 3.2 | 18.5 |
| 0800 | 2008 FA108 | 2457692.38892 | 1:20 | 15.6 | 20.43 | 0.11±0.04 | 0.24±0.05 | ≥0.14 | ≥80 | 0.772 | 0.000 | 0.000 | 0.228 | S | 3.1 | 13.8 |
| 0801 | 2008 FH67 | 2457692.38892 | 1:20 | 15.8 | 20.21 | -0.10±0.04 | -0.14±0.05 | ≥0.17 | ≥80 | 0.000 | 0.009 | 0.991 | 0.000 | C | 3.2 | 8.5 |
| 0802 | 2008 FM80 | 2457690.38513 | 0:48 | 15.4 | 19.46 | -0.04±0.05 | -0.12±0.09 | ≥0.12 | ≥49 | 0.013 | 0.088 | 0.898 | 0.001 | C | 3.2 | 16.6 |
| 0803 | 2008 FQ107 | 2457690.38513 | 0:48 | 15.4 | 20.16 | 0.04±0.09 | -0.03±0.11 | ≥0.25 | ≥49 | 0.321 | 0.164 | 0.483 | 0.031 | - | 3.2 | 11.1 |
| 0804 | 2008 FR15 | 2457691.46127 | 1:58 | 15.5 | 20.23 | 0.01±0.05 | -0.02±0.06 | ≥0.26 | ≥119 | 0.174 | 0.219 | 0.605 | 0.002 | C | 3.1 | 10.6 |
| 0805 | 2008 FW136 | 2457802.39311 | 0:45 | 16.1 | 19.51 | 0.04±0.05 | 0.09±0.05 | ≥0.29 | ≥46 | 0.510 | 0.354 | 0.033 | 0.103 | S | 2.8 | 10.3 |
| 0806 | 2008 HR36 | 2457722.38400 | 1:47 | 17.5 | 20.84 | 0.13±0.07 | 0.11±0.08 | ≥0.44 | ≥107 | 0.881 | 0.032 | 0.028 | 0.059 | S | 2.3 | 8.1 |
| 0807 | 2008 PH11 | 2457691.37970 | 1:54 | 17.4 | 20.06 | 0.08±0.04 | 0.09±0.04 | ≥0.27 | ≥114 | 0.874 | 0.080 | 0.005 | 0.041 | S | 2.5 | 1.0 |
| 0808 | 2008 SD24 | 2457691.37970 | 1:54 | 17.8 | 20.26 | 0.05±0.06 | 0.14±0.06 | ≥0.30 | ≥114 | 0.528 | 0.179 | 0.014 | 0.279 | S | 2.5 | 1.2 |
| 0809 | 2008 SJ184 | 2457691.38961 | 1:40 | 17.3 | 19.46 | 0.09±0.03 | 0.07±0.03 | ≥0.15 | ≥100 | 0.971 | 0.024 | 0.002 | 0.002 | S | 2.6 | 5.0 |
| 0810 | 2008 SV264 | 2457690.33738 | 1:57 | 16.9 | 20.06 | 0.09±0.04 | 0.12±0.04 | ≥0.37 | ≥118 | 0.870 | 0.043 | 0.001 | 0.086 | S | 2.5 | 8.3 |
| 0811 | 2008 TF94 | 2457691.46127 | 2:00 | 17.5 | 20.42 | 0.06±0.06 | 0.20±0.09 | ≥0.36 | ≥121 | 0.473 | 0.067 | 0.007 | 0.453 | - | 2.5 | 5.9 |
| 0812 | 2008 TZ35 | 2457692.38892 | 1:20 | 16.9 | 20.56 | 0.04±0.07 | -0.13±0.09 | ≥0.17 | ≥80 | 0.099 | 0.023 | 0.878 | 0.000 | C | 2.5 | 5.8 |
| 0813 | 2008 VR23 | 2457691.46127 | 2:00 | 17.4 | 20.52 | 0.18±0.06 | 0.18±0.07 | ≥0.37 | ≥121 | 0.973 | 0.002 | 0.000 | 0.025 | S | 2.5 | 8.9 |
| 0814 | 2008 WK83 | 2457722.38400 | 1:47 | 17.3 | 19.06 | -0.05±0.02 | -0.06±0.03 | ≥0.15 | ≥107 | 0.000 | 0.105 | 0.895 | 0.000 | C | 2.5 | 4.3 |
| 0815 | 2008 WX66 | 2457692.38892 | 1:20 | 16.9 | 19.49 | 0.06±0.02 | 0.09±0.03 | ≥0.12 | ≥80 | 0.858 | 0.118 | 0.004 | 0.020 | S | 2.6 | 1.6 |
| 0816 | 2008 XG8 | 2457687.38022 | 3:03 | 16.4 | 19.59 | 0.07±0.02 | 0.09±0.03 | ≥0.24 | ≥183 | 0.919 | 0.075 | 0.001 | 0.005 | S | 2.5 | 12.6 |
| 0817 | 2008 YH9 | 2457691.46735 | 1:51 | 17.2 | 21.00 | 0.13±0.10 | 0.18±0.15 | ≥0.67 | ≥112 | 0.754 | 0.042 | 0.048 | 0.157 | S | 2.6 | 6.2 |
| 0818 | 2009 BO64 | 2457691.46411 | 1:56 | 16.8 | 20.52 | 0.11±0.07 | 0.19±0.09 | ≥0.46 | ≥117 | 0.724 | 0.035 | 0.007 | 0.233 | S | 2.7 | 6.4 |
| 0819 | 2009 BR183 | 2457691.54720 | 0:52 | 16.5 | 20.06 | -0.01±0.07 | 0.15±0.14 | ≥0.17 | ≥52 | 0.175 | 0.249 | 0.124 | 0.453 | - | 2.6 | 19.1 |
| 0820 | 2009 DD82 | 2457692.45553 | 0:46 | 16.6 | 19.96 | 0.01±0.06 | -0.02±0.08 | ≥0.24 | ≥47 | 0.184 | 0.240 | 0.554 | 0.023 | C | 2.8 | 4.5 |
| 0821 | 2009 DO94 | 2457691.50828 | 0:52 | 16.5 | 20.30 | -0.10±0.11 | -0.06±0.10 | ≥0.30 | ≥53 | 0.049 | 0.211 | 0.732 | 0.008 | C | 2.8 | 6.7 |
| 0822 | 2009 FE38 | 2457691.37970 | 1:54 | 16.7 | 20.70 | 0.03±0.12 | -0.01±0.12 | ≥0.27 | ≥114 | 0.345 | 0.186 | 0.407 | 0.061 | - | 2.8 | 6.5 |
| 0823 | 2009 FK3 | 2457722.39284 | 1:34 | 16.8 | 20.98 | 0.10±0.09 | -0.01±0.11 | ≥0.31 | ≥94 | 0.557 | 0.081 | 0.332 | 0.029 | S | 2.7 | 11.0 |
| 0824 | 2009 FR24 | 2457691.46127 | 2:00 | 16.2 | 19.93 | -0.06±0.04 | 0.09±0.06 | ≥0.25 | ≥121 | 0.017 | 0.829 | 0.047 | 0.106 | X | 2.8 | 5.8 |
| 0825 | 2009 FR5 | 2457690.33738 | 1:57 | 16.1 | 19.79 | 0.01±0.03 | 0.06±0.04 | ≥0.41 | ≥118 | 0.299 | 0.565 | 0.115 | 0.021 | X | 2.8 | 4.2 |
| 0826 | 2009 HX83 | 2457691.37970 | 1:54 | 15.9 | 20.27 | 0.01±0.05 | 0.08±0.06 | ≥0.40 | ≥114 | 0.343 | 0.479 | 0.088 | 0.090 | - | 3.1 | 15.2 |

*Table 1 continued on next page*

**Table 1** *(continued)*

| No.[a] | Object | Obs. Start | Obs. Duration | H[b] | V | V-R[c] | V-I[c] | Amplitude[d] | Rot. Period[d] | S-type | X-type | C-type | D-type | Tax. | a[b] | i[b] |
|---|---|---|---|---|---|---|---|---|---|---|---|---|---|---|---|---|
| | | (JD) | (h:mm) | (mag) | (mag) | (mag) | (mag) | (mag) | (min) | (probability) | | | | | (au) | (deg) |
| 0827 | 2009 SH352 | 2457692.44645 | 1:00 | 17.9 | 19.37 | 0.05±0.04 | 0.09±0.05 | ≥0.14 | ≥60 | 0.635 | 0.211 | 0.037 | 0.117 | S | 2.3 | 5.8 |
| 0828 | 2009 SV333 | 2457690.33738 | 1:57 | 17.3 | 20.04 | 0.03±0.04 | -0.14±0.05 | ≥0.26 | ≥118 | 0.008 | 0.001 | 0.991 | 0.000 | C | 2.3 | 5.1 |
| 0829 | 2009 SV352 | 2457690.33738 | 1:57 | 18.1 | 20.31 | -0.01±0.05 | 0.02±0.06 | ≥0.34 | ≥118 | 0.156 | 0.455 | 0.376 | 0.013 | - | 2.3 | 4.0 |
| 0830 | 2009 UC26 | 2457690.38513 | 0:48 | 17.8 | 19.47 | 0.05±0.05 | 0.03±0.07 | ≥0.21 | ≥49 | 0.538 | 0.167 | 0.255 | 0.040 | S | 2.3 | 3.1 |
| 0831 | 2009 UJ34 | 2457692.44645 | 1:00 | 18.8 | 20.11 | 0.07±0.05 | 0.10±0.07 | ≥0.16 | ≥60 | 0.714 | 0.123 | 0.040 | 0.123 | S | 2.3 | 4.8 |
| 0832 | 2009 UN83 | 2457692.44645 | 1:00 | 18.0 | 19.26 | 0.04±0.05 | 0.17±0.05 | ≥0.07 | ≥60 | 0.398 | 0.116 | 0.002 | 0.485 | - | 2.3 | 1.9 |
| 0833 | 2010 FR47 | 2457690.33738 | 1:57 | 16.1 | 19.66 | -0.08±0.03 | -0.12±0.05 | ≥0.23 | ≥118 | 0.000 | 0.015 | 0.985 | 0.000 | C | 2.8 | 7.5 |
| 0834 | 2010 GG156 | 2457724.34161 | 0:57 | 16.3 | 20.34 | 0.27±0.12 | 0.05±0.14 | ≥0.68 | ≥58 | 0.915 | 0.009 | 0.065 | 0.010 | S | 2.6 | 14.8 |
| 0835 | 2010 HP25 | 2457722.38400 | 1:47 | 16.3 | 20.51 | -0.04±0.06 | -0.03±0.07 | ≥0.87 | ≥107 | 0.054 | 0.323 | 0.619 | 0.004 | C | 2.8 | 5.3 |
| 0836 | 2010 LW82 | 2457690.33738 | 1:57 | 15.7 | 19.87 | -0.09±0.07 | -0.05±0.05 | ≥0.31 | ≥118 | 0.011 | 0.182 | 0.807 | 0.000 | C | 3.2 | 25.5 |
| 0837 | 2010 NH45 | 2457691.46127 | 2:00 | 16.4 | 20.61 | -0.08±0.10 | -0.02±0.11 | ≥0.52 | ≥121 | 0.074 | 0.323 | 0.567 | 0.036 | C | 3.2 | 6.7 |
| 0838 | 2010 OG15 | 2457690.42183 | 2:19 | 15.8 | 19.90 | 0.02±0.05 | 0.10±0.06 | ≥0.40 | ≥139 | 0.346 | 0.421 | 0.055 | 0.178 | - | 3.2 | 24.8 |
| 0839 | 2010 UJ99 | 2457730.42038 | 1:30 | 15.8 | 19.49 | 0.09±0.04 | 0.13±0.04 | ≥0.19 | ≥90 | 0.876 | 0.033 | 0.000 | 0.091 | S | 3.1 | 16.7 |
| 0840 | 2010 VN39 | 2457730.42038 | 1:30 | 15.4 | 20.11 | 0.13±0.05 | 0.18±0.06 | ≥0.21 | ≥90 | 0.885 | 0.004 | 0.000 | 0.111 | S | 3.2 | 9.5 |
| 0841 | 2011 GV32 | 2457692.39079 | 1:17 | 17.3 | 20.85 | 0.04±0.07 | 0.18±0.09 | ≥0.28 | ≥77 | 0.421 | 0.134 | 0.014 | 0.430 | - | 2.3 | 7.2 |
| 0842 | 2011 QV5 | 2457722.38400 | 1:47 | 16.7 | 20.85 | 0.09±0.08 | 0.17±0.08 | ≥0.27 | ≥107 | 0.657 | 0.084 | 0.011 | 0.247 | S | 2.6 | 10.1 |
| 0843 | 2011 SK90 | 2457724.38316 | 1:08 | 16.4 | 20.19 | 0.09±0.06 | 0.17±0.07 | ≥0.37 | ≥69 | 0.699 | 0.051 | 0.005 | 0.244 | S | 2.8 | 4.9 |
| 0844 | 2011 SP232 | 2457724.38316 | 1:08 | 16.5 | 20.53 | 0.06±0.06 | 0.11±0.07 | ≥0.34 | ≥69 | 0.607 | 0.167 | 0.043 | 0.183 | S | 2.8 | 7.1 |
| 0845 | 2011 SY154 | 2457692.38892 | 1:20 | 16.7 | 20.33 | 0.10±0.04 | 0.14±0.05 | ≥0.20 | ≥80 | 0.882 | 0.019 | 0.001 | 0.098 | S | 2.8 | 2.9 |
| 0846 | 2011 SY166 | 2457690.34053 | 1:53 | 16.9 | 20.64 | 0.01±0.10 | -0.06±0.11 | ≥0.33 | ≥113 | 0.214 | 0.151 | 0.614 | 0.022 | C | 2.9 | 3.1 |
| 0847 | 2011 UK178 | 2457690.33738 | 1:57 | 16.5 | 19.48 | -0.05±0.03 | -0.00±0.03 | ≥0.27 | ≥118 | 0.002 | 0.616 | 0.382 | 0.000 | X | 3.0 | 3.0 |
| 0848 | 2011 UU176 | 2457691.37970 | 1:54 | 16.6 | 21.00 | 0.08±0.10 | -0.07±0.11 | ≥0.26 | ≥114 | 0.397 | 0.068 | 0.524 | 0.011 | C | 3.0 | 11.6 |
| 0849 | 2011 UW102 | 2457692.38892 | 0:35 | 16.0 | 18.99 | 0.02±0.03 | 0.10±0.05 | ≥0.04 | ≥35 | 0.331 | 0.443 | 0.043 | 0.184 | - | 3.0 | 10.9 |
| 0850 | 2011 UZ327 | 2457692.44645 | 1:00 | 16.5 | 19.64 | -0.06±0.04 | -0.04±0.06 | ≥0.22 | ≥60 | 0.008 | 0.317 | 0.674 | 0.001 | C | 3.0 | 1.6 |
| 0851 | 2011 WJ33 | 2457692.38892 | 1:20 | 15.5 | 20.18 | 0.02±0.04 | 0.10±0.04 | ≥0.16 | ≥80 | 0.364 | 0.468 | 0.019 | 0.148 | - | 3.0 | 9.9 |
| 0852 | 2011 WM120 | 2457687.38022 | 3:03 | 16.7 | 19.77 | 0.08±0.02 | 0.19±0.03 | ≥0.30 | ≥183 | 0.715 | 0.000 | 0.000 | 0.285 | S | 3.0 | 12.7 |
| 0853 | 2011 YG27 | 2457691.46127 | 2:00 | 15.9 | 20.44 | 0.15±0.07 | 0.21±0.07 | ≥0.42 | ≥121 | 0.883 | 0.005 | 0.001 | 0.112 | S | 3.2 | 9.5 |
| 0854 | 2012 RG40 | 2457690.33738 | 1:57 | 17.2 | 19.83 | 0.06±0.04 | 0.07±0.05 | ≥0.25 | ≥118 | 0.788 | 0.134 | 0.045 | 0.034 | S | 2.4 | 2.3 |
| 0855 | 2012 SK1 | 2457690.38513 | 0:48 | 17.5 | 19.61 | 0.09±0.05 | 0.06±0.11 | ≥0.38 | ≥49 | 0.700 | 0.042 | 0.184 | 0.074 | S | 2.5 | 2.2 |
| 0856 | 2012 SR46 | 2457692.38892 | 1:20 | 18.1 | 19.84 | 0.01±0.03 | 0.01±0.04 | ≥0.14 | ≥80 | 0.188 | 0.336 | 0.476 | 0.001 | - | 2.5 | 1.7 |
| 0857 | 2012 TV155 | 2457691.37970 | 1:54 | 17.3 | 20.84 | 0.06±0.07 | 0.02±0.12 | ≥0.33 | ≥114 | 0.453 | 0.120 | 0.344 | 0.084 | - | 2.5 | 5.5 |
| 0858 | 2012 TV257 | 2457724.38316 | 1:08 | 17.6 | 20.16 | -0.01±0.05 | 0.10±0.10 | ≥0.25 | ≥69 | 0.170 | 0.397 | 0.136 | 0.297 | - | 2.5 | 5.3 |
| 0859 | 2012 TZ200 | 2457691.37970 | 1:52 | 17.0 | 20.56 | 0.08±0.10 | -0.05±0.10 | ≥0.25 | ≥113 | 0.435 | 0.087 | 0.466 | 0.011 | - | 2.5 | 7.1 |
| 0860 | 2012 UL58 | 2457691.37970 | 1:54 | 17.7 | 20.28 | -0.08±0.05 | -0.14±0.06 | ≥0.31 | ≥114 | 0.000 | 0.020 | 0.979 | 0.000 | C | 2.6 | 9.3 |
| 0861 | 2012 VA91 | 2457724.38316 | 1:08 | 17.6 | 20.41 | 0.00±0.06 | 0.04±0.07 | ≥0.18 | ≥69 | 0.258 | 0.412 | 0.267 | 0.063 | - | 2.5 | 3.1 |
| 0862 | 2012 VR96 | 2457692.44645 | 1:00 | 17.2 | 19.21 | 0.05±0.04 | 0.00±0.06 | ≥0.16 | ≥60 | 0.473 | 0.118 | 0.403 | 0.006 | - | 2.6 | 3.4 |
| 0863 | 2012 XC136 | 2457730.42038 | 1:30 | 17.4 | 19.91 | 0.07±0.04 | 0.07±0.05 | ≥0.25 | ≥90 | 0.782 | 0.142 | 0.049 | 0.026 | S | 2.6 | 15.5 |
| 0864 | 2012 XL53 | 2457722.38400 | 1:47 | 17.2 | 20.35 | 0.10±0.05 | 0.09±0.06 | ≥0.41 | ≥107 | 0.860 | 0.056 | 0.025 | 0.058 | S | 2.6 | 6.5 |
| 0865 | 2012 XP94 | 2457690.42183 | 2:19 | 16.7 | 20.46 | 0.17±0.07 | 0.20±0.08 | ≥0.46 | ≥139 | 0.925 | 0.004 | 0.001 | 0.070 | S | 2.6 | 11.2 |
| 0866 | 2012 XT114 | 2457724.39556 | 0:50 | 17.0 | 19.33 | -0.01±0.03 | -0.05±0.04 | ≥0.27 | ≥51 | 0.020 | 0.114 | 0.867 | 0.000 | C | 2.6 | 5.1 |
| 0867 | 2012 XU131 | 2457722.38400 | 1:47 | 16.8 | 20.27 | 0.07±0.05 | 0.09±0.07 | ≥0.54 | ≥107 | 0.722 | 0.120 | 0.052 | 0.106 | S | 2.6 | 4.5 |
| 0868 | 2013 AD91 | 2457692.44645 | 1:00 | 16.8 | 19.56 | 0.06±0.05 | 0.19±0.06 | ≥0.15 | ≥60 | 0.441 | 0.052 | 0.000 | 0.507 | D | 2.7 | 10.3 |
| 0869 | 2013 AH121 | 2457691.50542 | 0:57 | 15.9 | 20.60 | -0.09±0.10 | -0.09±0.14 | ≥0.46 | ≥57 | 0.041 | 0.190 | 0.743 | 0.025 | C | 3.1 | 7.0 |
| 0870 | 2013 AH24 | 2457724.34608 | 0:51 | 17.6 | 20.50 | 0.36±0.12 | -0.10±0.27 | ≥0.51 | ≥51 | 0.801 | 0.000 | 0.197 | 0.001 | S | 2.6 | 3.0 |
| 0871 | 2013 AU6 | 2457691.54720 | 1:27 | 16.0 | 20.45 | 0.09±0.06 | 0.04±0.08 | ≥0.25 | ≥87 | 0.731 | 0.066 | 0.170 | 0.033 | S | 2.7 | 23.0 |
| 0872 | 2013 CG7 | 2457724.38626 | 1:04 | 16.4 | 20.12 | -0.04±0.05 | 0.03±0.05 | ≥0.15 | ≥64 | 0.084 | 0.631 | 0.275 | 0.010 | X | 2.8 | 3.2 |

*Table 1 continued on next page*

**Table 1** *(continued)*

| No.[a] | Object | Obs. Start (JD) | Obs. Duration (h:mm) | H[b] (mag) | V (mag) | V-R[c] (mag) | V-I[c] (mag) | Amplitude[d] (mag) | Rot. Period[d] (min) | S-type (probability) | X-type | C-type | D-type | Tax. | a[b] (au) | i[b] (deg) |
|---|---|---|---|---|---|---|---|---|---|---|---|---|---|---|---|---|
| 0873 | 2014 DH97 | 2457722.38400 | 1:45 | 16.5 | 20.53 | 0.05±0.06 | -0.07±0.09 | ≥0.32 | ≥105 | 0.236 | 0.064 | 0.696 | 0.004 | C | 2.5 | 5.2 |
| 0874 | 2014 ET18 | 2457691.46127 | 2:00 | 17.0 | 20.25 | -0.04±0.06 | -0.01±0.07 | ≥0.50 | ≥121 | 0.076 | 0.379 | 0.536 | 0.009 | C | 2.6 | 5.6 |
| 0875 | 2014 HH169 | 2457691.46127 | 2:00 | 16.0 | 20.68 | 0.03±0.10 | 0.04±0.12 | ≥0.42 | ≥121 | 0.363 | 0.206 | 0.312 | 0.120 | - | 3.1 | 11.8 |
| 0876 | 2014 HS12 | 2457692.38892 | 1:20 | 16.8 | 19.95 | -0.00±0.03 | 0.01±0.04 | ≥0.23 | ≥80 | 0.135 | 0.371 | 0.493 | 0.001 | - | 2.7 | 10.5 |
| 0877 | 2014 HV150 | 2457692.38892 | 1:20 | 16.3 | 20.43 | -0.10±0.04 | -0.09±0.06 | ≥0.19 | ≥80 | 0.000 | 0.099 | 0.901 | 0.000 | C | 3.2 | 10.3 |
| 0878 | 2014 JD45 | 2457722.38400 | 1:47 | 16.1 | 19.99 | 0.01±0.03 | -0.02±0.06 | ≥0.32 | ≥107 | 0.138 | 0.206 | 0.651 | 0.005 | C | 3.0 | 10.7 |
| 0879 | 2014 KQ6 | 2457692.38892 | 1:20 | 16.5 | 20.21 | -0.00±0.04 | -0.01±0.04 | ≥0.15 | ≥80 | 0.098 | 0.253 | 0.649 | 0.000 | C | 2.8 | 4.9 |
| 0880 | 2014 KX31 | 2457722.38400 | 1:47 | 16.4 | 20.80 | 0.03±0.07 | 0.02±0.09 | ≥0.59 | ≥107 | 0.359 | 0.233 | 0.344 | 0.064 | - | 3.1 | 9.4 |
| 0881 | 2015 KE75 | 2457691.46127 | 2:00 | 17.1 | 19.80 | 0.07±0.04 | 0.14±0.05 | ≥0.23 | ≥121 | 0.695 | 0.089 | 0.002 | 0.214 | S | 2.6 | 12.0 |
| 0882 | 2015 KK10 | 2457691.58506 | 0:32 | 16.9 | 20.63 | 0.17±0.10 | 0.23±0.13 | ≥0.18 | ≥33 | 0.840 | 0.019 | 0.009 | 0.133 | S | 2.6 | 14.4 |
| 0883 | 2015 PX173 | 2457691.46127 | 2:00 | 16.0 | 20.61 | 0.01±0.09 | 0.06±0.11 | ≥0.41 | ≥121 | 0.318 | 0.281 | 0.258 | 0.143 | - | 3.2 | 9.1 |
| 0884 | 2016 PB75 | 2457690.33738 | 1:55 | 17.7 | 20.11 | -0.06±0.09 | -0.03±0.07 | ≥0.31 | ≥116 | 0.082 | 0.282 | 0.634 | 0.002 | C | 2.4 | 2.3 |
| 0885 | 2016 RP29 | 2457690.33738 | 1:57 | 16.3 | 20.25 | -0.09±0.04 | -0.02±0.06 | ≥0.42 | ≥118 | 0.000 | 0.452 | 0.546 | 0.002 | C | 3.2 | 5.0 |
| 0886 | 2016 SK21 | 2457690.33738 | 1:57 | 17.9 | 20.35 | 0.13±0.06 | 0.19±0.06 | ≥0.38 | ≥118 | 0.878 | 0.006 | 0.000 | 0.115 | S | 2.4 | 5.3 |
| 0887 | 2016 UA23 | 2457692.39380 | 1:11 | 19.4 | 20.83 | 0.08±0.07 | 0.06±0.10 | ≥0.15 | ≥71 | 0.649 | 0.094 | 0.174 | 0.083 | S | 2.3 | 2.3 |
| 0888 | 2016 UA90 | 2457692.38892 | 1:20 | 17.9 | 19.92 | 0.05±0.03 | 0.07±0.04 | ≥0.17 | ≥80 | 0.707 | 0.215 | 0.046 | 0.032 | S | 2.4 | 3.7 |
| 0889 | 2016 UF23 | 2457692.38892 | 1:20 | 16.9 | 20.40 | -0.00±0.04 | -0.02±0.06 | ≥0.22 | ≥80 | 0.126 | 0.251 | 0.622 | 0.002 | C | 2.9 | 1.6 |
| 0890 | 2016 UG69 | 2457692.39079 | 1:13 | 17.9 | 20.43 | -0.09±0.05 | -0.10±0.07 | ≥0.22 | ≥73 | 0.000 | 0.096 | 0.903 | 0.000 | C | 3.2 | 15.3 |
| 0891 | 2016 UG72 | 2457692.39079 | 1:17 | 17.9 | 20.72 | -0.07±0.06 | 0.02±0.08 | ≥0.27 | ≥77 | 0.030 | 0.582 | 0.352 | 0.035 | X | 2.7 | 4.4 |
| 0892 | 2016 UG75 | 2457692.38892 | 1:20 | 17.2 | 20.87 | -0.02±0.05 | -0.05±0.07 | ≥0.55 | ≥80 | 0.061 | 0.206 | 0.731 | 0.002 | C | 2.8 | 1.8 |
| 0893 | 2016 UG83 | 2457692.38892 | 1:20 | 18.8 | 20.22 | 0.03±0.04 | 0.05±0.05 | ≥0.18 | ≥80 | 0.421 | 0.380 | 0.166 | 0.032 | - | 2.4 | 6.2 |
| 0894 | 2016 UH69 | 2457692.39079 | 1:17 | 17.0 | 20.52 | 0.10±0.05 | 0.20±0.06 | ≥0.14 | ≥77 | 0.720 | 0.014 | 0.000 | 0.266 | S | 2.6 | 7.5 |
| 0895 | 2016 UM15 | 2457692.38892 | 1:20 | 17.4 | 20.69 | 0.05±0.05 | 0.11±0.07 | ≥0.20 | ≥80 | 0.568 | 0.204 | 0.037 | 0.191 | S | 3.1 | 17.0 |
| 0896 | 2016 UO69 | 2457692.38892 | 1:20 | 17.7 | 20.49 | 0.01±0.04 | 0.02±0.07 | ≥0.40 | ≥80 | 0.235 | 0.364 | 0.368 | 0.033 | - | 3.1 | 4.1 |
| 0897 | 2016 UR73 | 2457692.39380 | 1:12 | 18.7 | 20.45 | 0.00±0.05 | 0.04±0.06 | ≥0.36 | ≥73 | 0.230 | 0.452 | 0.278 | 0.041 | - | 2.6 | 14.1 |
| 0898 | 2016 US77 | 2457692.38892 | 1:20 | 18.4 | 20.40 | -0.00±0.04 | -0.00±0.05 | ≥0.17 | ≥80 | 0.146 | 0.333 | 0.518 | 0.003 | C | 2.4 | 7.9 |
| 0899 | 2016 US92 | 2457692.39079 | 1:17 | 19.7 | 20.73 | 0.06±0.07 | 0.09±0.07 | ≥0.13 | ≥77 | 0.604 | 0.207 | 0.073 | 0.116 | S | 2.2 | 2.2 |
| 0900 | 2016 UV69 | 2457692.38892 | 1:20 | 17.2 | 20.02 | 0.02±0.04 | 0.06±0.05 | ≥0.32 | ≥80 | 0.360 | 0.465 | 0.144 | 0.032 | - | 2.6 | 15.7 |
| 0901 | 2016 UW67 | 2457724.38316 | 1:08 | 16.2 | 20.37 | -0.05±0.06 | -0.03±0.07 | ≥0.71 | ≥69 | 0.033 | 0.338 | 0.625 | 0.005 | C | 3.2 | 11.1 |
| 0902 | 2016 UY57 | 2457692.38892 | 1:20 | 19.1 | 20.70 | 0.03±0.05 | 0.05±0.06 | ≥0.53 | ≥80 | 0.462 | 0.302 | 0.190 | 0.046 | - | 2.3 | 1.6 |
| 0903 | 2016 VB15 | 2457722.38400 | 1:47 | 15.7 | 19.43 | -0.03±0.03 | 0.03±0.03 | 0.94±0.06 | 214±44 | 0.040 | 0.708 | 0.252 | 0.000 | X | 3.1 | 18.6 |
| 0904 | 2017 BG12 | 2457802.39311 | 0:45 | 16.7 | 20.13 | 0.09±0.05 | 0.09±0.10 | ≥0.18 | ≥46 | 0.759 | 0.051 | 0.083 | 0.107 | S | 2.6 | 21.3 |
| 0905 | 2619 T-3 | 2457691.42249 | 0:52 | 14.5 | 19.55 | -0.01±0.05 | 0.05±0.06 | ≥0.13 | ≥53 | 0.197 | 0.515 | 0.242 | 0.047 | X | 3.1 | 18.9 |
| 0906 | 3114 T-3 | 2457724.28943 | 1:03 | 15.4 | 19.58 | -0.07±0.04 | -0.08±0.05 | ≥0.15 | ≥64 | 0.001 | 0.102 | 0.898 | 0.000 | C | 3.1 | 0.2 |
| 0907 | 3147 T-2 | 2457690.33738 | 1:57 | 14.1 | 18.90 | -0.05±0.06 | -0.03±0.04 | ≥0.26 | ≥118 | 0.040 | 0.291 | 0.668 | 0.000 | C | 3.1 | 3.4 |
| 0908 | 3355 T-3 | 2457798.50513 | 1:09 | 14.9 | 19.48 | 0.29±0.13 | 0.31±0.12 | ≥0.36 | ≥69 | 0.962 | 0.003 | 0.000 | 0.035 | S | 2.8 | 8.7 |
| 0909 | 4109 T-2 | 2457690.42183 | 2:19 | 16.3 | 19.47 | 0.03±0.04 | 0.03±0.04 | ≥0.33 | ≥139 | 0.433 | 0.295 | 0.265 | 0.007 | - | 2.5 | 7.8 |
| 0910 | 4198 T-3 | 2457765.28694 | 1:01 | 13.8 | 19.44 | -0.07±0.06 | -0.10±0.07 | ≥0.22 | ≥62 | 0.005 | 0.075 | 0.920 | 0.000 | C | 3.1 | 16.0 |
| 0911 | 4303 T-3 | 2457802.50174 | 0:55 | 14.7 | 19.69 | -0.11±0.06 | -0.04±0.10 | ≥0.21 | ≥55 | 0.004 | 0.343 | 0.640 | 0.014 | C | 3.1 | 9.5 |
| 0912 | 4311 T-1 | 2457768.50255 | 2:00 | 15.4 | 18.54 | 0.02±0.16 | 0.08±0.11 | ≥0.64 | ≥120 | 0.423 | 0.265 | 0.185 | 0.128 | - | 2.6 | 5.6 |
| 0913 | 9587 P-L | 2457692.44645 | 1:00 | 15.0 | 18.52 | -0.07±0.04 | -0.07±0.06 | ≥0.09 | ≥60 | 0.001 | 0.198 | 0.801 | 0.000 | C | 2.8 | 5.2 |
| 0914 | Aberdonia | 2457692.44645 | 1:00 | 12.6 | 16.35 | 0.05±0.04 | 0.10±0.04 | ≥0.16 | ≥60 | 0.678 | 0.216 | 0.015 | 0.092 | S | 2.8 | 1.5 |
| 0915 | Andronikov | 2457796.48277 | 0:51 | 11.7 | 16.23 | -0.02±0.05 | 0.02±0.05 | ≥0.29 | ≥52 | 0.112 | 0.524 | 0.356 | 0.007 | X | 2.6 | 6.3 |
| 0916 | Annettelee | 2457692.44645 | 1:00 | 15.1 | 17.78 | 0.00±0.14 | 0.10±0.05 | ≥0.24 | ≥60 | 0.400 | 0.483 | 0.031 | 0.085 | - | 2.5 | 1.7 |
| 0917 | Ara | 2457724.34161 | 0:57 | 8.1 | 13.26 | -0.04±0.03 | 0.04±0.04 | ≥0.02 | ≥58 | 0.008 | 0.855 | 0.135 | 0.002 | X | 3.1 | 19.5 |
| 0918 | Arthuradel | 2457687.38022 | 0:58 | 12.6 | 16.17 | 0.09±0.03 | 0.14±0.03 | ≥0.17 | ≥58 | 0.900 | 0.010 | 0.000 | 0.090 | S | 2.5 | 13.5 |

*Table 1 continued on next page*

**Table 1** *(continued)*

| No.[a] | Object | Obs. Start | Obs. Duration | H[b] | V | V-R[c] | V-I[c] | Amplitude[d] | Rot. Period[d] | S-type | X-type | C-type | D-type | Tax. | a[b] | i[b] |
|---|---|---|---|---|---|---|---|---|---|---|---|---|---|---|---|---|
| | | (JD) | (h:mm) | (mag) | (mag) | (mag) | (mag) | (mag) | (min) | (probability) | | | | | (au) | (deg) |
| 0919 | Balbastre | 2457802.50174 | 0:55 | 14.8 | 17.86 | -0.06±0.02 | -0.08±0.04 | ≥0.15 | ≥55 | 0.000 | 0.043 | 0.957 | 0.000 | C | 2.2 | 6.1 |
| 0920 | Banpeiyu | 2457722.38400 | 1:47 | 15.2 | 18.64 | 0.08±0.02 | 0.10±0.03 | 0.15±0.06 | 214±52 | 0.971 | 0.019 | 0.000 | 0.010 | S | 2.3 | 3.7 |
| 0921 | Benjaminlu | 2457690.33738 | 1:57 | 15.1 | 18.66 | 0.08±0.02 | 0.03±0.02 | ≥0.31 | ≥118 | 0.991 | 0.002 | 0.007 | 0.000 | S | 2.3 | 6.4 |
| 0922 | Billmclaughlin | 2457802.39910 | 0:37 | 13.1 | 16.45 | -0.00±0.02 | 0.03±0.03 | ≥0.04 | ≥37 | 0.092 | 0.531 | 0.377 | 0.000 | X | 2.4 | 8.9 |
| 0923 | Bokor | 2457690.33738 | 1:05 | 14.5 | 17.61 | 0.07±0.02 | 0.13±0.04 | 0.14±0.04 | 103±37 | 0.850 | 0.033 | 0.000 | 0.117 | S | 2.3 | 2.5 |
| 0924 | Bonus | 2457724.28943 | 1:08 | 14.2 | 18.49 | 0.06±0.02 | 0.13±0.03 | ≥0.08 | ≥68 | 0.708 | 0.076 | 0.000 | 0.217 | S | 2.4 | 1.6 |
| 0925 | Bounty | 2457692.44645 | 0:40 | 12.3 | 16.99 | -0.12±0.15 | -0.03±0.07 | ≥0.12 | ≥40 | 0.105 | 0.231 | 0.660 | 0.003 | C | 3.2 | 1.0 |
| 0926 | Bridgetoei | 2457691.37970 | 1:54 | 14.7 | 18.36 | 0.11±0.02 | -0.05±0.02 | ≥0.15 | ≥114 | 0.728 | 0.000 | 0.272 | 0.000 | S | 2.4 | 4.5 |
| 0927 | Brueghel | 2457724.38316 | 1:08 | 13.3 | 18.51 | -0.07±0.02 | -0.14±0.03 | ≥0.07 | ≥69 | 0.000 | 0.000 | 1.000 | 0.000 | C | 3.2 | 2.3 |
| 0928 | Cauchy | 2457692.44645 | 1:00 | 14.3 | 18.33 | 0.03±0.03 | 0.07±0.05 | ≥0.09 | ≥60 | 0.441 | 0.405 | 0.106 | 0.048 | - | 2.8 | 3.0 |
| 0929 | Chrisrussell | 2457692.44645 | 1:00 | 14.3 | 17.36 | 0.05±0.04 | 0.09±0.04 | ≥0.07 | ≥60 | 0.674 | 0.228 | 0.021 | 0.077 | S | 2.3 | 7.1 |
| 0930 | Consadole | 2457724.28943 | 1:08 | 14.6 | 18.06 | 0.06±0.02 | 0.10±0.03 | ≥0.05 | ≥68 | 0.774 | 0.156 | 0.001 | 0.070 | S | 2.3 | 4.0 |
| 0931 | Dastidar | 2457722.38400 | 1:47 | 14.8 | 18.79 | -0.04±0.02 | -0.08±0.03 | ≥0.12 | ≥107 | 0.000 | 0.022 | 0.978 | 0.000 | C | 2.3 | 5.6 |
| 0932 | Davidgans | 2457690.42183 | 2:19 | 14.7 | 19.04 | 0.06±0.03 | 0.09±0.04 | 0.65±0.09 | 279±62 | 0.792 | 0.163 | 0.006 | 0.039 | S | 3.1 | 9.4 |
| 0933 | Dawilliams | 2457728.49022 | 0:57 | 14.0 | 17.59 | 0.02±0.05 | -0.02±0.06 | ≥0.32 | ≥58 | 0.237 | 0.198 | 0.560 | 0.005 | C | 2.4 | 6.9 |
| 0934 | Disalvo | 2457691.46127 | 2:00 | 14.8 | 18.44 | 0.07±0.03 | 0.12±0.04 | ≥0.14 | ≥121 | 0.821 | 0.071 | 0.001 | 0.107 | S | 2.5 | 6.9 |
| 0935 | Duncan-Lewis | 2457692.44645 | 0:31 | 15.9 | 19.14 | 0.03±0.05 | 0.14±0.07 | ≥0.16 | ≥32 | 0.395 | 0.213 | 0.027 | 0.365 | - | 2.3 | 3.0 |
| 0936 | Elyna | 2457798.46669 | 0:44 | 10.9 | 15.21 | 0.05±0.03 | 0.12±0.04 | ≥0.09 | ≥45 | 0.645 | 0.175 | 0.003 | 0.177 | S | 3.0 | 8.5 |
| 0937 | Erwingroten | 2457691.37970 | 1:54 | 14.6 | 18.46 | 0.09±0.02 | 0.12±0.03 | ≥0.11 | ≥114 | 0.994 | 0.001 | 0.000 | 0.005 | S | 2.4 | 3.0 |
| 0938 | Evanchen | 2457800.27020 | 1:07 | 13.5 | 18.98 | -0.00±0.11 | 0.07±0.06 | ≥0.18 | ≥68 | 0.368 | 0.489 | 0.088 | 0.054 | - | 3.0 | 10.3 |
| 0939 | Felicienrops | 2457796.44027 | 0:58 | 13.2 | 17.11 | -0.03±0.04 | -0.02±0.04 | ≥0.05 | ≥59 | 0.037 | 0.343 | 0.620 | 0.000 | C | 2.6 | 12.5 |
| 0940 | Floss | 2457798.46669 | 0:53 | 15.0 | 18.05 | 0.08±0.04 | 0.11±0.05 | ≥0.25 | ≥53 | 0.868 | 0.053 | 0.004 | 0.075 | S | 2.2 | 4.4 |
| 0941 | Gardon | 2457728.49022 | 0:57 | 14.0 | 18.41 | 0.12±0.05 | 0.06±0.07 | ≥0.40 | ≥58 | 0.918 | 0.022 | 0.046 | 0.014 | S | 2.7 | 6.5 |
| 0942 | Gelfond | 2457690.38513 | 0:48 | 16.0 | 18.52 | 0.05±0.04 | 0.02±0.06 | ≥0.25 | ≥49 | 0.532 | 0.155 | 0.298 | 0.015 | S | 2.2 | 6.9 |
| 0943 | Gringauz | 2457691.46127 | 1:01 | 15.0 | 16.90 | 0.01±0.04 | 0.03±0.05 | ≥0.29 | ≥61 | 0.253 | 0.444 | 0.293 | 0.010 | - | 2.2 | 4.9 |
| 0944 | Gryphia | 2457724.28943 | 1:08 | 11.6 | 14.78 | 0.08±0.02 | 0.12±0.03 | ≥0.03 | ≥68 | 0.984 | 0.001 | 0.000 | 0.015 | S | 2.2 | 3.8 |
| 0945 | Hempel | 2457724.34161 | 0:57 | 15.0 | 19.05 | 0.05±0.08 | 0.10±0.07 | ≥0.24 | ≥58 | 0.556 | 0.267 | 0.048 | 0.129 | S | 2.4 | 0.3 |
| 0946 | Hirayama | 2457768.50255 | 2:00 | 11.0 | 15.47 | -0.09±0.03 | -0.09±0.03 | ≥0.21 | ≥120 | 0.000 | 0.007 | 0.993 | 0.000 | C | 3.1 | 12.5 |
| 0947 | Houzeau | 2457768.50564 | 1:55 | 11.4 | 17.13 | -0.01±0.04 | -0.03±0.04 | ≥0.11 | ≥116 | 0.092 | 0.205 | 0.703 | 0.000 | C | 3.1 | 0.8 |
| 0948 | Ienli | 2457802.42659 | 1:45 | 15.4 | 19.30 | 0.06±0.02 | 0.02±0.03 | ≥0.10 | ≥106 | 0.864 | 0.010 | 0.126 | 0.000 | S | 2.3 | 5.3 |
| 0949 | Iguanodon | 2457798.50513 | 1:09 | 14.8 | 17.16 | 0.07±0.03 | 0.09±0.03 | ≥0.04 | ≥69 | 0.855 | 0.116 | 0.001 | 0.028 | S | 2.3 | 5.4 |
| 0950 | Jamesjones | 2457724.34310 | 0:55 | 13.5 | 18.34 | -0.05±0.04 | -0.08±0.04 | ≥0.05 | ≥55 | 0.001 | 0.057 | 0.942 | 0.000 | C | 3.1 | 0.2 |
| 0951 | Jandecleir | 2457724.28943 | 1:08 | 14.7 | 18.84 | 0.17±0.02 | 0.19±0.04 | 0.89±0.04 | 136±33 | 1.000 | 0.000 | 0.000 | 0.000 | S | 2.6 | 3.1 |
| 0952 | Juliegrady | 2457691.37970 | 1:54 | 14.6 | 19.25 | 0.07±0.03 | 0.11±0.03 | ≥0.42 | ≥114 | 0.861 | 0.084 | 0.000 | 0.055 | S | 2.7 | 2.8 |
| 0953 | Junepatterson | 2457692.38892 | 1:20 | 12.0 | 16.50 | -0.06±0.02 | -0.08±0.03 | ≥0.04 | ≥80 | 0.000 | 0.011 | 0.989 | 0.000 | C | 3.2 | 1.5 |
| 0954 | Karelzeman | 2457722.38400 | 1:47 | 12.9 | 18.15 | 0.07±0.02 | 0.18±0.03 | 0.60±0.04 | 214±52 | 0.680 | 0.001 | 0.000 | 0.319 | S | 3.2 | 16.3 |
| 0955 | Katherynshi | 2457691.46127 | 2:00 | 15.5 | 19.24 | 0.12±0.04 | 0.26±0.04 | ≥0.12 | ≥121 | 0.870 | 0.000 | 0.000 | 0.130 | S | 2.6 | 6.0 |
| 0956 | Katuhikoikeda | 2457692.44645 | 1:00 | 13.9 | 16.49 | 0.06±0.03 | 0.11±0.05 | ≥0.08 | ≥60 | 0.734 | 0.137 | 0.007 | 0.122 | S | 2.2 | 6.4 |
| 0957 | Kaztaniguchi | 2457691.46127 | 2:00 | 14.9 | 18.65 | 0.07±0.02 | 0.14±0.03 | ≥0.14 | ≥121 | 0.833 | 0.018 | 0.000 | 0.149 | S | 2.7 | 7.0 |
| 0958 | Kentrobinson | 2457802.50174 | 0:55 | 14.7 | 19.84 | -0.06±0.07 | -0.16±0.09 | ≥0.21 | ≥55 | 0.005 | 0.034 | 0.960 | 0.000 | C | 3.1 | 9.1 |
| 0959 | Kravtsov | 2457691.37970 | 1:54 | 13.5 | 16.10 | -0.08±0.01 | -0.10±0.02 | ≥0.21 | ≥114 | 0.000 | 0.000 | 1.000 | 0.000 | C | 2.6 | 6.6 |
| 0960 | Kubrick | 2457724.28943 | 1:08 | 13.9 | 17.94 | 0.14±0.02 | 0.11±0.03 | ≥0.08 | ≥68 | 1.000 | 0.000 | 0.000 | 0.000 | S | 2.4 | 7.5 |
| 0961 | Lem | 2457692.44645 | 1:00 | 13.9 | 16.76 | 0.09±0.04 | 0.22±0.05 | ≥0.03 | ≥60 | 0.599 | 0.003 | 0.000 | 0.398 | S | 2.2 | 2.0 |
| 0962 | Lifeson | 2457798.46669 | 0:53 | 13.4 | 17.75 | 0.08±0.03 | 0.13±0.04 | ≥0.11 | ≥53 | 0.818 | 0.043 | 0.001 | 0.138 | S | 2.6 | 12.3 |
| 0963 | Lindsayleona | 2457692.44645 | 1:00 | 15.8 | 17.30 | 0.06±0.04 | 0.14±0.04 | ≥0.11 | ≥60 | 0.633 | 0.096 | 0.002 | 0.270 | S | 2.2 | 3.0 |
| 0964 | Lochner | 2457692.38892 | 1:20 | 14.8 | 17.87 | 0.06±0.02 | 0.09±0.03 | ≥0.13 | ≥80 | 0.892 | 0.098 | 0.000 | 0.011 | S | 2.6 | 3.4 |

*Table 1 continued on next page*

**Table 1** *(continued)*

| No.[a] | Object | Obs. Start | Obs. Duration | H[b] | V | V-R[c] | V-I[c] | Amplitude[d] | Rot. Period[d] | S-type | X-type | C-type | D-type | Tax. | a[b] | i[b] |
|---|---|---|---|---|---|---|---|---|---|---|---|---|---|---|---|---|
| | | (JD) | (h:mm) | (mag) | (mag) | (mag) | (mag) | (mag) | (min) | (probability) | | | | | (au) | (deg) |
| 0965 | Mapihsia | 2457691.46127 | 2:00 | 15.8 | 18.66 | -0.06±0.03 | -0.14±0.03 | ≥0.08 | ≥121 | 0.000 | 0.000 | 1.000 | 0.000 | C | 3.0 | 6.3 |
| 0966 | Matebezdek | 2457724.34161 | 0:57 | 15.2 | 18.77 | 0.03±0.05 | 0.11±0.07 | ≥0.14 | ≥58 | 0.374 | 0.308 | 0.047 | 0.272 | - | 2.3 | 3.0 |
| 0967 | McLaughlin | 2457724.38316 | 1:08 | 12.9 | 15.35 | 0.13±0.02 | 0.17±0.03 | ≥0.04 | ≥69 | 1.000 | 0.000 | 0.000 | 0.000 | S | 2.3 | 7.3 |
| 0968 | Miahelena | 2457766.49959 | 1:05 | 11.7 | 16.50 | 0.09±0.03 | 0.11±0.04 | ≥0.50 | ≥66 | 0.933 | 0.027 | 0.000 | 0.040 | S | 2.8 | 9.6 |
| 0969 | Miahelena | 2457768.50255 | 2:00 | 11.7 | 16.14 | 0.09±0.03 | 0.13±0.04 | 0.65±0.06 | 241±51 | 0.898 | 0.023 | 0.000 | 0.079 | S | 2.8 | 9.6 |
| 0970 | Mikelin | 2457724.38316 | 1:08 | 15.0 | 18.92 | -0.04±0.02 | -0.02±0.04 | ≥0.07 | ≥69 | 0.001 | 0.402 | 0.598 | 0.000 | C | 2.7 | 1.9 |
| 0971 | Mikewagner | 2457692.44645 | 1:00 | 14.5 | 18.10 | 0.06±0.04 | 0.08±0.05 | ≥0.13 | ≥60 | 0.710 | 0.186 | 0.053 | 0.051 | S | 2.7 | 2.4 |
| 0972 | Miles | 2457730.42354 | 1:25 | 11.7 | 16.16 | 0.05±0.04 | 0.10±0.05 | ≥0.05 | ≥86 | 0.645 | 0.198 | 0.020 | 0.137 | S | 2.9 | 13.0 |
| 0973 | Minitti | 2457802.50174 | 0:55 | 11.7 | 17.31 | 0.03±0.02 | 0.15±0.04 | 0.41±0.04 | 111±21 | 0.353 | 0.166 | 0.001 | 0.480 | - | 3.0 | 10.3 |
| 0974 | Negrelli | 2457692.38892 | 1:20 | 14.5 | 17.92 | 0.08±0.02 | 0.15±0.03 | ≥0.15 | ≥80 | 0.907 | 0.001 | 0.000 | 0.092 | S | 2.7 | 2.8 |
| 0975 | Nickthomas | 2457796.48277 | 0:41 | 13.8 | 18.18 | 0.12±0.12 | 0.18±0.13 | ≥0.14 | ≥41 | 0.689 | 0.074 | 0.034 | 0.202 | S | 2.2 | 4.9 |
| 0976 | Noelleoas | 2457802.50174 | 0:55 | 14.2 | 18.07 | 0.07±0.03 | -0.03±0.04 | ≥0.19 | ≥55 | 0.432 | 0.003 | 0.566 | 0.000 | C | 2.4 | 6.9 |
| 0977 | Okabayashi | 2457691.42249 | 0:52 | 13.7 | 16.88 | 0.06±0.02 | 0.08±0.03 | ≥0.10 | ≥53 | 0.860 | 0.125 | 0.004 | 0.011 | S | 2.3 | 0.8 |
| 0978 | Pauls | 2457690.33738 | 1:05 | 16.4 | 20.01 | 0.02±0.04 | 0.04±0.05 | ≥0.17 | ≥66 | 0.365 | 0.370 | 0.248 | 0.017 | - | 2.7 | 12.3 |
| 0979 | Pepikzicha | 2457798.46669 | 0:53 | 13.8 | 17.88 | 0.04±0.03 | 0.11±0.05 | ≥0.10 | ≥53 | 0.527 | 0.291 | 0.017 | 0.165 | S | 2.8 | 4.3 |
| 0980 | Peteworden | 2457690.33738 | 1:57 | 15.3 | 19.21 | -0.05±0.07 | -0.01±0.03 | 0.45±0.07 | 151±38 | 0.073 | 0.371 | 0.555 | 0.000 | C | 2.4 | 3.2 |
| 0981 | Petruskoning | 2457802.50174 | 0:55 | 13.8 | 18.15 | 0.01±0.03 | 0.13±0.04 | ≥0.13 | ≥55 | 0.203 | 0.439 | 0.004 | 0.354 | - | 2.6 | 11.1 |
| 0982 | Pohjola | 2457800.44661 | 0:54 | 12.3 | 17.22 | 0.09±0.04 | 0.17±0.06 | ≥0.12 | ≥54 | 0.757 | 0.017 | 0.000 | 0.226 | S | 2.6 | 12.4 |
| 0983 | Rawls | 2457724.28943 | 1:08 | 13.7 | 18.53 | -0.06±0.02 | -0.06±0.03 | ≥0.08 | ≥68 | 0.000 | 0.131 | 0.870 | 0.000 | C | 2.9 | 14.2 |
| 0984 | Rossini | 2457802.42659 | 1:45 | 13.0 | 17.17 | -0.10±0.02 | -0.07±0.03 | ≥0.31 | ≥106 | 0.000 | 0.015 | 0.985 | 0.000 | C | 2.7 | 5.1 |
| 0985 | Saint-Veran | 2457692.44645 | 1:00 | 14.9 | 18.52 | 0.04±0.04 | 0.14±0.04 | ≥0.13 | ≥60 | 0.463 | 0.175 | 0.001 | 0.360 | - | 2.7 | 4.1 |
| 0986 | Sanashareef | 2457796.48277 | 0:51 | 14.5 | 18.33 | 0.01±0.10 | 0.08±0.10 | ≥0.54 | ≥52 | 0.356 | 0.317 | 0.152 | 0.175 | - | 2.3 | 3.8 |
| 0987 | Sandralitvin | 2457724.28943 | 1:08 | 15.2 | 19.73 | 0.10±0.05 | 0.12±0.06 | ≥0.35 | ≥68 | 0.834 | 0.047 | 0.010 | 0.110 | S | 2.4 | 3.1 |
| 0988 | Sherwoodrowland | 2457798.52750 | 0:37 | 13.5 | 18.19 | 0.09±0.06 | 0.13±0.06 | ≥0.44 | ≥37 | 0.747 | 0.094 | 0.009 | 0.150 | S | 3.1 | 9.1 |
| 0989 | Shipka | 2457800.27321 | 1:01 | 12.2 | 18.01 | 0.02±0.03 | 0.03±0.04 | ≥0.05 | ≥61 | 0.339 | 0.373 | 0.284 | 0.004 | - | 3.0 | 10.1 |
| 0990 | Subramanian | 2457724.34161 | 0:57 | 15.3 | 19.95 | 0.06±0.07 | -0.15±0.10 | ≥0.33 | ≥58 | 0.117 | 0.017 | 0.865 | 0.001 | C | 2.6 | 0.4 |
| 0991 | Suvanto | 2457691.37970 | 0:59 | 12.0 | 16.00 | 0.10±0.02 | 0.16±0.03 | ≥0.23 | ≥60 | 0.960 | 0.000 | 0.000 | 0.040 | S | 2.7 | 13.4 |
| 0992 | Suzaku | 2457724.38316 | 0:35 | 14.9 | 18.62 | 0.04±0.03 | 0.06±0.05 | ≥0.27 | ≥36 | 0.570 | 0.281 | 0.111 | 0.038 | S | 2.4 | 2.3 |
| 0993 | Tjeerd | 2457690.33738 | 1:57 | 13.6 | 17.57 | 0.09±0.02 | 0.11±0.05 | 0.28±0.04 | 235±48 | 0.980 | 0.001 | 0.001 | 0.019 | S | 2.8 | 10.7 |
| 0994 | Unilandes | 2457692.38892 | 1:20 | 13.5 | 16.73 | -0.07±0.02 | -0.10±0.03 | ≥0.09 | ≥80 | 0.000 | 0.003 | 0.997 | 0.000 | C | 3.2 | 2.4 |
| 0995 | Vadimsimona | 2457796.45814 | 0:33 | 13.0 | 15.89 | 0.00±0.04 | 0.07±0.06 | ≥0.03 | ≥33 | 0.257 | 0.513 | 0.142 | 0.088 | X | 2.7 | 10.4 |
| 0996 | Vass | 2457796.44027 | 0:33 | 14.5 | 18.89 | -0.00±0.16 | 0.10±0.14 | ≥0.26 | ≥34 | 0.342 | 0.256 | 0.195 | 0.207 | - | 2.6 | 14.7 |
| 0997 | Venezuela | 2457768.50255 | 2:00 | 13.7 | 18.60 | 0.11±0.09 | 0.18±0.09 | ≥0.37 | ≥120 | 0.691 | 0.065 | 0.009 | 0.235 | S | 2.9 | 1.0 |
| 0998 | Viseggi | 2457730.42038 | 1:30 | 14.9 | 19.70 | -0.02±0.05 | -0.01±0.05 | ≥0.47 | ≥90 | 0.104 | 0.324 | 0.570 | 0.002 | C | 3.0 | 8.2 |
| 0999 | Waynedwards | 2457692.44645 | 1:00 | 14.4 | 19.08 | -0.07±0.04 | -0.04±0.06 | ≥0.08 | ≥60 | 0.005 | 0.322 | 0.672 | 0.000 | C | 3.1 | 7.0 |
| 1000 | Weihai | 2457692.44645 | 1:00 | 16.5 | 19.90 | 0.07±0.05 | 0.06±0.06 | ≥0.09 | ≥60 | 0.683 | 0.160 | 0.115 | 0.043 | S | 2.6 | 2.6 |
| 1001 | Wikipedia | 2457730.43806 | 1:02 | 16.9 | 20.31 | 0.11±0.08 | 0.06±0.08 | ≥0.30 | ≥63 | 0.782 | 0.074 | 0.098 | 0.046 | S | 2.4 | 6.7 |
| 1002 | Witt | 2457728.49321 | 0:53 | 11.9 | 15.88 | 0.11±0.04 | 0.13±0.08 | ≥0.10 | ≥53 | 0.891 | 0.014 | 0.013 | 0.082 | S | 2.8 | 6.5 |
| 1003 | Wumengchao | 2457692.38892 | 1:20 | 14.2 | 17.08 | -0.05±0.02 | -0.07±0.03 | ≥0.25 | ≥80 | 0.000 | 0.036 | 0.964 | 0.000 | C | 2.6 | 13.6 |

[a] This table has 1003 entries but there are four MBAs that were by chance observed twice. They are 1981 SU2, 1999 RE190, 2000 UM3 and Miahelena. See Section 6.4 for more detail on duplicate observations.

[b] $H$ magnitude, semi-major axis ($a$) and orbital inclination ($i$) were obtained from https://ssd.jpl.nasa.gov/horizons.cgi

[c] Colors have been corrected for solar colors by subtracting the respective $V-R = 0.41$ and $V-I = 0.75$ solar colors (Binney & Merrifield 1998).

[d] Lower limit shown in cases where observational duration was insufficient to observe entire light-curve period